\documentclass[11pt]{article}
\usepackage{amsfonts}
\usepackage{amsthm}
\usepackage{amsmath}
\usepackage{amscd}
\usepackage{anyfontsize}
\usepackage{tabularx,booktabs}
\usepackage[mathscr]{eucal}
\usepackage{graphicx}
\usepackage{graphics}
\usepackage{pict2e}
\usepackage{epic}
\usepackage{siunitx}
\usepackage{standalone}
\usepackage{subcaption}
\usepackage[title]{appendix}

\numberwithin{equation}{section}

\usepackage{algorithm}
\usepackage{algorithmicx}
\usepackage{algpseudocode}

\usepackage[utf8]{inputenc}
\usepackage[T1]{fontenc}
\usepackage{ae}
\usepackage{aecompl}
\usepackage{lmodern}
\usepackage[canadian]{babel}
\usepackage[babel=true]{microtype}
\usepackage{scrextend}
\usepackage[twoside=semi, DIV=default, headinclude]{typearea}

\usepackage{epstopdf} 
\usepackage{mathtools}
\usepackage{commath}
\usepackage{amssymb}  

\usepackage{newtxmath}
\usepackage{float}
\usepackage{amstext} % for \text macro
\usepackage{array}   % for \newcolumntype macro

\usepackage[dvipsnames,svgnames]{xcolor}
\definecolor{darkblue}{rgb}{0.0,0.0,0.6}
\definecolor{dc143c}{RGB}{220, 20, 60}
\definecolor{ba55d3}{RGB}{186, 85, 211}
\definecolor{4169e1}{RGB}{65, 105, 225}
\usepackage[pdflang=en-UK, colorlinks, urlcolor=darkblue, linkcolor=darkblue, citecolor=darkblue]{hyperref}

\usepackage{cleveref} % for referencing 
\newcolumntype{L}{>{$}l<{$}} % math-mode version of "l" column type

\usepackage{tikz-cd}
\usetikzlibrary{quantikz2}
\usepackage[round]{natbib} 
\usepackage{fullpage}

\tikzstyle{mutable}=[inner sep=0.5mm,circle,draw,minimum size=2mm]
\tikzstyle{frozen}=[inner sep=0.5mm,rectangle,draw]
\tikzstyle{marked}=[inner sep=0.5mm,circle,draw,fill=black!50]

\usetikzlibrary{arrows,decorations.pathmorphing,backgrounds,positioning,fit,trees,shapes,arrows.meta}

\usepackage{xcolor}
\usepackage[dvipsnames]{xcolor}

\definecolor{defcolour}{rgb}{0.6,0.3,1}

\definecolor{MC1}{HTML}{4ed1c8}

\newcommand{\db}[1]{\textcolor{black}{#1}} %colour for Bob
\newcommand{\dB}[1]{\textcolor{black}{#1}} %colour for Bob
\makeatletter
\renewcommand{\boxed}[1]{\text{ \kern 0.01em \fboxsep=.2em\fbox{\m@th$\displaystyle#1$}}}
\makeatother

\theoremstyle{plain}
\newtheorem{Th}{Theorem}[section]

\theoremstyle{definition}
\newtheorem{Def}[Th]{Definition}

\algblockdefx[MuliIf]{MultiIf}{EndMultiIf}[0]{\textbf{if} $($}{$)$}
\algcblockdefx[MultiThen]{MultiIf}{MultiThen}{EndMultiThen}{$)$ $\{$}{$\}$}

\newcommand{\DCC}{DisCoCirc}

\title{Efficient Generation of Parameterised Quantum Circuits\\from Large Texts}
\author{
Colin Krawchuk$^*$~~~
Nikhil Khatri$^*$~~~         
Neil John Ortega~~~
Dimitri Kartsaklis\vspace{0.2cm} \\ 
\small{($^*$Contributed equally to this work)}\vspace{0.2cm} \\
\textit{Quantinuum}\\
\textit{17 Beaumont Street, Oxford, OX1 2NA, UK}\vspace{0.2cm} \\
\texttt{\normalsize \{firstname.lastname\}@quantinuum.com} \\ }
        
\begin{document}

\date{}
\maketitle

\begin{abstract}
Quantum approaches to natural language processing (NLP) are redefining how linguistic information is represented and processed. While traditional hybrid quantum-classical models rely heavily on classical neural networks, recent advancements propose a novel framework, DisCoCirc, capable of directly encoding entire documents as parameterised quantum circuits (PQCs), besides enjoying some additional interpretability and compositionality benefits. 

Following these ideas, this paper introduces an efficient methodology for converting large-scale texts into quantum circuits using tree-like representations of pregroup diagrams. Exploiting the compositional parallels between language and quantum mechanics, grounded in symmetric monoidal categories, our approach enables faithful and efficient encoding of syntactic and discourse relationships in long and complex texts (up to 6410 words in our experiments) to quantum circuits. The developed system is provided to the community as part of the \db{augmented} open-source quantum NLP package \db{\texttt{lambeq} Gen II}.
\end{abstract}

%%%%%%%%%%%%%%%%%%%
\section{Introduction}
%%%%%%%%%%%%%%%%%%%

Quantum approaches in natural language processing (NLP) are emerging as a transformative paradigm, leveraging principles from quantum mechanics to address complex linguistic challenges \citep{GrammarAware, Lorenz_2023, mehrnoosh}.  Compositional models of this form are capable of encoding sentences into trainable quantum circuits, introducing a paradigm shift from traditional hybrid quantum machine learning pipelines, which often rely on large classical neural network components for the bulk of their processing. They also have some other unique advantages when compared to black-box statistical approaches, such as compatibility and scalability to quantum hardware and interpretability \citep{tull}, while at the same time provide the required theoretical depth for interesting linguistic analysis---see, for example, \cite{bankova_2019}. Recent research extends these advances to the discourse level, presenting a framework (referred to as \textit{DisCoCirc} -- DIStributional COmpositional CIRCuits) for encoding entire documents into parameterised quantum circuits (PQCs), ready to run and train on a quantum computer \citep{discocirc, Duneau2024ScalableAI}, \db{with the additional promise of enhanced performance on quantum hardware \citep{Laakkonen_2024}}.

%The aim of Natural Language Processing (NLP) is to enable computers to understand, interpret, and generate human language. The employment of machine learning methods in this field, including large language models and neural networks, has led to significant advancements in NLP applications such as translation, sentiment analysis, speech recognition, and conversational agents \ck{Add citations}\citep{}. Recently, new approaches to natural language processing using quantum circuits have allowed these computationally intensive tasks to be performed on quantum computers \citep{GrammarAware, Lorenz_2023, Duneau2024ScalableAI, lambeq, Quixer}. 

Generating a large-scale quantum circuit that encodes all the necessary syntactic and discourse relationships to faithfully represent an arbitrary real-world document entails significant technical challenges that may affect performance and coverage. In this article we introduce a methodology, as well as a tool, to address these issues with efficiency. In particular, we describe an efficient algorithm for converting large-scale texts into structured parameterised quantum circuits using tree-like representations of pregroup diagrams \citep{lambek}, which we call \emph{pregroup trees}. These grammatical structures allow us to capitalise on the compositional nature of the DisCoCirc framework, which is shared between language and quantum mechanics due to the rigorous mathematical foundation of symmetric monoidal categories and string diagrams. The result is an efficient encoding of large-scale texts as quantum circuits that can be trained and composed to solve NLP tasks.

% In this article, we introduce a methodology to support the implementation and training of natural language experiments using quantum circuits. In particular, we describe an efficient algorithm for converting long texts into structured parameterised quantum circuits (PQCs) using tree-like representations of pregroup diagrams \citep{lambek}, which we call \emph{pregroup trees}. These grammatical structures allow us to capitalise on a shared compositional model for language and quantum mechanics using symmetric monoidal categories \citep{discocirc}. The result is a natural encoding of text as quantum circuits that can be trained and composed to form large scale language models. \nk{todo neaten}

To make DisCoCirc models publicly available to the quantum computing research community, we rely on the structures and components provided by the open-source Quantum NLP (QNLP) toolkit \texttt{lambeq} \citep{lambeq}\footnote{\url{https://docs.quantinuum.com/lambeq}.}, which until now has been used for generating and training sentence-level QNLP models. With this work, we introduce a new \texttt{lambeq} module,\footnote{\texttt{lambeq.experimental.discocirc}.} which exploits the compositional nature of the underlying models to construct large-scale text circuits from the derivations of the individual sentences. 

We show that our algorithm outperforms existing approaches \citep{jono} based on Combinatory Categorial Grammar (CCG) \citep{syntactic_process} derivations by orders of magnitude, both in language coverage and efficiency. Specifically, our crash tests demonstrate that the new methodology is capable of successfully parsing documents as long as 6410 words (13 standard 500-word pages of text) into quantum circuit form. In addition, we provide utilities to support language-based experiments including semantic rewrites, state filtering and ansatz-dependent conversion to quantum circuits. The interested reader can also find in Appendix \ref{sec:appendix-training} a proof-of-concept training example for a small-scale NLP task.

% The code introduced in this paper presents the necessary tools to design and implement NLP experiments on discourse using quantum circuits, and is available as part of the Quantum NLP toolkit lambeq \citep{lambeq}\footnote{\url{https://docs.quantinuum.com/lambeq}.}.

The paper is organised as follows: Section \ref{sec:background} summarises the necessary background on pregroup grammars and {\DCC} that forms the theoretical basis for our work. In Section \ref{sec:text2diagram} we outline the procedure for converting text into diagrams. First we introduce a novel pregroup parsing method for extracting grammatical structure from text. Then we describe how these individual trees can be translated into local circuits and stitched together into a {\DCC} diagram. In order to apply these circuits to QNLP experiments, we implement several useful tools for constructing trainable PQCs, which are discussed in Section \ref{sec:simplifications}. Sections \ref{subsec:sandwich} and \ref{subsec:PQCs} describe the final steps for generating quantum circuits from the text. In Section \ref{sec:results}, we provide coverage and efficiency statistics on our algorithm by testing it on a natural language dataset. Additionally, we demonstrate its ability to generate large-scale circuits and perform comparisons against an existing model. Finally, Section \ref{sec:conclusion} discusses future work and conclusions.

\section{Background}\label{sec:background}
%%%%%%%%%%%%%%%%%%%

This section provides a short introduction to the DisCoCirc framework, as well as a summary of some important existing implementation approaches. For a more detailed description, please refer to the original papers provided in the text.

\subsection{DisCoCirc}

\begin{figure}[h]
\centering
\resizebox{!}{0.20\textwidth}{
    % When embedding into a *.tex file, uncomment and include the following lines:
\pgfdeclarelayer{nodelayer}
\pgfdeclarelayer{edgelayer}
\pgfdeclarelayer{labellayer}
\pgfsetlayers{nodelayer, edgelayer, labellayer}
\begin{tikzpicture}
	\begin{pgfonlayer}{nodelayer}
		\node [] (0) at (0, 0.5) {};
		\node [] (1) at (-0.5, 1.75) {};
		\node [] (2) at (0.5, 1.75) {};
		\node [] (3) at (0.5, 2.25) {};
		\node [] (4) at (-0.5, 2.25) {};
		\node [] (6) at (2, 1.75) {};
		\node [] (7) at (3, 1.75) {};
		\node [] (8) at (3, 2.25) {};
		\node [] (9) at (2, 2.25) {};
		\node [] (11) at (-0.5, -0.5) {};
		\node [] (12) at (3, -0.5) {};
		\node [] (13) at (3, 1.25) {};
		\node [] (14) at (-0.5, 1.25) {};
		\node [] (16) at (0.25, -0.25) {};
		\node [] (17) at (2.25, -0.25) {};
		\node [] (18) at (2.25, 0.75) {};
		\node [] (19) at (0.25, 0.75) {};
		\node [] (20) at (0.75, 0) {};
		\node [] (21) at (1.75, 0) {};
		\node [] (22) at (1.75, 0.5) {};
		\node [] (23) at (0.75, 0.5) {};
		\node [] (25) at (0, 1.75) {};
		\node [] (26) at (0, 1.25) {};
		\node [] (27) at (2.5, 1.75) {};
		\node [] (28) at (2.5, 1.25) {};
		\node [] (29) at (1.25, 0.75) {};
		\node [] (30) at (1.25, 0.5) {};
		\node [] (31) at (1.25, 0) {};
		\node [] (32) at (1.25, -0.25) {};
		\node [] (33) at (0, -0.5) {};
		\node [] (34) at (0, -1.25) {};
		\node [] (35) at (2.5, -0.5) {};
		\node [] (36) at (2.5, -1.25) {};
	\end{pgfonlayer}
	\begin{pgfonlayer}{edgelayer}
		\draw [-, fill={rgb,255: red,224; green,224; blue,224}, line width=1.0pt] (1.center)
			 to (2.center)
			 to (3.center)
			 to (4.center)
			 to cycle;
		\draw [-, fill={rgb,255: red,224; green,224; blue,224}, line width=1.0pt] (6.center)
			 to (7.center)
			 to (8.center)
			 to (9.center)
			 to cycle;
		\draw [-, fill={rgb,255: red,251; green,232; blue,231}, line width=1.0pt] (11.center)
			 to (12.center)
			 to (13.center)
			 to (14.center)
			 to cycle;
		\draw [-, fill={rgb,255: red,255; green,255; blue,255}, line width=1.0pt] (16.center)
			 to (17.center)
			 to (18.center)
			 to (19.center)
			 to cycle;
		\draw [-, fill={rgb,255: red,224; green,224; blue,224}, line width=1.0pt] (20.center)
			 to (21.center)
			 to (22.center)
			 to (23.center)
			 to cycle;
		\draw [-, draw={rgb,255: red,156; green,84; blue,14}, line width=1.0pt, in=90, out=-90] (25.center) to (26.center);
		\draw [-, draw={rgb,255: red,244; green,169; blue,64}, line width=1.0pt, in=90, out=-90] (27.center) to (28.center);
		\draw [-, draw={rgb,255: red,0; green,0; blue,0}, line width=1.0pt, in=90, out=-90] (29.center) to (30.center);
		\draw [-, draw={rgb,255: red,0; green,0; blue,0}, line width=1.0pt, in=90, out=-90] (31.center) to (32.center);
		\draw [-, draw={rgb,255: red,156; green,84; blue,14}, line width=1.0pt, in=90, out=-90] (33.center) to (34.center);
		\draw [-, draw={rgb,255: red,244; green,169; blue,64}, line width=1.0pt, in=90, out=-90] (35.center) to (36.center);
	\end{pgfonlayer}
	\begin{pgfonlayer}{labellayer}
		\node [] (5) at (0, 2) {Alice};
		\node [] (10) at (2.5, 2) {bikes};
		\node [] (15) at (1.25, 1) {loves};
		\node [] (24) at (1.25, 0.25) {fast};
	\end{pgfonlayer}
\end{tikzpicture}
}
\caption{A DisCoCirc diagram for the sentence ``Alice loves fast bikes''.}
\label{fig:diagram}
\end{figure}
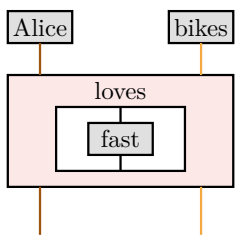

In the DisCoCirc framework \citep{discocirc}, the narrative structure of text is modelled by \textit{string diagrams} generated from states, boxes and frames. The states represent nouns that appear in the text. The information encoded in states is transmitted through the wires in the diagram and acted upon by boxes. These boxes correspond to words in the text that update states or allow them to interact. For example, transitive verbs describe the action of a pair of nouns, while adjectives contextualise individual nouns (Figure \ref{fig:diagram}).

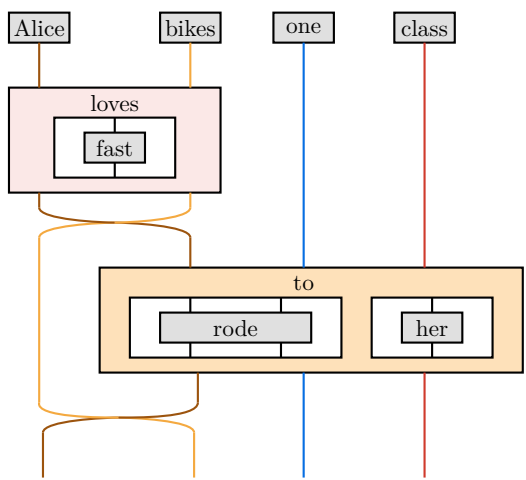
\begin{figure}[h]
\vspace{1cm}
\centering
\resizebox{!}{0.40\textwidth}{
    % When embedding into a *.tex file, uncomment and include the following lines:
\pgfdeclarelayer{nodelayer}
\pgfdeclarelayer{edgelayer}
\pgfdeclarelayer{labellayer}
\pgfsetlayers{nodelayer, edgelayer, labellayer}
\begin{tikzpicture}
	\begin{pgfonlayer}{nodelayer}
		\node [] (0) at (0, 1.75) {};
		\node [] (1) at (-0.5, 7.125) {};
		\node [] (2) at (0.5, 7.125) {};
		\node [] (3) at (0.5, 7.625) {};
		\node [] (4) at (-0.5, 7.625) {};
		\node [] (6) at (2, 7.125) {};
		\node [] (7) at (3, 7.125) {};
		\node [] (8) at (3, 7.625) {};
		\node [] (9) at (2, 7.625) {};
		\node [] (11) at (3.875, 7.125) {};
		\node [] (12) at (4.875, 7.125) {};
		\node [] (13) at (4.875, 7.625) {};
		\node [] (14) at (3.875, 7.625) {};
		\node [] (16) at (5.875, 7.125) {};
		\node [] (17) at (6.875, 7.125) {};
		\node [] (18) at (6.875, 7.625) {};
		\node [] (19) at (5.875, 7.625) {};
		\node [] (21) at (-0.5, 4.625) {};
		\node [] (22) at (3, 4.625) {};
		\node [] (23) at (3, 6.375) {};
		\node [] (24) at (-0.5, 6.375) {};
		\node [] (26) at (0.25, 4.875) {};
		\node [] (27) at (2.25, 4.875) {};
		\node [] (28) at (2.25, 5.875) {};
		\node [] (29) at (0.25, 5.875) {};
		\node [] (30) at (0.75, 5.125) {};
		\node [] (31) at (1.75, 5.125) {};
		\node [] (32) at (1.75, 5.625) {};
		\node [] (33) at (0.75, 5.625) {};
		\node [] (35) at (1, 1.625) {};
		\node [] (36) at (8, 1.625) {};
		\node [] (37) at (8, 3.375) {};
		\node [] (38) at (1, 3.375) {};
		\node [] (40) at (1.5, 1.875) {};
		\node [] (41) at (5, 1.875) {};
		\node [] (42) at (5, 2.875) {};
		\node [] (43) at (1.5, 2.875) {};
		\node [] (44) at (2, 2.125) {};
		\node [] (45) at (4.5, 2.125) {};
		\node [] (46) at (4.5, 2.625) {};
		\node [] (47) at (2, 2.625) {};
		\node [] (49) at (5.5, 1.875) {};
		\node [] (50) at (7.5, 1.875) {};
		\node [] (51) at (7.5, 2.875) {};
		\node [] (52) at (5.5, 2.875) {};
		\node [] (53) at (6, 2.125) {};
		\node [] (54) at (7, 2.125) {};
		\node [] (55) at (7, 2.625) {};
		\node [] (56) at (6, 2.625) {};
		\node [] (58) at (0, 7.125) {};
		\node [] (59) at (0, 6.375) {};
		\node [] (60) at (2.5, 7.125) {};
		\node [] (61) at (2.5, 6.375) {};
		\node [] (62) at (1.25, 5.875) {};
		\node [] (63) at (1.25, 5.625) {};
		\node [] (64) at (1.25, 5.125) {};
		\node [] (65) at (1.25, 4.875) {};
		\node [] (66) at (0, 4.625) {};
		\node [] (67) at (0, 4.375) {};
		\node [] (68) at (2.5, 4.625) {};
		\node [] (69) at (2.5, 4.375) {};
		\node [] (70) at (2.5, 3.875) {};
		\node [] (71) at (2.5, 3.375) {};
		\node [] (72) at (4.375, 7.125) {};
		\node [] (73) at (4.375, 3.375) {};
		\node [] (74) at (6.375, 7.125) {};
		\node [] (75) at (6.375, 3.375) {};
		\node [] (76) at (2.5, 2.875) {};
		\node [] (77) at (2.5, 2.625) {};
		\node [] (78) at (4, 2.875) {};
		\node [] (79) at (4, 2.625) {};
		\node [] (80) at (2.5, 2.125) {};
		\node [] (81) at (2.5, 1.875) {};
		\node [] (82) at (4, 2.125) {};
		\node [] (83) at (4, 1.875) {};
		\node [] (84) at (6.5, 2.875) {};
		\node [] (85) at (6.5, 2.625) {};
		\node [] (86) at (6.5, 2.125) {};
		\node [] (87) at (6.5, 1.875) {};
		\node [] (88) at (0, 3.875) {};
		\node [] (89) at (0, 1.125) {};
		\node [] (90) at (2.625, 1.625) {};
		\node [] (91) at (2.625, 1.125) {};
		\node [] (92) at (0.0625, 0.625) {};
		\node [] (93) at (0.0625, -0.125) {};
		\node [] (94) at (2.5625, 0.625) {};
		\node [] (95) at (2.5625, -0.125) {};
		\node [] (96) at (4.375, 1.625) {};
		\node [] (97) at (4.375, -0.125) {};
		\node [] (98) at (6.375, 1.625) {};
		\node [] (99) at (6.375, -0.125) {};
		\node [] (100) at (1.25, 4.125) {};
		\node [] (101) at (1.3125, 0.875) {};
	\end{pgfonlayer}
	\begin{pgfonlayer}{edgelayer}
		\draw [-, fill={rgb,255: red,224; green,224; blue,224}, line width=1.0pt] (1.center)
			 to (2.center)
			 to (3.center)
			 to (4.center)
			 to cycle;
		\draw [-, fill={rgb,255: red,224; green,224; blue,224}, line width=1.0pt] (6.center)
			 to (7.center)
			 to (8.center)
			 to (9.center)
			 to cycle;
		\draw [-, fill={rgb,255: red,224; green,224; blue,224}, line width=1.0pt] (11.center)
			 to (12.center)
			 to (13.center)
			 to (14.center)
			 to cycle;
		\draw [-, fill={rgb,255: red,224; green,224; blue,224}, line width=1.0pt] (16.center)
			 to (17.center)
			 to (18.center)
			 to (19.center)
			 to cycle;
		\draw [-, fill={rgb,255: red,251; green,232; blue,231}, line width=1.0pt] (21.center)
			 to (22.center)
			 to (23.center)
			 to (24.center)
			 to cycle;
		\draw [-, fill={rgb,255: red,255; green,255; blue,255}, line width=1.0pt] (26.center)
			 to (27.center)
			 to (28.center)
			 to (29.center)
			 to cycle;
		\draw [-, fill={rgb,255: red,224; green,224; blue,224}, line width=1.0pt] (30.center)
			 to (31.center)
			 to (32.center)
			 to (33.center)
			 to cycle;
		\draw [-, fill={rgb,255: red,254; green,225; blue,186}, line width=1.0pt] (35.center)
			 to (36.center)
			 to (37.center)
			 to (38.center)
			 to cycle;
		\draw [-, fill={rgb,255: red,255; green,255; blue,255}, line width=1.0pt] (40.center)
			 to (41.center)
			 to (42.center)
			 to (43.center)
			 to cycle;
		\draw [-, fill={rgb,255: red,224; green,224; blue,224}, line width=1.0pt] (44.center)
			 to (45.center)
			 to (46.center)
			 to (47.center)
			 to cycle;
		\draw [-, fill={rgb,255: red,255; green,255; blue,255}, line width=1.0pt] (49.center)
			 to (50.center)
			 to (51.center)
			 to (52.center)
			 to cycle;
		\draw [-, fill={rgb,255: red,224; green,224; blue,224}, line width=1.0pt] (53.center)
			 to (54.center)
			 to (55.center)
			 to (56.center)
			 to cycle;
		\draw [-, draw={rgb,255: red,156; green,84; blue,14}, line width=1.0pt, in=90, out=-90] (58.center) to (59.center);
		\draw [-, draw={rgb,255: red,244; green,169; blue,64}, line width=1.0pt, in=90, out=-90] (60.center) to (61.center);
		\draw [-, draw={rgb,255: red,0; green,0; blue,0}, line width=1.0pt, in=90, out=-90] (62.center) to (63.center);
		\draw [-, draw={rgb,255: red,0; green,0; blue,0}, line width=1.0pt, in=90, out=-90] (64.center) to (65.center);
		\draw [-, draw={rgb,255: red,156; green,84; blue,14}, line width=1.0pt, in=90, out=-90] (66.center) to (67.center);
		\draw [-, draw={rgb,255: red,244; green,169; blue,64}, line width=1.0pt, in=90, out=-90] (68.center) to (69.center);
		\draw [-, draw={rgb,255: red,156; green,84; blue,14}, line width=1.0pt, in=90, out=-90, looseness=0.75] (70.center) to (71.center);
		\draw [-, draw={rgb,255: red,6; green,110; blue,226}, line width=1.0pt, in=90, out=-90] (72.center) to (73.center);
		\draw [-, draw={rgb,255: red,208; green,59; blue,45}, line width=1.0pt, in=90, out=-90] (74.center) to (75.center);
		\draw [-, draw={rgb,255: red,0; green,0; blue,0}, line width=1.0pt, in=90, out=-90] (76.center) to (77.center);
		\draw [-, draw={rgb,255: red,0; green,0; blue,0}, line width=1.0pt, in=90, out=-90] (78.center) to (79.center);
		\draw [-, draw={rgb,255: red,0; green,0; blue,0}, line width=1.0pt, in=90, out=-90] (80.center) to (81.center);
		\draw [-, draw={rgb,255: red,0; green,0; blue,0}, line width=1.0pt, in=90, out=-90] (82.center) to (83.center);
		\draw [-, draw={rgb,255: red,0; green,0; blue,0}, line width=1.0pt, in=90, out=-90] (84.center) to (85.center);
		\draw [-, draw={rgb,255: red,0; green,0; blue,0}, line width=1.0pt, in=90, out=-90] (86.center) to (87.center);
		\draw [-, draw={rgb,255: red,244; green,169; blue,64}, line width=1.0pt, in=90, out=-90] (88.center) to (89.center);
		\draw [-, draw={rgb,255: red,156; green,84; blue,14}, line width=1.0pt, in=90, out=-90] (90.center) to (91.center);
		\draw [-, draw={rgb,255: red,156; green,84; blue,14}, line width=1.0pt, in=90, out=-90] (92.center) to (93.center);
		\draw [-, draw={rgb,255: red,244; green,169; blue,64}, line width=1.0pt, in=90, out=-90] (94.center) to (95.center);
		\draw [-, draw={rgb,255: red,6; green,110; blue,226}, line width=1.0pt, in=90, out=-90] (96.center) to (97.center);
		\draw [-, draw={rgb,255: red,208; green,59; blue,45}, line width=1.0pt, in=90, out=-90] (98.center) to (99.center);
		\draw [-, draw={rgb,255: red,244; green,169; blue,64}, line width=1.0pt, in=90, out=180, looseness=0.41] (100.center) to (88.center);
		\draw [-, draw={rgb,255: red,156; green,84; blue,14}, line width=1.0pt, in=90, out=0, looseness=0.41] (100.center) to (70.center);
		\draw [-, draw={rgb,255: red,156; green,84; blue,14}, line width=1.0pt, in=180, out=-90, looseness=0.41] (67.center) to (100.center);
		\draw [-, draw={rgb,255: red,244; green,169; blue,64}, line width=1.0pt, in=0, out=-90, looseness=0.41] (69.center) to (100.center);
		\draw [-, draw={rgb,255: red,156; green,84; blue,14}, line width=1.0pt, in=90, out=180, looseness=0.41] (101.center) to (92.center);
		\draw [-, draw={rgb,255: red,244; green,169; blue,64}, line width=1.0pt, in=90, out=0, looseness=0.41] (101.center) to (94.center);
		\draw [-, draw={rgb,255: red,244; green,169; blue,64}, line width=1.0pt, in=180, out=-105, looseness=0.50] (89.center) to (101.center);
		\draw [-, draw={rgb,255: red,156; green,84; blue,14}, line width=1.0pt, in=0, out=-90, looseness=0.50] (91.center) to (101.center);
	\end{pgfonlayer}
    \begin{pgfonlayer}{labellayer}
        \node [] (5) at (0, 7.375) {Alice};
		\node [] (10) at (2.5, 7.375) {bikes};
		\node [] (15) at (4.375, 7.375) {one};
		\node [] (20) at (6.375, 7.375) {class};
		\node [] (25) at (1.25, 6.125) {loves};
		\node [] (34) at (1.25, 5.375) {fast};
		\node [] (39) at (4.375, 3.125) {to};
		\node [] (48) at (3.25, 2.375) {rode};
		\node [] (57) at (6.5, 2.375) {her};
    \end{pgfonlayer}
\end{tikzpicture}
}
\caption{DisCoCirc diagram for the sentence ``Alice loves fast bikes. She rode one to her class.''}
\label{fig:diagram-frames}
\end{figure}

Frames depict higher-order maps that transform boxes, sub-diagrams or other frames. For example, these include adverbs that modify verb boxes, prepositions that group linguistic elements into phrases, and conjunctions that combine diagrams arising from clauses. Just as sentences may be combined to form a text, diagrams can be composed along wires that correspond to the same noun. Functionally, the information conveyed by a DisCoCirc diagram depends only on its connectivity and a temporal ordering corresponding to the narrative progression of the text (Figure \ref{fig:diagram-frames}).

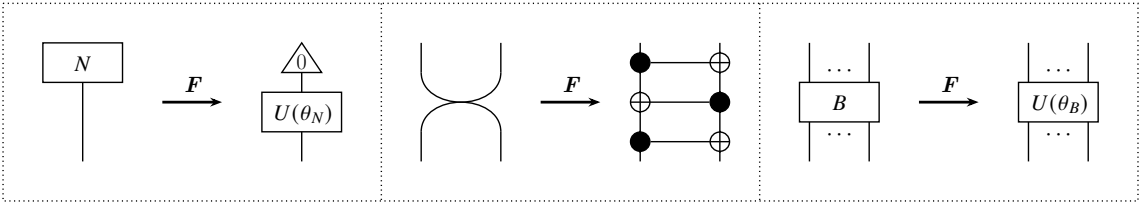
\begin{figure}
\hspace{-1cm}
\centering
\tikzset{every picture/.style={line width=1.0pt, anchor=base}} %set default line width to 0.75pt
\tikzset{every node/.style={node font=\fontsize{16}{16}\selectfont}}

\tikzstyle{box}=[-, fill={rgb,255: red,255; green,255; blue,255}, line width=1.0pt]
\tikzstyle{black_wire}=[-, draw={rgb,255: red,0; green,0; blue,0}, line width=1.0pt]
\tikzstyle{dashed}=[-, dash pattern={on 0.84pt off 2.51pt}, line width=1.0pt]
\tikzstyle{arrow}=[-{Stealth[length=3mm]}, draw={rgb,255: red,0; green,0; blue,0}, line width=2.0pt]
\tikzstyle{black_dot}=[fill=black, draw=black, shape=circle, line width=1.0pt]
\tikzstyle{white_dot}=[fill=white, draw=black, shape=circle, line width=1.0pt]

% When embedding into a *.tex file, uncomment and include the following lines:
\pgfdeclarelayer{nodelayer}
\pgfdeclarelayer{edgelayer}
\pgfdeclarelayer{labellayer}
\pgfsetlayers{nodelayer, edgelayer, labellayer}
\begin{tikzpicture}
	\begin{pgfonlayer}{nodelayer}
		\node [] (1) at (-4.5, 1.5) {};
		\node [] (2) at (-4.5, 0.75) {};
		\node [] (3) at (-2.5, 1.5) {};
		\node [] (4) at (-2.5, 0.75) {};
		\node [] (5) at (-4.5, -0.75) {};
		\node [] (6) at (-4.5, -1.5) {};
		\node [] (7) at (-2.5, -0.75) {};
		\node [] (8) at (-2.5, -1.5) {};
		\node [] (9) at (-3.5, 0) {};
		\node [] (10) at (1, 0) {};
		\node [style={black_dot}] (11) at (1, 1) {};
		\node [style={white_dot}] (12) at (3, 1) {};
		\node [] (13) at (3.4125, 1) {};
		\node [] (14) at (3, 0.75) {};
		\node [] (15) at (3, 1.25) {};
		\node [] (16) at (1, 0.75) {};
		\node [] (17) at (1, 1.25) {};
		\node [style={black_dot}] (18) at (3, 0) {};
		\node [style={white_dot}] (19) at (1, 0) {};
		\node [] (20) at (0.5875, 0) {};
		\node [] (21) at (1, -0.25) {};
		\node [] (22) at (1, 0.25) {};
		\node [] (23) at (3, -0.25) {};
		\node [] (24) at (3, 0.25) {};
		\node [style={black_dot}] (25) at (1, -1) {};
		\node [style={white_dot}] (26) at (3, -1) {};
		\node [] (27) at (3.4125, -1) {};
		\node [] (28) at (3, -1.25) {};
		\node [] (29) at (3, -0.75) {};
		\node [] (30) at (1, -1.25) {};
		\node [] (31) at (1, -0.75) {};
		\node [] (32) at (1, 1.5) {};
		\node [] (33) at (3, 1.5) {};
		\node [] (34) at (1, -1.5) {};
		\node [] (35) at (3, -1.5) {};
		\node [] (36) at (-1.5, 0) {};
		\node [] (37) at (0, 0) {};
		\node [] (38) at (-15, 2.5) {};
		\node [] (39) at (13.5, 2.5) {};
		\node [] (40) at (13.5, -2.5) {};
		\node [] (41) at (-15, -2.5) {};
		\node [] (42) at (-0.5, 0.25) {};
		\node [] (44) at (4, -2.5) {};
		\node [] (45) at (4, 2.5) {};
		\node [] (47) at (5, -0.5) {};
		\node [] (48) at (7, -0.5) {};
		\node [] (49) at (7, 0.5) {};
		\node [] (50) at (5, 0.5) {};
		\node [] (51) at (5.25, 1.5) {};
		\node [] (52) at (5.25, 0.5) {};
		\node [] (53) at (6.75, 1.5) {};
		\node [] (54) at (6.75, 0.5) {};
		\node [] (55) at (5.25, -0.5) {};
		\node [] (56) at (5.25, -1.5) {};
		\node [] (57) at (6.75, -0.5) {};
		\node [] (58) at (6.75, -1.5) {};
		\node [] (60) at (8, 0) {};
		\node [] (61) at (9.5, 0) {};
		\node [] (62) at (9, 0.25) {};
		\node [] (64) at (10.5, -0.5) {};
		\node [] (65) at (12.5, -0.5) {};
		\node [] (66) at (12.5, 0.5) {};
		\node [] (67) at (10.5, 0.5) {};
		\node [] (68) at (10.75, 1.5) {};
		\node [] (69) at (10.75, 0.5) {};
		\node [] (70) at (12.25, 1.5) {};
		\node [] (71) at (12.25, 0.5) {};
		\node [] (72) at (10.75, -0.5) {};
		\node [] (73) at (10.75, -1.5) {};
		\node [] (74) at (12.25, -0.5) {};
		\node [] (75) at (12.25, -1.5) {};
		\node [] (81) at (-5.5, -2.5) {};
		\node [] (82) at (-5.5, 2.5) {};
		\node [] (84) at (-14, 0.5) {};
		\node [] (85) at (-12, 0.5) {};
		\node [] (86) at (-12, 1.5) {};
		\node [] (87) at (-14, 1.5) {};
		\node [] (88) at (-13, 0.5) {};
		\node [] (89) at (-13, -1.5) {};
		\node [] (91) at (-11, 0) {};
		\node [] (92) at (-9.5, 0) {};
		\node [] (93) at (-10.5, 0.25) {};
		\node [] (96) at (-8, 0.75) {};
		\node [] (97) at (-7, 0.75) {};
		\node [] (98) at (-7.5, 1.5) {};
		\node [] (99) at (-8.5, -0.75) {};
		\node [] (100) at (-6.5, -0.75) {};
		\node [] (101) at (-6.5, 0.25) {};
		\node [] (102) at (-8.5, 0.25) {};
		\node [] (111) at (-7.5, 0.75) {};
		\node [] (112) at (-7.5, 0.25) {};
		\node [] (113) at (-7.5, -0.75) {};
		\node [] (114) at (-7.5, -1.5) {};
	\end{pgfonlayer}
	\begin{pgfonlayer}{edgelayer}
		\draw [style={black_wire}, in=90, out=-90] (1.center) to (2.center);
		\draw [style={black_wire}, in=90, out=-90] (3.center) to (4.center);
		\draw [style={black_wire}, in=90, out=-90] (5.center) to (6.center);
		\draw [style={black_wire}, in=90, out=-90] (7.center) to (8.center);
		\draw [style={black_wire}, in=90, out=180] (9.center) to (5.center);
		\draw [style={black_wire}, in=90, out=0] (9.center) to (7.center);
		\draw [style={black_wire}, in=180, out=-90] (2.center) to (9.center);
		\draw [style={black_wire}, in=0, out=-90] (4.center) to (9.center);
		\draw [style={black_wire}, in=90, out=-90] (14.center) to (15.center);
		\draw [style={black_wire}, in=90, out=-90] (16.center) to (17.center);
		\draw [style={black_wire}, in=180, out=0] (11) to (13);
		\draw [style={black_wire}, in=90, out=-90] (21.center) to (22.center);
		\draw [style={black_wire}, in=90, out=-90] (23.center) to (24.center);
		\draw [style={black_wire}, in=0, out=180] (18) to (20);
		\draw [style={black_wire}, in=90, out=-90] (28.center) to (29.center);
		\draw [style={black_wire}, in=90, out=-90] (30.center) to (31.center);
		\draw [style={black_wire}, in=180, out=0] (25) to (27);
		\draw [style={black_wire}, in=90, out=-90] (32.center) to (17.center);
		\draw [style={black_wire}, in=90, out=-90] (33.center) to (15.center);
		\draw [style={black_wire}, in=90, out=-90] (16.center) to (22.center);
		\draw [style={black_wire}, in=90, out=-90] (14.center) to (24.center);
		\draw [style={black_wire}, in=90, out=-90] (21.center) to (31.center);
		\draw [style={black_wire}, in=90, out=-90] (23.center) to (29.center);
		\draw [style={black_wire}, in=90, out=-90] (30.center) to (34.center);
		\draw [style={black_wire}, in=90, out=-90] (28.center) to (35.center);
		\draw [style={arrow}] (36.center) to (37.center);
		\draw [style={dashed}] (38.center)
            to (39.center)
            to (40.center)
            to (41.center)
            to cycle;
		\draw [style={dashed}] (45.center) to (44.center);
		\draw [style={box}] (47.center)
			to (48.center)
			to (49.center)
			to (50.center)
			to cycle;
		\draw [style={black_wire}, in=90, out=-90] (51.center) to (52.center);
		\draw [style={black_wire}, in=90, out=-90] (53.center) to (54.center);
		\draw [style={black_wire}, in=90, out=-90] (55.center) to (56.center);
		\draw [style={black_wire}, in=90, out=-90] (57.center) to (58.center);
		\draw [style={arrow}] (60.center) to (61.center);
		\draw [style={box}] (64.center)
            to (65.center)
            to (66.center)
            to (67.center)
            to cycle;
		\draw [style={black_wire}, in=90, out=-90] (68.center) to (69.center);
		\draw [style={black_wire}, in=90, out=-90] (70.center) to (71.center);
		\draw [style={black_wire}, in=90, out=-90] (72.center) to (73.center);
		\draw [style={black_wire}, in=90, out=-90] (74.center) to (75.center);
		\draw [style={dashed}] (82.center) to (81.center);
		\draw [style={box}] (84.center)
			 to (85.center)
			 to (86.center)
			 to (87.center)
			 to cycle;
		\draw [style={black_wire}, in=90, out=-90] (88.center) to (89.center);
		\draw [style={arrow}] (91.center) to (92.center);
		\draw [style={box}] (96.center)
			 to (97.center)
			 to (98.center)
			 to cycle;
		\draw [style={box}] (99.center)
			 to (100.center)
			 to (101.center)
			 to (102.center)
			 to cycle;
		\draw [style={black_wire}, in=90, out=-90] (111.center) to (112.center);
		\draw [style={black_wire}, in=90, out=-90] (113.center) to (114.center);
	\end{pgfonlayer}
    \begin{pgfonlayer}{labellayer}
        \node [] (119) at (-7.5, 0.85) {0};
		\node [] (76) at (11.55, -0.2) {$U(\theta_{B})$};
		\node [] (90) at (-13, 0.8) {$N$};
		\node [] (94) at (-10.25, 0.25) {$\textit{\textbf{F}}$};
		\node [] (77) at (6.05, 0.65) {$\cdots$};
		\node [] (80) at (6.05, -0.95) {$\cdots$};
		\node [] (78) at (11.55, 0.65) {$\cdots$};
		\node [] (79) at (11.55, -0.95) {$\cdots$};
		\node [] (120) at (-7.45, -0.45) {$U(\theta_{N})$};
		\node [] (59) at (6, -0.2) {$B$};
		\node [] (63) at (8.75, 0.25) {$\textit{\textbf{F}}$};
		\node [] (43) at (-0.75, 0.25) {$\textit{\textbf{F}}$};
    \end{pgfonlayer}
\end{tikzpicture}
\caption{The semantic functor $F$ assigns atomic components of a {\DCC} diagram  to elements of a parameterised quantum circuit.}
\label{fig:semanticFunctor}
\end{figure}

Each atomic component of the {\DCC} framework has a natural analogy in terms of quantum circuits. First, each wire in the string diagram is assigned a fixed number of qubits. Then, nouns are encoded within quantum states and boxes correspond to unitary operators that act on these states (see Figure \ref{fig:semanticFunctor}). This is achieved through a \textit{semantic functor} that relies on a choice of ansatz for the circuit. The details of this functor are described in Section \ref{subsec:PQCs}.

%%%%%%%%%%%%%%%%%%%%
%\begin{center}
%\includestandalone[width =0.55\textwidth]{background/disco}
%\end{center}
%%%%%%%%%%%%%%%%%%%%

%%%%%%%%%%%%%%%%%%%
%%%%%%%%%%%%%%%%%%%

Unlike nouns and boxes, frames representing higher-order linguistic constructions cannot be mapped directly to quantum circuits. This is because the internal states of a frame are not guaranteed to map bijectively to the ouput states, such as with conjunctions or reflexive verbs. Instead, we can functorially assign a circuit corresponding to the frame and its components, as described in \citep{Laakkonen_2024}. An explanation of the mechanics of this functor can be found in Section \ref{subsec:sandwich}.

\subsection{Pregroup grammars}

The transition from raw text to a computational framework for meaning is enabled by \textit{functorial semantics}. That is, a choice of a syntax category that provides linguistic rules and structure, a semantics category that assigns context, and a functor between them that allows a meaningful interpretation of text. An appropriate grammatical formalism to represent the syntactic aspect of the model is \textit{pregroup grammar} \citep{lambek}. The mathematical properties of a pregroup grammar, rooted in its algebraic structure, naturally align with tensor manipulations in vector spaces, enabling a seamless translation of grammatical relationships into compositional operations within quantum or classical computational frameworks \citep{coecke_2010}. Below, we provide some necessary background of how this is achieved.

%%%%%%%%%%%%%%%%%%%
\begin{Def}
    A \textit{pregroup} $(P, \leq, \cdot, 1, (-)^\ell,(-)^r)$ consists of a partially ordered monoid $(P, \leq, \cdot, 1)$ where each element $p \in P$ has a left adjoint $p^\ell$ and a right adjoint $p^r$. Moreover, these elements are required to satisfy conditions:
    \[
    p^\ell \cdot p \leq 1 \leq p \cdot p^\ell  \text{ and } p \cdot p^r \leq 1 \leq p^r \cdot p.
    \]
\label{def:pregroup}
\end{Def}
%%%%%%%%%%%%%%%%%%%

We shall use a pregroup generated from the primitive types $n$ and $s$ associated to noun phrases and sentence constructions respectively. In this system, transitive verbs are assigned the type $n^r \cdot s \cdot n^\ell$ as they require a subject and an object in order to return a complete sentence. For example, the sentence ``Alice reads books'' can be modelled with the following expression:

\[
    \text{Alice} \hspace{20pt}  \text{reads} \hspace{20pt} \text{books} \]
\[
     n \hspace{20pt} n^r \cdot s \cdot n^\ell  \hspace{20pt} n 
\]

Applying the grammatical rules to this expression gives the reductions
\[
    n \cdot n^r \cdot s \cdot n^\ell \cdot n  \leq 1 \cdot s \cdot n^\ell \cdot n \leq s \cdot 1 \leq s
\]
which show that the sentence is grammatically correct. We can now assign meaning to these formal expressions by specifying a compact closed symmetric monoidal category as our semantics category.  In this case we elect to use the category of finite dimensional vector spaces over a specific field where reductions of pregroup grammar can be viewed as contractions of an associated string diagram. This can be done by mapping primitive grammar types to vector spaces and compound types to tensor products of the corresponding spaces. 
\[
n \mapsto N \hspace{25 pt} s \mapsto S \hspace{25pt} n^r \cdot s \cdot n^\ell \mapsto N \otimes S \otimes N
\]
From this viewpoint reductions of the pregroup grammar types correspond to tensor contractions in the associated spaces. 

%%%%%%%%%%%%%%%%%%%      
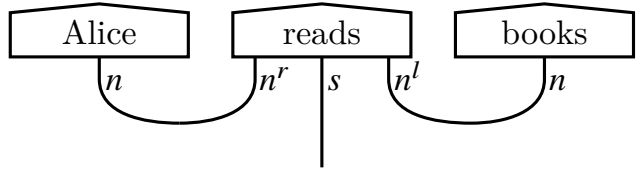
\begin{figure}[]
\hspace{-0.5cm}
\centering
\tikzset{every picture/.style={line width=0.75pt}, anchor=base} %set default line width to 0.75pt   

\tikzstyle{black_wire}=[-, draw={rgb,255: red,0; green,0; blue,0}, line width=1.0pt]
\tikzstyle{box}=[-, fill={rgb,255: red,255; green,255; blue,255}, line width=1.0pt]

% When embedding into a *.tex file, uncomment and include the following lines:
\pgfdeclarelayer{nodelayer}
\pgfdeclarelayer{edgelayer}
\pgfdeclarelayer{labellayer}
\pgfsetlayers{nodelayer, edgelayer, labellayer}
\begin{tikzpicture}
	\begin{pgfonlayer}{nodelayer}
		\node [] (1) at (4.25, 0.375) {};
		\node [] (2) at (4.25, 0.125) {};
		\node [] (4) at (6, 0.375) {};
		\node [] (5) at (6, 0.125) {};
		\node [] (7) at (1, 0.375) {};
		\node [] (8) at (1, 0.125) {};
		\node [] (10) at (2.75, 0.375) {};
		\node [] (11) at (2.75, 0.125) {};
		\node [] (13) at (3.5, 0.375) {};
		\node [] (14) at (3.5, -0.875) {};
		\node [] (16) at (5.175, -0.375) {};
		\node [] (17) at (1.9, -0.375) {};
		\node [] (18) at (0, 0.375) {};
		\node [] (19) at (2, 0.375) {};
		\node [] (20) at (2, 0.875) {};
		\node [] (21) at (1, 0.975) {};
		\node [] (22) at (0, 0.875) {};
		\node [] (24) at (2.5, 0.375) {};
		\node [] (25) at (4.5, 0.375) {};
		\node [] (26) at (4.5, 0.875) {};
		\node [] (27) at (3.5, 0.975) {};
		\node [] (28) at (2.5, 0.875) {};
		\node [] (30) at (5, 0.375) {};
		\node [] (31) at (7, 0.375) {};
		\node [] (32) at (7, 0.875) {};
		\node [] (33) at (6, 0.975) {};
		\node [] (34) at (5, 0.875) {};
	\end{pgfonlayer}
	\begin{pgfonlayer}{edgelayer}
		\draw [style={black_wire}, in=90, out=-90] (1.center) to (2.center);
		\draw [style={black_wire}, in=90, out=-90] (4.center) to (5.center);
		\draw [style={black_wire}, in=90, out=-90] (7.center) to (8.center);
		\draw [style={black_wire}, in=90, out=-90] (10.center) to (11.center);
		\draw [style={black_wire}, in=90, out=-90] (13.center) to (14.center);
		\draw [style={black_wire}, in=180, out=-90] (2.center) to (16.center);
		\draw [style={black_wire}, in=0, out=-90] (5.center) to (16.center);
		\draw [style={black_wire}, in=180, out=-90] (8.center) to (17.center);
		\draw [style={black_wire}, in=0, out=-90] (11.center) to (17.center);
		\draw [style=box] (18.center)
			 to (19.center)
			 to (20.center)
			 to (21.center)
			 to (22.center)
			 to cycle;
		\draw [style=box] (24.center)
			 to (25.center)
			 to (26.center)
			 to (27.center)
			 to (28.center)
			 to cycle;
		\draw [style=box] (30.center)
			 to (31.center)
			 to (32.center)
			 to (33.center)
			 to (34.center)
			 to cycle;
	\end{pgfonlayer}
    \begin{pgfonlayer}{labellayer}
        \node [] (3) at (4.47, 0) {$n^l$};
		\node [] (6) at (6.15, 0) {$n$};
		\node [] (9) at (1.17, 0) {$n$};
		\node [] (12) at (2.98, 0) {$n^r$};
		\node [] (15) at (3.64, 0) {$s$};
		\node [] (23) at (1, 0.525) {Alice};
		\node [] (29) at (3.5, 0.525) {reads};
		\node [] (35) at (6, 0.525) {books};
    \end{pgfonlayer}
\end{tikzpicture}
\caption{Pregroup diagram for the sentence ``Alice reads books''.}
\label{fig:pregroup}
\end{figure}
%%%%%%%%%%%%%%%%%%%

String diagrams of this form that represent sentences are the building blocks we use to generate discourse-level, large-scale DisCoCirc diagrams. In Section \ref{sec:text2diagram} we shall explain the process for doing this with details.

\subsection{Related work}

Compositional \db{models for natural language meaning} based on string diagrams have been around for the past 15 years, originated in \citep{coecke_2010} that described \dB{the DisCoCat} mathematical framework based on monoidal categories capable of translating grammatical derivations of sentences into tensor manipulations in vector spaces. Variations of this model have been used for conceptual linguistic tasks such as sentence entailment \citep{lewis_2019, bankova_2019}, while \cite{piedeleu_2015} detail a variation based on density matrices for handling lexical ambiguity. In \citep{yeung_2021}, a passage of the grammatical formalism from pregroups to CCG has been detailed, which made possible the creation of wide-coverage compositional NLP tools such as \texttt{lambeq}. 

The \texttt{lambeq} toolkit \citep{lambeq} is a high-level open-source software in Python for implementing compositional models of meaning based on string diagrams. It provides parsing features and several ansätze for the direct conversion of the diagrams into quantum circuits, ready to run on quantum hardware. It also comes with support for the most popular machine learning (ML) and Quantum ML (QML) packages such as PyTorch and Pennylane \citep{pennylane}, allowing efficient  supervised learning pipelines. While until now \texttt{lambeq} operated at the sentence-level, in this work we extend its functionality to support quantum circuits at the level of discourse.

The DisCoCirc model originated in \citep{discocirc}, and more details \dB{and additional features were provided in \citep{coecke_2021b, vincent, Urdu}}. An implementation method of preparing grammar-based discourse diagrams was proposed by \cite{jono}, achieved by mapping text to simply-typed $\lambda$-calculus terms using a CCG parser. However, due to the complexity of the composition and unary rules introduced by the underlying CCG structure, this model achieved limited coverage and usability in practice. Further \db{complexity-theoretical results were presented in \citep{Laakkonen_2024}}, while \citep{Duneau2024ScalableAI} presented a complete experiment on trapped ions hardware using custom small-scale datasets. Until now, none of the existing approaches have been able to provide wide coverage on arbitrary large-scale `real-world' text data. This is addressed by the present work.

\paragraph{Compositionality and interpretability} In a recent extended study, \cite{tull} introduce the notion of \textit{compositionally interpretable} (CI) models and demonstrate that DisCoCirc-like models serve as an instance of this framework. They highlight the explainability benefits of CI models by showing how their compositional structure enables the computation of additional quantities of interest and facilitates diagrammatic explanations of their behavior. These explanations can be based on influence constraints, diagram surgery, and rewrite rules. The authors suggest that DisCoCirc models could apply strict logical reasoning to large texts, even when trained only on small fragments of data, as demonstrated in the quantum model proposed by \citep{Duneau2024ScalableAI}.

\paragraph{Other QML approaches} Following a slightly different path from the structured models mentioned above, there are some notable QML approaches that aim to adapt and extend architectures from classical machine learning to leverage the unique capabilities of quantum computing for solving NLP tasks. For example, \cite{tamburini-2017} present a quantum version of a recursive neural network (RNN) trained to solve a phone recognition task. \cite{xu2024quantum} describe a hybrid quantum-classical RNN and present results competitive with classical baselines for a sentiment classification task. Further work has been done on implementing quantum models inspired by the transformer architecture. \cite{quixer} implement a quantum version of the transformer architecture utilising as building blocks linear combination of unitaries and the quantum singular value transform, tackling a language modelling task. Also, \cite{liao2024gpt} and \cite{guo2024quantum} present quantum circuit implementations of various individual architectural components of the transformer.

\section{Encoding text into diagrams}\label{sec:text2diagram}
%%%%%%%%%%%

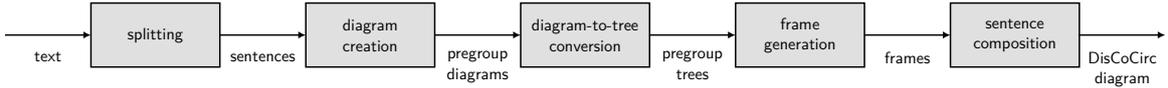
\begin{figure}
\hspace{-0.5cm}
\centering
\tikzset{every picture/.style={line width=1.0pt, font=\sffamily}} %set default line width to 1.0pt

\tikzstyle{gray_box}=[-, fill={rgb,255: red,224; green,224; blue,224}, line width=1.0pt]
\tikzstyle{arrow}=[-{Latex[width=0pt 5, length=5pt]}, draw={rgb,255: red,0; green,0; blue,0}, line width=1.0pt]

% When embedding into a *.tex file, uncomment and include the following lines:
\pgfdeclarelayer{nodelayer}
\pgfdeclarelayer{edgelayer}
\pgfdeclarelayer{labellayer}
\pgfsetlayers{nodelayer, edgelayer, labellayer}
\begin{tikzpicture}
	\begin{pgfonlayer}{nodelayer}
		\node [] (0) at (1, 3.25) {};
		\node [] (1) at (4, 3.25) {};
		\node [] (2) at (1, 1.75) {};
		\node [] (3) at (4, 1.75) {};
		\node [] (6) at (6, 3.25) {};
		\node [] (7) at (9, 3.25) {};
		\node [] (8) at (6, 1.75) {};
		\node [] (9) at (9, 1.75) {};
		\node [] (10) at (7.5, 2.5) {};
		\node [] (12) at (4, 2.5) {};
		\node [] (13) at (6, 2.5) {};
		\node [] (14) at (-1, 2.5) {};
		\node [] (15) at (1, 2.5) {};
		\node [] (17) at (9, 2.5) {};
		\node [] (18) at (11, 2.5) {};
		\node [] (20) at (11, 3.25) {};
		\node [] (21) at (14, 3.25) {};
		\node [] (22) at (11, 1.75) {};
		\node [] (23) at (14, 1.75) {};
		\node [] (24) at (12.5, 2.5) {};
		\node [] (28) at (16, 3.25) {};
		\node [] (29) at (19, 3.25) {};
		\node [] (30) at (16, 1.75) {};
		\node [] (31) at (19, 1.75) {};
		\node [] (32) at (17.5, 2.5) {};
		\node [] (34) at (14, 2.5) {};
		\node [] (35) at (16, 2.5) {};
		\node [] (45) at (19, 2.5) {};
		\node [] (46) at (21, 2.5) {};
		\node [] (47) at (-4, 3.25) {};
		\node [] (48) at (-1, 3.25) {};
		\node [] (49) at (-4, 1.75) {};
		\node [] (50) at (-1, 1.75) {};
		\node [] (53) at (-6, 2.5) {};
		\node [] (54) at (-4, 2.5) {};
	\end{pgfonlayer}
	\begin{pgfonlayer}{edgelayer}
		\draw [style={gray_box}] (0.center)
			 to (1.center)
			 to (3.center)
			 to (2.center)
			 to cycle;
		\draw [style={gray_box}] (6.center)
			 to (7.center)
			 to (9.center)
			 to (8.center)
			 to cycle;
		\draw [style={arrow}] (12.center) to (13.center);
		\draw [style={arrow}] (14.center) to (15.center);
		\draw [style={arrow}] (17.center) to (18.center);
		\draw [style={gray_box}] (20.center)
            to (21.center)
            to (23.center)
            to (22.center)
            to cycle;
		\draw [style={gray_box}] (28.center)
            to (29.center)
            to (31.center)
            to (30.center)
            to cycle;
		\draw [style={arrow}] (34.center) to (35.center);
		\draw [style={arrow}] (45.center) to (46.center);
		\draw [style={gray_box}] (47.center)
			 to (48.center)
			 to (50.center)
			 to (49.center)
			 to cycle;
		\draw [style={arrow}] (53.center) to (54.center);
	\end{pgfonlayer}
    \begin{pgfonlayer}{labellayer}
        \node [align=center] (4) at (2.5, 2.5) {diagram\\creation};
		\node [align=center, anchor=north] (5) at (5, 2.25) {pregroup\\diagrams};
		\node [align=center] (11) at (7.5, 2.5) {diagram-to-tree\\conversion};
		\node [align=center, anchor=north] (16) at (10, 2.25) {pregroup\\trees};
		\node [align=center, anchor=north] (19) at (0, 2.25) {sentences};
		\node [align=center] (25) at (12.5, 2.5) {frame\\generation};
		\node [align=center, anchor=north] (26) at (15, 2.25) {frames};
		\node [align=center] (33) at (17.5, 2.5) {sentence\\composition};
		\node [align=center, anchor=north] (38) at (20, 2.25) {DisCoCirc\\diagram};
		\node [align=center] (51) at (-2.5, 2.5) {splitting};
		\node [align=center, anchor=north] (55) at (-5, 2.25) {text};
    \end{pgfonlayer}
\end{tikzpicture}
\caption{The process of converting long texts into quantum circuits.}
\label{fig:disco-process}
\end{figure}

In this section, we introduce a novel procedure for converting text of arbitrary length into a DisCoCirc diagram. Our approach builds a tree-like structure from the pregroup derivation of each sentence in the text, transforms the tree into a local diagram and then composes these diagrams along coreferenced wires. While we make use of \texttt{lambeq}'s parsing and grammatical backends, our algorithm is entirely original. The high-level process is shown in Figure \ref{fig:disco-process}.

%%%%%%%%%%%
\subsection{Pregroup trees}\label{subsec:pregroup-trees}
%%%%%%%%%%%

Although a pregroup diagram is an efficient grammatical representation of a sentence, it still includes extraneous topology and redundant information that can be inferred from the context. To enhance computational efficiency and simplify the underlying algorithmic processes, we transform each string diagram into a structure we call the \textit{pregroup tree}.

\begin{figure}[h]
\centering
\begin{subfigure}{0.57\textwidth}
    \tikzset{every picture/.style={line width=0.75pt, anchor=base}} %set default line width to 0.75pt

% When embedding into a *.tex file, uncomment and include the following lines:
\pgfdeclarelayer{nodelayer}
\pgfdeclarelayer{edgelayer}
\pgfdeclarelayer{labellayer}
\pgfsetlayers{nodelayer, edgelayer, labellayer}
\begin{tikzpicture}
	\begin{pgfonlayer}{nodelayer}
		\node [] (1) at (4.35, 0.75) {};
		\node [] (2) at (4.35, 0.25) {};
		\node [] (4) at (6, 0.75) {};
		\node [] (5) at (6, 0.25) {};
		\node [] (7) at (8.83333, 0.75) {};
		\node [] (8) at (8.83333, 0.25) {};
		\node [] (10) at (11, 0.75) {};
		\node [] (11) at (11, 0.25) {};
		\node [] (13) at (3.78333, 0.75) {};
		\node [] (14) at (3.78333, -0.25) {};
		\node [] (16) at (8.16667, 0.75) {};
		\node [] (17) at (8.16667, -0.25) {};
		\node [] (19) at (1, 0.75) {};
		\node [] (20) at (1, 0.25) {};
		\node [] (22) at (2.65, 0.75) {};
		\node [] (23) at (2.65, 0.25) {};
		\node [] (25) at (3.21667, 0.75) {};
		\node [] (26) at (3.21667, -1.25) {};
		\node [] (28) at (5.175, 0) {};
		\node [] (29) at (9.91667, 0) {};
		\node [] (30) at (5.93333, -0.5) {};
		\node [] (31) at (1.825, 0) {};
		\node [] (32) at (0, 0.75) {};
		\node [] (33) at (2, 0.75) {};
		\node [] (34) at (2, 1.25) {};
		\node [] (35) at (1, 1.35) {};
		\node [] (36) at (0, 1.25) {};
		\node [] (38) at (2.5, 0.75) {};
		\node [] (39) at (4.5, 0.75) {};
		\node [] (40) at (4.5, 1.25) {};
		\node [] (41) at (3.5, 1.35) {};
		\node [] (42) at (2.5, 1.25) {};
		\node [] (44) at (5, 0.75) {};
		\node [] (45) at (7, 0.75) {};
		\node [] (46) at (7, 1.25) {};
		\node [] (47) at (6, 1.35) {};
		\node [] (48) at (5, 1.25) {};
		\node [] (50) at (7.5, 0.75) {};
		\node [] (51) at (9.5, 0.75) {};
		\node [] (52) at (9.5, 1.25) {};
		\node [] (53) at (8.5, 1.35) {};
		\node [] (54) at (7.5, 1.25) {};
		\node [] (56) at (10, 0.75) {};
		\node [] (57) at (12, 0.75) {};
		\node [] (58) at (12, 1.25) {};
		\node [] (59) at (11, 1.35) {};
		\node [] (60) at (10, 1.25) {};
	\end{pgfonlayer}
	\begin{pgfonlayer}{edgelayer}
		\draw [-, draw={rgb,255: red,0; green,0; blue,0}, line width=1.0pt, in=90, out=-90] (1.center) to (2.center);
		\draw [-, draw={rgb,255: red,0; green,0; blue,0}, line width=1.0pt, in=90, out=-90] (4.center) to (5.center);
		\draw [-, draw={rgb,255: red,0; green,0; blue,0}, line width=1.0pt, in=90, out=-90] (7.center) to (8.center);
		\draw [-, draw={rgb,255: red,0; green,0; blue,0}, line width=1.0pt, in=90, out=-90] (10.center) to (11.center);
		\draw [-, draw={rgb,255: red,0; green,0; blue,0}, line width=1.0pt, in=90, out=-90] (13.center) to (14.center);
		\draw [-, draw={rgb,255: red,0; green,0; blue,0}, line width=1.0pt, in=90, out=-90] (16.center) to (17.center);
		\draw [-, draw={rgb,255: red,0; green,0; blue,0}, line width=1.0pt, in=90, out=-90] (19.center) to (20.center);
		\draw [-, draw={rgb,255: red,0; green,0; blue,0}, line width=1.0pt, in=90, out=-90] (22.center) to (23.center);
		\draw [-, draw={rgb,255: red,0; green,0; blue,0}, line width=1.0pt, in=90, out=-90] (25.center) to (26.center);
		\draw [-, draw={rgb,255: red,0; green,0; blue,0}, line width=1.0pt, in=-180, out=-90, looseness=0.75] (2.center) to (28.center);
		\draw [-, draw={rgb,255: red,0; green,0; blue,0}, line width=1.0pt, in=0, out=-90, looseness=0.75] (5.center) to (28.center);
		\draw [-, draw={rgb,255: red,0; green,0; blue,0}, line width=1.0pt, in=180, out=-90, looseness=0.47] (8.center) to (29.center);
		\draw [-, draw={rgb,255: red,0; green,0; blue,0}, line width=1.0pt, in=0, out=-90, looseness=0.47] (11.center) to (29.center);
		\draw [-, draw={rgb,255: red,0; green,0; blue,0}, line width=1.0pt, in=180, out=-90, looseness=0.23] (14.center) to (30.center);
		\draw [-, draw={rgb,255: red,0; green,0; blue,0}, line width=1.0pt, in=0, out=-90, looseness=0.23] (17.center) to (30.center);
		\draw [-, draw={rgb,255: red,0; green,0; blue,0}, line width=1.0pt, in=180, out=-90, looseness=0.75] (20.center) to (31.center);
		\draw [-, draw={rgb,255: red,0; green,0; blue,0}, line width=1.0pt, in=0, out=-90, looseness=0.75] (23.center) to (31.center);
		\draw [-, fill={rgb,255: red,255; green,255; blue,255}, line width=1.0pt] (32.center)
			 to (33.center)
			 to (34.center)
			 to (35.center)
			 to (36.center)
			 to cycle;
		\draw [-, fill={rgb,255: red,255; green,255; blue,255}, line width=1.0pt] (38.center)
			 to (39.center)
			 to (40.center)
			 to (41.center)
			 to (42.center)
			 to cycle;
		\draw [-, fill={rgb,255: red,255; green,255; blue,255}, line width=1.0pt] (44.center)
			 to (45.center)
			 to (46.center)
			 to (47.center)
			 to (48.center)
			 to cycle;
		\draw [-, fill={rgb,255: red,255; green,255; blue,255}, line width=1.0pt] (50.center)
			 to (51.center)
			 to (52.center)
			 to (53.center)
			 to (54.center)
			 to cycle;
		\draw [-, fill={rgb,255: red,255; green,255; blue,255}, line width=1.0pt] (56.center)
			 to (57.center)
			 to (58.center)
			 to (59.center)
			 to (60.center)
			 to cycle;
	\end{pgfonlayer}
	\begin{pgfonlayer}{labellayer}
		\node [] (3) at (4.55, 0.3) {$n^l$};
		\node [] (6) at (6.15, 0.3) {$n$};
		\node [] (9) at (9.05, 0.3) {$n^l$};
		\node [] (12) at (11.15, 0.3) {$n$};
		\node [] (15) at (4.0, 0.3) {$n^l$};
		\node [] (18) at (8.31667, 0.3) {$n$};
		\node [] (21) at (1.15, 0.3) {$n$};
		\node [] (24) at (2.85, 0.3) {$n^r$};
		\node [] (27) at (3.35, 0.3) {$s$};
		\node [] (37) at (1, 0.9) {Bob};
		\node [] (43) at (3.5, 0.9) {gave};
		\node [] (49) at (6, 0.9) {Alice};
		\node [] (55) at (8.5, 0.9) {a};
		\node [] (61) at (11, 0.9) {book};
    \end{pgfonlayer}
\end{tikzpicture}
    \caption{}
\end{subfigure}%
\hfill
\begin{subfigure}{0.43\textwidth}
    \tikzset{every picture/.style={line width=1.0pt, anchor=center}} %set default line width to 0.75pt

\tikzstyle{black_wire}=[-, draw={rgb,255: red,0; green,0; blue,0}, line width=1.0pt]
\tikzstyle{treeNode}=[rectangle, fill=white, line width=1.0pt]
\tikzstyle{dotted}=[-, dash pattern=on 0.84pt off 2.51pt, line width=1.0pt]
\tikzstyle{prune}=[-, draw={rgb,255: red,208: green,48: blue,32}, line width=3.0pt]

% When embedding into a *.tex file, uncomment and include the following lines:
\pgfdeclarelayer{nodelayer}
\pgfdeclarelayer{edgelayer}
\pgfdeclarelayer{labellayer}
\pgfsetlayers{nodelayer, edgelayer, labellayer}
\begin{tikzpicture}
	\begin{pgfonlayer}{nodelayer}
		\node [style=treeNode] (4) at (15.5, -0.6) {Bob$_n$};
		\node [style=treeNode] (9) at (18.5, 0.9) {gave$_s$};
		\node [style=treeNode] (14) at (18.5, -0.6) {Alice$_n$};
		\node [style=treeNode] (19) at (21.5, -0.6) {a$_n$};
		\node [style=treeNode] (24) at (21.5, -2.1) {book$_n$};
	\end{pgfonlayer}
	\begin{pgfonlayer}{edgelayer}
		\draw [style=black_wire] (4) to (9);
		\draw [style=black_wire] (14) to (9);
		\draw [style=black_wire] (19) to (9);
		\draw [style=black_wire] (24) to (19);
	\end{pgfonlayer}
\end{tikzpicture}
    \caption{}
\end{subfigure}%
\caption{A pregroup diagram (a) and its corresponding tree representation (b). The nodes contain both (1) the assigned type during the tree derivation, shown here as subscripts, and (2) the word index in the sentence, not shown here for conciseness. The compound type of a node is derivable by the type of the node extended with the adjoints of the types of its children as dictated by the order of words, e.g. $n \cdot n^l$ for the determiner ``a''.}
\label{fig:pregroup-tree}
\end{figure}
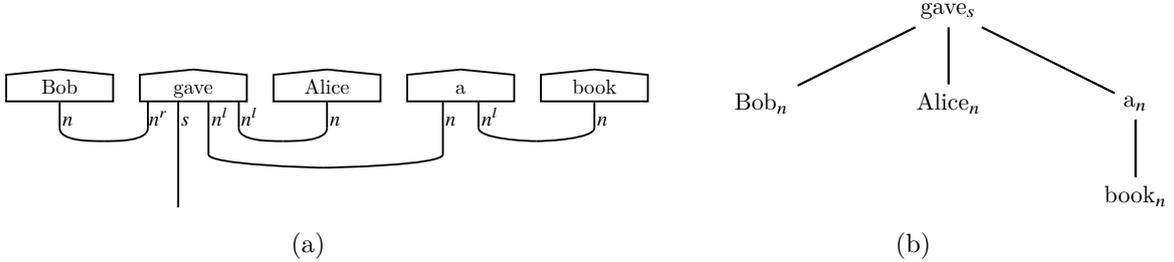

A pregroup tree is a compact, tree-like representation of a pregroup diagram, designed to maximize processing efficiency. Each node corresponds to a token in the sentence and is annotated with the pregroup type of the output wire(s) associated with that word---e.g., $s$ for a verb, $n$ for an adjective, or $n^r \cdot s$ for an adverb. The root of the tree represents the sentence’s head word---typically a word with free wires, often an $s$, that encapsulates the sentence’s state after composition. The branches of the tree denote cups, which identify input wires connecting to the parent node. This way, the diagram is converted into a tree of pregroup types, indicating the result expected from each token (Figure \ref{fig:pregroup-tree}). The complete compound type of a token is fully recoverable from the mathematical properties of pregroups (Definition \ref{def:pregroup}) by taking into account the types of the children and the order of words in the sentence.

% Following this procedure, every pregroup diagram can be trivially converted into a pregroup tree (Figure \ref{fig:pregroup-tree}).

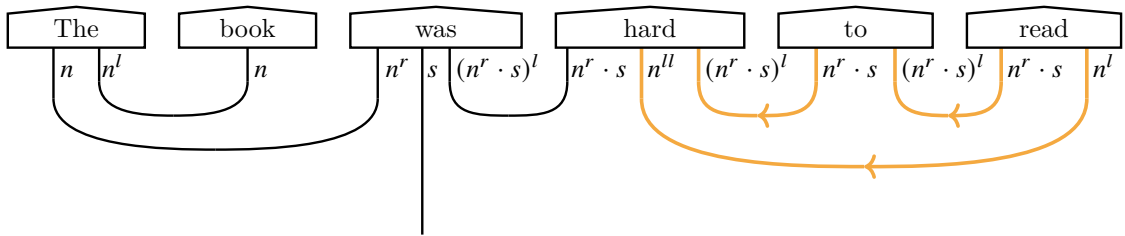
\begin{figure}[h]
\hspace{-1cm}
\centering
    \tikzset{every picture/.style={line width=0.75pt, anchor=base}} %set default line width to 0.75pt

\tikzstyle{black_wire}=[-, draw={rgb,255: red,0; green,0; blue,0}, line width=1.0pt]
\tikzstyle{box}=[-, fill={rgb,255: red,255; green,255; blue,255}, line width=1.0pt]
\tikzstyle{f4a940_wire}=[-, draw={rgb,255: red,244; green,169; blue,64}, line width=1.5pt]
\tikzstyle{f4a940_wire_arrow}=[->, draw={rgb,255: red,244; green,169; blue,64}, line width=1.5pt]

% When embedding into a *.tex file, uncomment and include the following lines:
\pgfdeclarelayer{nodelayer}
\pgfdeclarelayer{edgelayer}
\pgfdeclarelayer{labellayer}
\pgfsetlayers{nodelayer, edgelayer, labellayer}
\begin{tikzpicture}
	\begin{pgfonlayer}{nodelayer}
		\node [] (1) at (0.83333, 1.125) {};
		\node [] (2) at (0.83333, 0.625) {};
		\node [] (4) at (3, 1.125) {};
		\node [] (5) at (3, 0.625) {};
		\node [] (11) at (14, 0.875) {};
		\node [] (13) at (12.45, 1.125) {};
		\node [] (14) at (12.45, 0.625) {};
		\node [] (16) at (14, 1.125) {};
		\node [] (17) at (14, 0.625) {};
		\node [] (25) at (9.58333, 1.125) {};
		\node [] (26) at (9.58333, 0.625) {};
		\node [] (28) at (11.3, 1.125) {};
		\node [] (29) at (11.3, 0.625) {};
		\node [] (31) at (8.75, 1.125) {};
		\node [] (32) at (8.75, 0.375) {};
		\node [] (34) at (15.25, 1.125) {};
		\node [] (35) at (15.25, 0.375) {};
		\node [] (43) at (5.95, 1.125) {};
		\node [] (44) at (5.95, 0.625) {};
		\node [] (46) at (7.66667, 1.125) {};
		\node [] (47) at (7.66667, 0.625) {};
		\node [] (49) at (0.166667, 1.125) {};
		\node [] (50) at (0.166667, 0.375) {};
		\node [] (52) at (4.9, 1.125) {};
		\node [] (53) at (4.9, 0.375) {};
		\node [] (55) at (5.55, 1.125) {};
		\node [] (56) at (5.55, -1.625) {};
		\node [] (58) at (1.91667, 0.125) {};
		\node [] (60) at (13.225, 0.125) {};
		\node [] (62) at (10.4417, 0.125) {};
		\node [] (63) at (12, -0.625) {};
		\node [] (65) at (6.80833, 0.125) {};
		\node [] (66) at (2.53333, -0.375) {};
		\node [] (67) at (-0.5, 1.125) {};
		\node [] (68) at (1.5, 1.125) {};
		\node [] (69) at (1.5, 1.625) {};
		\node [] (70) at (0.5, 1.725) {};
		\node [] (71) at (-0.5, 1.625) {};
		\node [] (73) at (2, 1.125) {};
		\node [] (74) at (4, 1.125) {};
		\node [] (75) at (4, 1.625) {};
		\node [] (76) at (3, 1.725) {};
		\node [] (77) at (2, 1.625) {};
		\node [] (79) at (4.5, 1.125) {};
		\node [] (80) at (7, 1.125) {};
		\node [] (81) at (7, 1.625) {};
		\node [] (82) at (5.75, 1.725) {};
		\node [] (83) at (4.5, 1.625) {};
		\node [] (85) at (7.5, 1.125) {};
		\node [] (86) at (10.25, 1.125) {};
		\node [] (87) at (10.25, 1.625) {};
		\node [] (88) at (8.875, 1.725) {};
		\node [] (89) at (7.5, 1.625) {};
		\node [] (91) at (10.75, 1.125) {};
		\node [] (92) at (13, 1.125) {};
		\node [] (93) at (13, 1.625) {};
		\node [] (94) at (11.875, 1.725) {};
		\node [] (95) at (10.75, 1.625) {};
		\node [] (97) at (13.5, 1.125) {};
		\node [] (98) at (15.75, 1.125) {};
		\node [] (99) at (15.75, 1.625) {};
		\node [] (100) at (14.625, 1.725) {};
		\node [] (101) at (13.5, 1.625) {};
	\end{pgfonlayer}
	\begin{pgfonlayer}{edgelayer}
		\draw [style={black_wire}, in=90, out=-90] (1.center) to (2.center);
		\draw [style={black_wire}, in=90, out=-90] (4.center) to (5.center);
		\draw [style={f4a940_wire}, in=90, out=-90] (13.center) to (14.center);
		\draw [style={f4a940_wire}, in=90, out=-90] (16.center) to (17.center);
		\draw [style={f4a940_wire}, in=90, out=-90] (25.center) to (26.center);
		\draw [style={f4a940_wire}, in=90, out=-90] (28.center) to (29.center);
		\draw [style={f4a940_wire}, in=90, out=-90] (31.center) to (32.center);
		\draw [style={f4a940_wire}, in=90, out=-90] (34.center) to (35.center);
		\draw [style={black_wire}, in=90, out=-90] (43.center) to (44.center);
		\draw [style={black_wire}, in=90, out=-90] (46.center) to (47.center);
		\draw [style={black_wire}, in=90, out=-90] (49.center) to (50.center);
		\draw [style={black_wire}, in=90, out=-90] (52.center) to (53.center);
		\draw [style={black_wire}, in=90, out=-90] (55.center) to (56.center);
		\draw [style={black_wire}, in=180, out=-90] (2.center) to (58.center);
		\draw [style={black_wire}, in=0, out=-90] (5.center) to (58.center);
		\draw [style={f4a940_wire}, in=-180, out=-90, looseness=1.25] (14.center) to (60.center);
		\draw [style={f4a940_wire_arrow}, in=0, out=-90, looseness=1.25] (17.center) to (60.center);
		\draw [style={f4a940_wire}, in=180, out=-90, looseness=1.25] (26.center) to (62.center);
		\draw [style={f4a940_wire_arrow}, in=0, out=-90, looseness=1.25] (29.center) to (62.center);
		\draw [style={f4a940_wire}, in=180, out=-90, looseness=0.75] (32.center) to (63.center);
		\draw [style={f4a940_wire_arrow}, in=0, out=-90, looseness=0.75] (35.center) to (63.center);
		\draw [style={black_wire}, in=180, out=-90, looseness=1.25] (44.center) to (65.center);
		\draw [style={black_wire}, in=0, out=-90, looseness=1.25] (47.center) to (65.center);
		\draw [style={black_wire}, in=-180, out=-90, looseness=0.75] (50.center) to (66.center);
		\draw [style={black_wire}, in=0, out=-90, looseness=0.75] (53.center) to (66.center);
		\draw [style=box] (67.center)
			 to (68.center)
			 to (69.center)
			 to (70.center)
			 to (71.center)
			 to cycle;
		\draw [style=box] (73.center)
			 to (74.center)
			 to (75.center)
			 to (76.center)
			 to (77.center)
			 to cycle;
		\draw [style=box] (79.center)
			 to (80.center)
			 to (81.center)
			 to (82.center)
			 to (83.center)
			 to cycle;
		\draw [style=box] (85.center)
			 to (86.center)
			 to (87.center)
			 to (88.center)
			 to (89.center)
			 to cycle;
		\draw [style=box] (91.center)
			 to (92.center)
			 to (93.center)
			 to (94.center)
			 to (95.center)
			 to cycle;
		\draw [style=box] (97.center)
			 to (98.center)
			 to (99.center)
			 to (100.center)
			 to (101.center)
			 to cycle;
	\end{pgfonlayer}
	\begin{pgfonlayer}{labellayer}
		\node [] (51) at (0.366667, 0.7) {$n$};
		\node [] (3) at (1.03333, 0.7) {$n^l$};
		\node [] (6) at (3.2, 0.7) {$n$};
		\node [] (54) at (5.2, 0.7) {$n^r$};
		\node [] (57) at (5.70, 0.7) {$s$};
		\node [] (45) at (6.65, 0.7) {$(n^r \cdot s)^l$};
		\node [] (42) at (8.13333, 0.7) {$n^r \cdot s$};
		\node [] (33) at (9.05, 0.7) {$n^{ll}$};
		\node [] (27) at (10.2833, 0.7) {$(n^r \cdot s)^l$};
		\node [] (24) at (11.80, 0.7) {$n^r \cdot s$};
		\node [] (15) at (13.15, 0.7) {$(n^r \cdot s)^l$};
		\node [] (12) at (14.5, 0.7) {$n^r \cdot s$};
		\node [] (36) at (15.50, 0.7) {$n^l$};
		\node [] (72) at (0.5, 1.275) {The};
		\node [] (78) at (3, 1.275) {book};
		\node [] (84) at (5.75, 1.275) {was};
		\node [] (90) at (8.875, 1.275) {hard};
		\node [] (96) at (11.875, 1.275) {to};
		\node [] (102) at (14.625, 1.275) {read};
	\end{pgfonlayer}
\end{tikzpicture}
    \caption{Pregroup diagram with a loop (arrows show the child-to-parent relationships). Note that node ``read'' has 2 parents, ``to'' and ``hard''. We break the loop by removing the longest-range dependency---in this case the one between ``hard'' and ``read''.}    
    % Although there are multiple cups between the words ``hard'' and ``to'', we encode this as a single edge with a compound type $n^r \cdot s$ in the pregroup tree. The same is true for the cups between ``to'' and ``take'', thus, this diagram has a single loop among the words these three words. The depth-first traversal of the diagram ensures that we capture the sequence of edges ``hard''---``to''---``take'' and we create a duplicate node for ``take'' with type $n^l$ to preserve the pregroup tree structure. Breaking the loop only requires removal of this shallower node for ``take''.}
    \label{fig:cycles}
\end{figure}

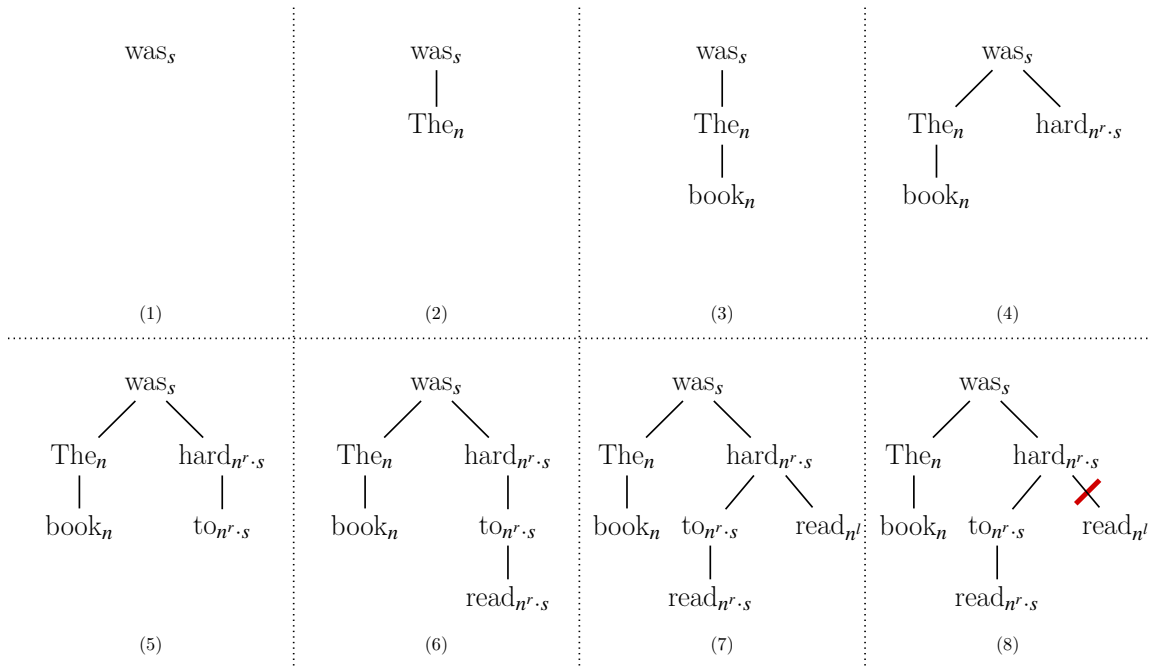
\begin{figure}[h]
\hspace{-1cm}
\centering
    \tikzset{every picture/.style={line width=1.0pt, anchor=center}} %set default line width to 0.75pt

\tikzstyle{black_wire}=[-, draw={rgb,255: red,0; green,0; blue,0}, line width=1.0pt]
\tikzstyle{treeNode}=[rectangle, fill=white, line width=1.0pt, font={\LARGE}]
\tikzstyle{dotted}=[-, dash pattern=on 0.84pt off 2.51pt, line width=1.0pt]
\tikzstyle{prune}=[-, draw={rgb,255: red,208: green,48: blue,32}, line width=3.0pt]

% When embedding into a *.tex file, uncomment and include the following lines:
\pgfdeclarelayer{nodelayer}
\pgfdeclarelayer{edgelayer}
\pgfdeclarelayer{labellayer}
\pgfsetlayers{nodelayer, edgelayer, labellayer}
\begin{tikzpicture}
	\begin{pgfonlayer}{nodelayer}
		\node [style=treeNode] (1) at (0, 2) {was$_{ s}$};
		\node [style=treeNode] (2) at (-1.5, 0.5) {The$_{ n}$};
		\node [style=treeNode] (3) at (-1.5, -1) {book$_{ n}$};
		\node [style=treeNode] (4) at (1.5, 0.5) {hard$_{ n^r \cdot s}$};
		\node [style=treeNode] (5) at (1.5, -1) {to$_{ n^r \cdot s}$};
		\node [style=treeNode] (6) at (1.5, -2.5) {read$_{ n^r \cdot s}$};
		\node [style=treeNode] (15) at (11.525, 2) {was$_{ s}$};
		\node [style=treeNode] (16) at (10.025, 0.5) {The$_{ n}$};
		\node [style=treeNode] (17) at (10.025, -1) {book$_{ n}$};
		\node [style=treeNode] (18) at (13.025, 0.5) {hard$_{ n^r \cdot s}$};
		\node [style=treeNode] (19) at (11.775, -1) {to$_{ n^r \cdot s}$};
		\node [style=treeNode] (20) at (11.775, -2.5) {read$_{ n^r \cdot s}$};
		\node [style=treeNode] (21) at (14.275, -1) {read$_{ n^l}$};
		\node [style=treeNode] (22) at (-6, 2) {was$_{ s}$};
		\node [style=treeNode] (23) at (-7.5, 0.5) {The$_{ n}$};
		\node [style=treeNode] (24) at (-7.5, -1) {book$_{ n}$};
		\node [style=treeNode] (25) at (-4.5, 0.5) {hard$_{ n^r \cdot s}$};
		\node [style=treeNode] (26) at (-4.5, -1) {to$_{ n^r \cdot s}$};
		\node [style=treeNode] (27) at (-6, 9) {was$_{ s}$};
		\node [] (33) at (-9, 3) {};
		\node [] (34) at (15, 3) {};
		\node [style=treeNode] (35) at (0, 9) {was$_{ s}$};
		\node [style=treeNode] (36) at (0, 7.5) {The$_{ n}$};
		\node [style=treeNode] (40) at (6, 9) {was$_{ s}$};
		\node [style=treeNode] (41) at (6, 7.5) {The$_{ n}$};
		\node [style=treeNode] (42) at (6, 6) {book$_{ n}$};
		\node [style=treeNode] (45) at (12, 9) {was$_{ s}$};
		\node [style=treeNode] (46) at (10.5, 7.5) {The$_{ n}$};
		\node [style=treeNode] (47) at (10.5, 6) {book$_{ n}$};
		\node [style=treeNode] (48) at (13.5, 7.5) {hard$_{ n^r \cdot s}$};
		\node [] (49) at (-3, 10) {};
		\node [] (50) at (-3, -4) {};
		\node [] (51) at (3, 10) {};
		\node [] (52) at (3, -4) {};
		\node [] (53) at (9, 10) {};
		\node [] (54) at (9, -4) {};
		\node [] (55) at (13.425, -0.5) {};
		\node [] (56) at (13.925, 0) {};
		\node [style=treeNode] (57) at (5.5, 2) {was$_{ s}$};
		\node [style=treeNode] (58) at (4, 0.5) {The$_{ n}$};
		\node [style=treeNode] (59) at (4, -1) {book$_{ n}$};
		\node [style=treeNode] (60) at (7, 0.5) {hard$_{ n^r \cdot s}$};
		\node [style=treeNode] (61) at (5.75, -1) {to$_{ n^r \cdot s}$};
		\node [style=treeNode] (62) at (5.75, -2.5) {read$_{ n^r \cdot s}$};
		\node [style=treeNode] (63) at (8.25, -1) {read$_{ n^l}$};
		\node [] (72) at (-6, 3.5) {(1)};
		\node [] (74) at (0, 3.5) {(2)};
		\node [] (75) at (6, 3.5) {(3)};
		\node [] (76) at (12, 3.5) {(4)};
		\node [] (77) at (-6, -3.5) {(5)};
		\node [] (78) at (0, -3.5) {(6)};
		\node [] (79) at (6, -3.5) {(7)};
		\node [] (80) at (12, -3.5) {(8)};
	\end{pgfonlayer}
	\begin{pgfonlayer}{edgelayer}
		\draw [style=dotted] (33.center) to (34.center);
		\draw [style=dotted] (49.center) to (50.center);
		\draw [style=dotted] (51.center) to (52.center);
		\draw [style=dotted] (53.center) to (54.center);
		\draw [style=prune] (55.center) to (56.center);
		\draw [style=black_wire] (36) to (35);
		\draw [style=black_wire] (41) to (40);
		\draw [style=black_wire] (42) to (41);
		\draw [style=black_wire] (46) to (45);
		\draw [style=black_wire] (47) to (46);
		\draw [style=black_wire] (48) to (45);
		\draw [style=black_wire] (23) to (22);
		\draw [style=black_wire] (24) to (23);
		\draw [style=black_wire] (25) to (22);
		\draw [style=black_wire] (26) to (25);
		\draw [style=black_wire] (2) to (1);
		\draw [style=black_wire] (3) to (2);
		\draw [style=black_wire] (4) to (1);
		\draw [style=black_wire] (5) to (4);
		\draw [style=black_wire] (6) to (5);
		\draw [style=black_wire] (58) to (57);
		\draw [style=black_wire] (59) to (58);
		\draw [style=black_wire] (60) to (57);
		\draw [style=black_wire] (61) to (60);
		\draw [style=black_wire] (62) to (61);
		\draw [style=black_wire] (63) to (60);
		\draw [style=black_wire] (16) to (15);
		\draw [style=black_wire] (17) to (16);
		\draw [style=black_wire] (18) to (15);
		\draw [style=black_wire] (19) to (18);
		\draw [style=black_wire] (20) to (19);
		\draw [style=black_wire] (21) to (18);
	\end{pgfonlayer}
\end{tikzpicture}
    \caption{The process of generating a tree from the pregroup diagram in Figure \ref{fig:cycles}. The node subscripts are the types assigned to the words. Beginning with the word that has free wires (the head of the sentence), we recursively add children by following the word order and performing a depth-first traversal of any dependencies. In Step 8, the loop highlighted in the diagram is resolved by removing the branch corresponding to the long-range dependency between ``hard'' and ``read''.}
    \label{fig:tree-steps}
\end{figure}

The parent-child relationships in the pregroup tree are recursively determined as follows: For each word, all arguments indicated in the pregroup derivation are assigned as children, following a strict left-to-right word order and a depth-first traversal of dependencies. This means that for arguments preceding the current word, processing follows an outer-to-inner order, starting from the outermost cup. Conversely, for arguments following the current word, processing occurs from the innermost cup outward. This process generates a tree structure that respects the modification relationships between the different parts of the sentence, and is necessary for the accurate representation of the frames later in the DisCoCirc diagram. An example is provided in Figures \ref{fig:cycles} and \ref{fig:tree-steps}, where the latter details the steps for deriving a tree from the pregroup diagram of the former.

While following the traversal process described above, certain grammatical relationships in the pregroup derivation can create loops (highlighted in a different colour in Figure \ref{fig:cycles}). In such cases, a node may have multiple parents, which is incompatible with a tree structure. To resolve this, we eliminate the longest-range dependency—--in Figure \ref{fig:cycles}, the link between ``hard'' and ``read''—--ensuring a properly structured representation.

%%%%%%%%%%%
\subsection{Generating frames}\label{subsec:processing-frames}
%%%%%%%%%%%

Once the pregroup tree for a sentence has been obtained, it is converted into a collection of nested frames. This conversion is done in a single post-order traversal of the tree. At each stage of the recursion, a collection of sub-diagrams and nouns that have been created so far are returned.

%%%%%%%%%%%
\begin{algorithm}
\caption{Tree to Frames Conversion}\label{alg:tree2frame}
\begin{algorithmic}[1]
\State \textbf{function} \texttt{tree\_to\_frame(node)}
\State\quad \textbf{Output:} Diagram representation as nested frames, list of nouns

\If{node is a \texttt{noun} and has no children}
    \State \textbf{Return:} \texttt{Id(noun)}, list containing \texttt{Box(node.word)}
\EndIf
\State \texttt{subdiags\_n\_nouns} $\gets$ \texttt{tree\_to\_frame(child)} for each \texttt{child} of \texttt{node}

\State \texttt{subdiags} $\gets$ list of non-empty diagrams from \texttt{subdiags\_n\_nouns}
\State \texttt{nouns} $\gets$ list of nouns from \texttt{subdiags\_n\_nouns}

\If{no subdiags}
    \State \textbf{Return:} \texttt{Box(node.word)}, \texttt{nouns}
\EndIf

\State \textbf{Return:} \texttt{Frame(node.word, subdiags)}, \texttt{nouns}
\end{algorithmic}
\end{algorithm}
%%%%%%%%%%%

Once a leaf node is visited, it is checked for noun type. If it has noun type, a box representing the corresponding state is returned. Otherwise, a box associated to the word is returned as a sub-diagram. In the recursive case, non-leaf nodes obtain the nouns and sub-diagrams corresponding to their children. If this collection of sub-diagrams is non-empty, a frame is created that is labelled by the current node and contains its sub-diagrams as an attribute (Algorithm \ref{alg:tree2frame}).

The output of this process is the frame corresponding to the root node in the pregroup tree, as well as the list of nouns appearing within the tree. The purpose of separating noun boxes from sub-diagrams is that it effectively incorporates \textit{dragging} nouns to the top of the diagram in a single step.

%%%%%%%%%%%
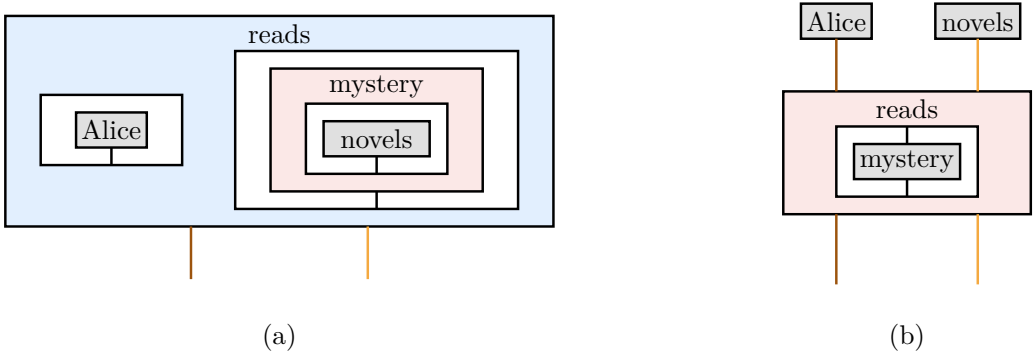
\begin{figure}[h]
\hspace{-0.75cm}
\centering
    \tikzset{every picture/.style={line width=0.75pt}} %set default line width to 0.75pt        
% When embedding into a *.tex file, uncomment and include the following lines:
\pgfdeclarelayer{nodelayer}
\pgfdeclarelayer{edgelayer}
\pgfdeclarelayer{labellayer}
\pgfsetlayers{nodelayer, edgelayer, labellayer}
\begin{tikzpicture}
	\begin{pgfonlayer}{nodelayer}
		\node [] (1) at (-1.5, 2.575) {};
		\node [] (2) at (6.25, 2.575) {};
		\node [] (3) at (6.25, 5.575) {};
		\node [] (4) at (-1.5, 5.575) {};
		\node [] (6) at (-1, 3.45) {};
		\node [] (7) at (1, 3.45) {};
		\node [] (8) at (1, 4.45) {};
		\node [] (9) at (-1, 4.45) {};
		\node [] (10) at (-0.5, 3.7) {};
		\node [] (11) at (0.5, 3.7) {};
		\node [] (12) at (0.5, 4.2) {};
		\node [] (13) at (-0.5, 4.2) {};
		\node [] (15) at (1.75, 2.825) {};
		\node [] (16) at (5.75, 2.825) {};
		\node [] (17) at (5.75, 5.075) {};
		\node [] (18) at (1.75, 5.075) {};
		\node [] (19) at (2.25, 3.075) {};
		\node [] (20) at (5.25, 3.075) {};
		\node [] (21) at (5.25, 4.825) {};
		\node [] (22) at (2.25, 4.825) {};
		\node [] (24) at (2.75, 3.325) {};
		\node [] (25) at (4.75, 3.325) {};
		\node [] (26) at (4.75, 4.325) {};
		\node [] (27) at (2.75, 4.325) {};
		\node [] (28) at (3, 3.575) {};
		\node [] (29) at (4.5, 3.575) {};
		\node [] (30) at (4.5, 4.075) {};
		\node [] (31) at (3, 4.075) {};
		\node [] (33) at (0, 3.7) {};
		\node [] (34) at (0, 3.45) {};
		\node [] (35) at (3.75, 3.575) {};
		\node [] (36) at (3.75, 3.325) {};
		\node [] (37) at (3.75, 3.075) {};
		\node [] (38) at (3.75, 2.825) {};
		\node [] (39) at (1.125, 2.575) {};
		\node [] (40) at (1.125, 1.825) {};
		\node [] (41) at (3.625, 2.575) {};
		\node [] (42) at (3.625, 1.825) {};
		\node [] (44) at (9.75, 5.25) {};
		\node [] (45) at (10.75, 5.25) {};
		\node [] (46) at (10.75, 5.75) {};
		\node [] (47) at (9.75, 5.75) {};
		\node [] (48) at (11.65, 5.25) {};
		\node [] (49) at (12.85, 5.25) {};
		\node [] (50) at (12.85, 5.75) {};
		\node [] (51) at (11.65, 5.75) {};
		\node [] (52) at (9.5, 2.75) {};
		\node [] (53) at (13, 2.75) {};
		\node [] (54) at (13, 4.5) {};
		\node [] (55) at (9.5, 4.5) {};
		\node [] (56) at (10.25, 3) {};
		\node [] (57) at (12.25, 3) {};
		\node [] (58) at (12.25, 4) {};
		\node [] (59) at (10.25, 4) {};
		\node [] (60) at (10.5, 3.25) {};
		\node [] (61) at (12, 3.25) {};
		\node [] (62) at (12, 3.75) {};
		\node [] (63) at (10.5, 3.75) {};
		\node [] (64) at (10.25, 5.25) {};
		\node [] (65) at (10.25, 4.5) {};
		\node [] (66) at (12.25, 5.25) {};
		\node [] (67) at (12.25, 4.5) {};
		\node [] (68) at (11.25, 4) {};
		\node [] (69) at (11.25, 3.75) {};
		\node [] (70) at (11.25, 3.25) {};
		\node [] (71) at (11.25, 3) {};
		\node [] (72) at (10.25, 2.75) {};
		\node [] (73) at (10.25, 1.75) {};
		\node [] (74) at (12.25, 2.75) {};
		\node [] (75) at (12.25, 1.75) {};
	\end{pgfonlayer}
	\begin{pgfonlayer}{edgelayer}
		\draw [-, fill={rgb,255: red,226; green,239; blue,254}, line width=1.0pt] (1.center)
			 to (2.center)
			 to (3.center)
			 to (4.center)
			 to cycle;
		\draw [-, fill={rgb,255: red,255; green,255; blue,255}, line width=1.0pt] (6.center)
			 to (7.center)
			 to (8.center)
			 to (9.center)
			 to cycle;
		\draw [-, fill={rgb,255: red,224; green,224; blue,224}, line width=1.0pt] (10.center)
			 to (11.center)
			 to (12.center)
			 to (13.center)
			 to cycle;
		\draw [-, fill={rgb,255: red,255; green,255; blue,255}, line width=1.0pt] (15.center)
			 to (16.center)
			 to (17.center)
			 to (18.center)
			 to cycle;
		\draw [-, fill={rgb,255: red,251; green,232; blue,231}, line width=1.0pt] (19.center)
			 to (20.center)
			 to (21.center)
			 to (22.center)
			 to cycle;
		\draw [-, fill={rgb,255: red,255; green,255; blue,255}, line width=1.0pt] (24.center)
			 to (25.center)
			 to (26.center)
			 to (27.center)
			 to cycle;
		\draw [-, fill={rgb,255: red,224; green,224; blue,224}, line width=1.0pt] (28.center)
			 to (29.center)
			 to (30.center)
			 to (31.center)
			 to cycle;
		\draw [-, draw={rgb,255: red,0; green,0; blue,0}, line width=1.0pt, in=90, out=-90] (33.center) to (34.center);
		\draw [-, draw={rgb,255: red,0; green,0; blue,0}, line width=1.0pt, in=90, out=-90] (35.center) to (36.center);
		\draw [-, draw={rgb,255: red,0; green,0; blue,0}, line width=1.0pt, in=90, out=-90] (37.center) to (38.center);
		\draw [-, draw={rgb,255: red,156; green,84; blue,14}, line width=1.0pt, in=90, out=-90] (39.center) to (40.center);
		\draw [-, draw={rgb,255: red,244; green,169; blue,64}, line width=1.0pt, in=90, out=-90] (41.center) to (42.center);
		\draw [-, fill={rgb,255: red,224; green,224; blue,224}, line width=1.0pt] (44.center)
			 to (45.center)
			 to (46.center)
			 to (47.center)
			 to cycle;
		\draw [-, fill={rgb,255: red,224; green,224; blue,224}, line width=1.0pt] (48.center)
			 to (49.center)
			 to (50.center)
			 to (51.center)
			 to cycle;
		\draw [-, fill={rgb,255: red,251; green,232; blue,231}, line width=1.0pt] (52.center)
			 to (53.center)
			 to (54.center)
			 to (55.center)
			 to cycle;
		\draw [-, fill={rgb,255: red,255; green,255; blue,255}, line width=1.0pt] (56.center)
			 to (57.center)
			 to (58.center)
			 to (59.center)
			 to cycle;
		\draw [-, fill={rgb,255: red,224; green,224; blue,224}, line width=1.0pt] (60.center)
			 to (61.center)
			 to (62.center)
			 to (63.center)
			 to cycle;
		\draw [-, draw={rgb,255: red,156; green,84; blue,14}, line width=1.0pt, in=90, out=-90] (64.center) to (65.center);
		\draw [-, draw={rgb,255: red,244; green,169; blue,64}, line width=1.0pt, in=90, out=-90] (66.center) to (67.center);
		\draw [-, draw={rgb,255: red,0; green,0; blue,0}, line width=1.0pt, in=90, out=-90] (68.center) to (69.center);
		\draw [-, draw={rgb,255: red,0; green,0; blue,0}, line width=1.0pt, in=90, out=-90] (70.center) to (71.center);
		\draw [-, draw={rgb,255: red,156; green,84; blue,14}, line width=1.0pt, in=90, out=-90] (72.center) to (73.center);
		\draw [-, draw={rgb,255: red,244; green,169; blue,64}, line width=1.0pt, in=90, out=-90] (74.center) to (75.center);
	\end{pgfonlayer}
	\begin{pgfonlayer}{labellayer}
		\node [] (5) at (2.375, 5.325) {reads};
		\node [] (14) at (0, 3.95) {Alice};
		\node [] (23) at (3.75, 4.575) {mystery};
		\node [] (32) at (3.75, 3.825) {novels};
		\node [] (76) at (10.25, 5.5) {Alice};
		\node [] (77) at (12.25, 5.5) {novels};
		\node [] (78) at (11.25, 4.25) {reads};
		\node [] (79) at (11.25, 3.5) {mystery};
		\node [] (78) at (11.25, 1) {(b)};
		\node [] (79) at (2.375, 1) {(a)};
	\end{pgfonlayer}
\end{tikzpicture}
    \caption{The diagram on the left depicts nouns before they are dragged to the top of the diagram, as illustrated on the right.}
    \label{fig:dragging_out}
\end{figure}
%%%%%%%%%%%

%%%%%%%%%%%
\subsection{Composing sentences}\label{subsec:composing_circuits}
%%%%%%%%%%%

A significant advantage of the {\DCC} framework is its compositional nature. This allows us to compose individual diagrams along wires corresponding to the same state. In this section, we describe the process of converting a text into a discourse diagram using the sentence-level diagrams obtained in Section \ref{subsec:processing-frames}.

Practically, shared references to a specific entity within a text can be obtained from a \textit{coreference resolver}.\footnote{We use the coreference models of the spaCy NLP package \citep{spacy}.} Starting with an initial text, the program prepares it for processing by removing extraneous formatting characters. Then, the coreference resolver returns the text indices associated to a given noun entity, as well as a tokenised version of the text.  Now, the {\DCC} diagram is generated dynamically by iterating over the individual sentences in the text. Throughout this iteration, the diagram \texttt{global\textunderscore diagram} and a list \texttt{global\textunderscore nouns} of the current nouns obtained so far are maintained. The algorithm described in Section \ref{subsec:processing-frames} is used to retrieve the sub-diagram \texttt{local\textunderscore diagram} and nouns \texttt{local\textunderscore nouns} for each sentence.

The next step is to append the sentence-level sub-circuit \texttt{local\textunderscore diagram} to the bottom of the compound diagram \texttt{global\textunderscore diagram} using the coreferencing information. If the current sentence does not share common nouns (up to coreferences) with the diagram so far, then additional wires are created and the local diagram is supported on these wires. Formally, this corresponds to tensoring the diagrams: 
\[
\texttt{global\textunderscore diagram}  = \texttt{global\textunderscore diagram} \otimes \texttt{local\textunderscore diagram} 
\]
Otherwise, if the diagrams share common coreferences, we use \texttt{unique[local\textunderscore nouns]} and \texttt{unique\-[global\textunderscore nouns]} to refer to the respective subsets of local and global nouns excluding shared coreferences (note that these may be empty in general). We construct a permutation between the following sets: 
\[
\texttt{global\textunderscore diagram} \cup  \texttt{unique[local\textunderscore nouns]} \longrightarrow \texttt{unique[global\textunderscore nouns]} \cup \texttt{local\textunderscore nouns}
\]
The permutation is converted into a diagram $\pi$ by composing adjacent transpositions of wires to obtain the desired bijection of states. If a single noun entity occurs more than once in \texttt{local\textunderscore diagram}, such as with reflexive verbs, a spider representing this duplication is concatenated to $\pi$. Finally, the sentence-level diagram is recursively appended to the global diagram.
%%%%%%%%%%%

\begin{align*}
\texttt{global\textunderscore diagram}  &= \textcolor{dc143c}{\texttt{global\textunderscore diagram}} \otimes \operatorname{id}(\texttt{unique[local\textunderscore nouns]}) \\
&>> \textcolor{ba55d3}{\pi\circ \operatorname{spiders}} \\
&>> \operatorname{id}(\texttt{unique[global\textunderscore nouns]}) \otimes \textcolor{4169e1}{\texttt{local\textunderscore diagram}})\\
&>> \textcolor{ba55d3}{(\pi\circ \operatorname{spiders})^{\dagger}}
\end{align*}
%%%%%%%%%%%
\begin{figure}[H]
    \centering
    \tikzset{every picture/.style={line width=0.75pt, anchor=base}} %set default line width to 0.75pt

% When embedding into a *.tex file, uncomment and include the following lines:
\pgfdeclarelayer{nodelayer}
\pgfdeclarelayer{edgelayer}
\pgfsetlayers{edgelayer, nodelayer}
\begin{tikzpicture}
	\begin{pgfonlayer}{nodelayer}
		\node [] (124) at (-0.75, 4.25) {};
		\node [] (125) at (6.5, 4.25) {};
		\node [] (126) at (-0.75, 5.5) {};
		\node [] (127) at (6.5, 5.5) {};
		\node [] (114) at (-0.75, 8.5) {};
		\node [] (115) at (6.5, 8.5) {};
		\node [] (116) at (-0.75, 7.25) {};
		\node [] (117) at (6.5, 7.25) {};
		\node [] (97) at (-0.75, 9.75) {};
		\node [] (98) at (3.25, 9.75) {};
		\node [] (99) at (-0.75, 9) {};
		\node [] (100) at (3.25, 9) {};
		\node [] (110) at (1.75, 6.75) {};
		\node [] (111) at (6.5, 6.75) {};
		\node [] (112) at (1.75, 6) {};
		\node [] (113) at (6.5, 6) {};
		\node [] (21) at (-0.5, 9.125) {};
		\node [] (22) at (3, 9.125) {};
		\node [] (23) at (3, 9.625) {};
		\node [] (24) at (-0.5, 9.625) {};
		\node [] (26) at (2, 6.125) {};
		\node [] (27) at (6.25, 6.125) {};
		\node [] (28) at (6.25, 6.625) {};
		\node [] (29) at (2, 6.625) {};
		\node [] (45) at (0, 10.125) {};
		\node [] (46) at (0, 9.625) {};
		\node [] (47) at (2.5, 10.125) {};
		\node [] (48) at (2.5, 9.625) {};
		\node [] (49) at (0, 9.125) {};
		\node [] (50) at (0, 7.875) {};
		\node [] (51) at (2.5, 9.125) {};
		\node [] (52) at (2.5, 7.875) {};
		\node [] (53) at (2.5, 7.375) {};
		\node [] (54) at (2.5, 6.625) {};
		\node [] (55) at (5, 10.125) {};
		\node [] (56) at (4.25, 6.625) {};
		\node [] (57) at (0, 7.375) {};
		\node [] (58) at (0, 5.375) {};
		\node [] (59) at (2.5, 6.125) {};
		\node [] (60) at (2.5, 5.375) {};
		\node [] (63) at (4.25, 6.125) {};
		\node [] (85) at (1.25, 7.625) {};
		\node [] (89) at (7, 8.75) {};
		\node [] (90) at (-1, 8.75) {};
		\node [] (93) at (7, 5.75) {};
		\node [] (94) at (-1, 5.75) {};
		\node [] (119) at (4.25, 7.75) {};
		\node [] (120) at (5.75, 7.75) {};
		\node [] (121) at (5.75, 6.625) {};
		\node [] (122) at (5.75, 6.125) {};
		\node [] (128) at (0, 4.875) {};
		\node [] (129) at (2.5, 4.875) {};
		\node [] (130) at (2.5, 5.375) {};
		\node [] (131) at (0, 5.375) {};
		\node [] (132) at (1.25, 5.125) {};
		\node [] (134) at (4.25, 5) {};
		\node [] (135) at (5.75, 5) {};
		\node [] (136) at (0, 3.375) {};
		\node [] (137) at (2.5, 3.375) {};
		\node [] (138) at (5, 3.375) {};
		\node [fill=black, draw=black, shape=circle, minimum size=4pt, inner sep=0.5ex] (118) at (5, 8.25) {};
		\node [fill=black, draw=black, shape=circle, minimum size=4pt, inner sep=0.5ex] (133) at (5, 4.5) {};
		\node [] (139) at (7, 7) {};
		\node [] (140) at (-1, 7) {};
	\end{pgfonlayer}
	\begin{pgfonlayer}{edgelayer}
		\draw [fill=ba55d3, draw=none, fill opacity=0.5, text opacity=1] (126.center)
			 to (124.center)
			 to (125.center)
			 to (127.center)
			 to cycle;
		\draw [fill=ba55d3, draw=none, fill opacity=0.5, text opacity=1] (116.center)
			 to (114.center)
			 to (115.center)
			 to (117.center)
			 to cycle;
		\draw [fill=dc143c, draw=none, fill opacity=0.5, text opacity=1] (99.center)
			 to (97.center)
			 to (98.center)
			 to (100.center)
			 to cycle;
		\draw [fill=4169e1, draw=none, fill opacity=0.5, text opacity=1] (112.center)
			 to (110.center)
			 to (111.center)
			 to (113.center)
			 to cycle;
		\draw [-, fill={rgb,255: red,224; green,224; blue,224}, line width=1.0pt] (21.center)
			 to (22.center)
			 to (23.center)
			 to (24.center)
			 to cycle;
		\draw [-, fill={rgb,255: red,224; green,224; blue,224}, line width=1.0pt] (26.center)
			 to (27.center)
			 to (28.center)
			 to (29.center)
			 to cycle;
		\draw [-, line width=1.0pt, in=90, out=-90] (45.center) to (46.center);
		\draw [-, line width=1.0pt, in=90, out=-90] (47.center) to (48.center);
		\draw [-, line width=1.0pt, in=90, out=-90] (49.center) to (50.center);
		\draw [-, line width=1.0pt, in=90, out=-90] (51.center) to (52.center);
		\draw [-, line width=1.0pt, in=90, out=-90] (53.center) to (54.center);
		\draw [-, line width=1.0pt, in=90, out=-90] (57.center) to (58.center);
		\draw [-, line width=1.0pt, in=90, out=-90] (59.center) to (60.center);
		\draw [-, line width=1.0pt, in=90, out=180, looseness=0.41] (85.center) to (57.center);
		\draw [-, line width=1.0pt, in=90, out=0, looseness=0.41] (85.center) to (53.center);
		\draw [-, line width=1.0pt, in=180, out=-90, looseness=0.41] (50.center) to (85.center);
		\draw [-, line width=1.0pt, in=0, out=-90, looseness=0.41] (52.center) to (85.center);
		\draw [-, dash pattern=on 0.84pt off 2.51pt] (90.center) to (89.center);
		\draw [-, dash pattern=on 0.84pt off 2.51pt] (94.center) to (93.center);
		\draw [-, dash pattern=on 0.84pt off 2.51pt] (140.center) to (139.center);
		\draw [-, line width=1.0pt] (55.center) to (118);
		\draw [-, line width=1.0pt, in=90, out=-180] (118) to (119.center);
		\draw [-, line width=1.0pt, in=90, out=0] (118) to (120.center);
		\draw [-, line width=1.0pt] (119.center) to (56.center);
		\draw [-, line width=1.0pt] (120.center) to (121.center);
		\draw [-, line width=1.0pt, in=-90, out=-180, looseness=0.41] (132.center) to (131.center);
		\draw [-, line width=1.0pt, in=-90, out=0, looseness=0.41] (132.center) to (130.center);
		\draw [-, line width=1.0pt, in=-180, out=90, looseness=0.41] (128.center) to (132.center);
		\draw [-, line width=1.0pt, in=0, out=90, looseness=0.41] (129.center) to (132.center);
		\draw [-, line width=1.0pt, in=-90, out=180] (133) to (134.center);
		\draw [-, line width=1.0pt, in=-90, out=0] (133) to (135.center);
		\draw [-, line width=1.0pt] (63.center) to (134.center);
		\draw [-, line width=1.0pt] (122.center) to (135.center);
		\draw [-, line width=1.0pt] (128.center) to (136.center);
		\draw [-, line width=1.0pt] (129.center) to (137.center);
		\draw [-, line width=1.0pt] (133) to (138.center);
	\end{pgfonlayer}
\end{tikzpicture}
    \label{fig:todo}
\end{figure}

\begin{figure}[]
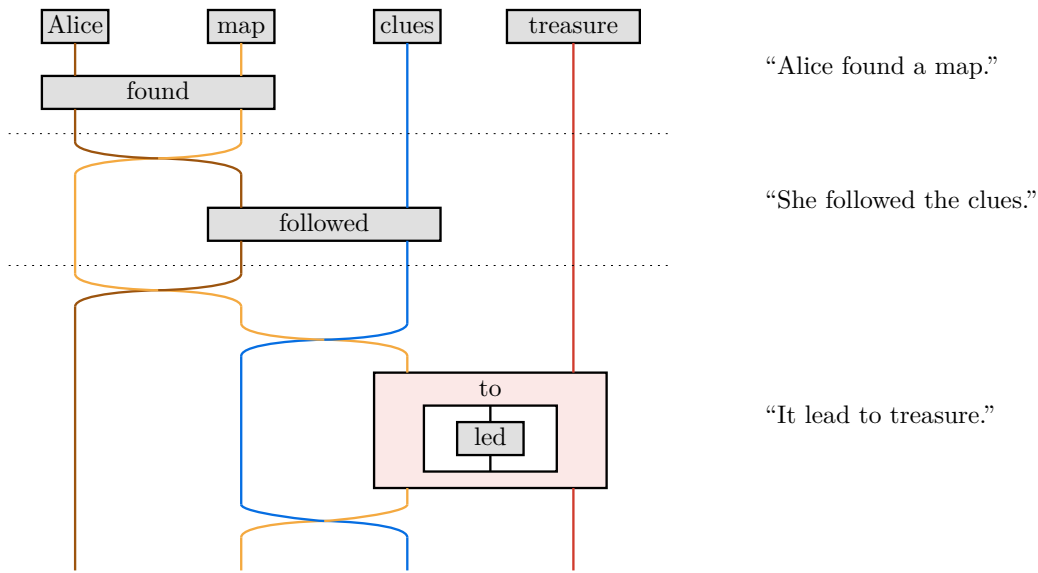

    \centering
    \includestandalone[width=0.9\textwidth]{pipeline/composedSentences}
    \caption{{\DCC} diagram constructed from the text: ``Alice found a map. She followed the clues. It led to treasure''. The partition into individual sentences is denoted by dotted lines.}
    \label{fig:text_diagram}
\end{figure}

We now have an initial version of a discourse diagram that effectively represents both the semantic relationships between the sentences in the text and the grammatical relationships between words within each sentence (Figure \ref{fig:text_diagram}). In the following sections, we shall explore how this diagram can be transformed into a quantum circuit.
 
%%%%%%%%%%%
\section{Simplifying the diagram}\label{sec:simplifications}
%%%%%%%%%%%

Beyond constructing diagrams from text, this work also introduces a range of practical tools for designing and training QNLP experiments. A key feature is the ability to simplify discourse diagrams in various ways, which is particularly valuable when working with large texts.

%%%%%%%%%%%
\subsection{Semantic rewrites}
%%%%%%%%%%%

In practical applications, hardware constraints can limit the depth and complexity of quantum circuits that can be effectively trained. One strategy to mitigate these limitations is through \textit{semantic rewrites}, which either remove or merge boxes in the diagram that satisfy specific grammatical or linguistic conditions.

The motivation behind these simplifications is that not all words contribute equally to task-based reasoning. When text is converted into a PQC, each word is assigned its own unitary operator whose weights are learned through training. Certain parts of speech, like articles (``a'', ``the'', ...) or auxiliary verbs (``has'', ``does'', ...), can appear frequently throughout a text and in varying contexts. Consequently, these words often increase the number of parameters in a circuit without significantly improving the predictive value of the model.  

A semantic rewrite rule is applied to a (word, type) pair to avoid removing polysemous words whose meaning depends on their grammatical type. For example, the word ``have'' functions as an auxiliary verb in the sentence ``I have finished my work'' (type $n^rss^ln$) and as a main verb in ``I have a bike'' (type $n^rsn^l$). In the first case, removing ``have'' would only alter the sentence’s timeline, but in the second case, its removal would render the sentence meaningless. To handle such distinctions, each rewrite rule specifies a list of words to be removed (\texttt{match\_words}) along with their corresponding grammatical types (\texttt{match\_types}).

In practice, these rewrites are implemented at the level of pregroup trees, allowing indexing information for each word to be preserved as an attribute of the tree node. The tree simplifications are performed in a post-order traversal by contracting branches up to a specified depth. If a node satisfies the required merge conditions and has a single child then it is replaced with the child and the number of merges is updated. Each rewrite rule may optionally alter the name of the merged node by concatenating the original node word with that of its child. The process is outlined in Algorithm \ref{alg:rewrites}.

%%%%%%%%%%%
\begin{algorithm}
\caption{Pregroup Tree Rewrites}\label{alg:rewrites}
\begin{algorithmic}[1]
\State \textbf{function} \texttt{rewrite\_tree(node, word\_merger)}
\State \textbf{Output:} Modified tree node after contraction

\State \texttt{word\_mergers} $\gets$ \{ 
    `merge': merge two words with a space, 
    `first': take the first word, 
    `last': take the second word \}

\State \texttt{node.children}, number of merges $\gets$ \texttt{rewrite\_tree(node, word\_merger)} for each \texttt{node.child}

\If{\texttt{node} matches \texttt{match\_type} and has exactly one child and child has the same type as \texttt{node}}
    \If{\texttt{node.word} occurs in \texttt{match\_words}}
        \If{number of merges < max depth}
            \State Replace \texttt{node.word} using appropriate strategy
            \State \textbf{Return:} Modified \texttt{node}, number of merges + 1
        \EndIf
    \EndIf
\EndIf
\State \textbf{Return:} \texttt{node} with updated children
\end{algorithmic}
\end{algorithm}
%%%%%%%%%%%

%%%%%%%%%%%
\begin{figure}[]
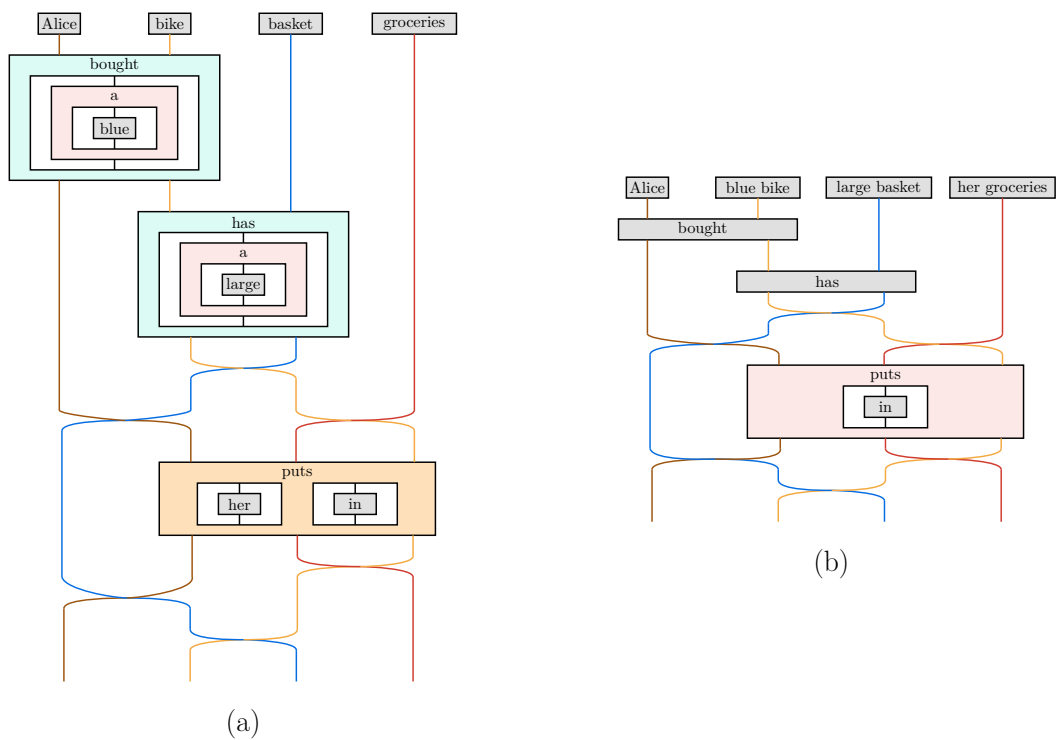

\vspace{-0.1cm}
\centering
    \includestandalone[width=0.9\textwidth]{rewrites/grammarRewrites}
    \caption{(a) The circuit for the text ``Alice bought a blue bike. It has a large basket. She puts her groceries in it''. (b) After applying the \texttt{determiner\textunderscore rule} and a rewrite \texttt{noun\textunderscore modification} that merges adjectives, we get this reduced circuit.}
    \label{fig:grammar_rewrites}
\end{figure}
%%%%%%%%%%%

The code provides a selection of optional rewrites including a \texttt{determiner\textunderscore rule} that removes articles, an \texttt{auxiliary\textunderscore rule} that discards auxiliary verbs, and a \texttt{noun\_modification} rule that concatenates chains of noun modifiers into a single token (e.g. ``beautiful blue bike'')---Figure \ref{fig:grammar_rewrites} provides examples. The performance benefits of these rewrites are task dependent and therefore we supply the flexibility to create tailored simplifications by inheriting an abstract \texttt{PregroupTreeRewriter} class. 

%%%%%%%%%%%
\subsection{Noun filtering}
%%%%%%%%%%%

While grammatical rewrites can be used to reduce the depth of circuits, they have no effect on the \textit{width} of it, or its number of states. When conducting experiments on quantum hardware, the number of states can significantly limit the model’s performance, and even compromise feasibily due to hardware constraints, as it directly determines the number of qubits required. To address these limitations, we provide functionality to \textit{prune} unwanted states in the circuit either manually or automatically by removing nouns that appear less frequently in the text. 

Given a subset of the nouns appearing in the input text (constructed automatically or by hand), our aim is to design a diagram where these nouns do not appear. In particular, we need to remove not only the specific nouns but also any coreferences to them, as well as any boxes that are solely supported on those wires. 

The first step is to calculate the indices of all occurrences of these nouns and their coreferences in the text. This is done by obtaining the first index of each noun and using the coreference resolver to find all related indices. The indexing information reflects the position of a noun within a sentence and the position of that sentence within the overall text. 

As each sentence is converted into a diagram (see Section \ref{subsec:processing-frames}), the indexing information associated to nouns we wish to remove from that sentence are passed to the diagram construction method. The procedure in Section \ref{subsec:composing_circuits} now checks if leaves are among the nouns we would like to prune, in which case an empty diagram is returned (Algorithm \ref{alg:noun_filtering}).

%%%%%%%%%%%
\begin{algorithm}
\caption{Tree to Frames Conversion with Noun Filtering}\label{alg:noun_filtering}
\begin{algorithmic}[1]
\State \textbf{Input:} A tree node \texttt{node}, a list \texttt{remove\textunderscore nouns}
\State \textbf{Output:} Diagram representation as nested frames

\If{node is a \texttt{noun} and has no children and its index is in  \texttt{remove\textunderscore nouns}}
    \State \textbf{Return:} Empty \texttt{Id}, empty list of \texttt{Box}
\EndIf

\[
    \vdots
\]

\end{algorithmic}
\end{algorithm}
%%%%%%%%%%%s

\begin{figure}[]
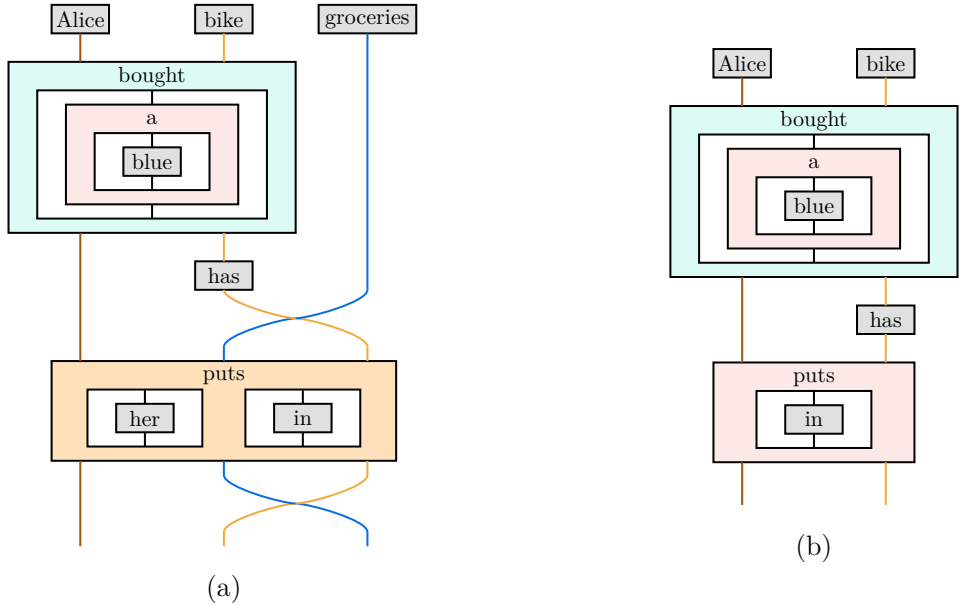

\hspace{-1.75cm}
\centering
\includestandalone[width=0.9\textwidth]{rewrites/nounPruning}
\caption{(a) Using the example in Figure \ref{fig:grammar_rewrites}, one might decide that the state \texttt{basket} has a marginal role in the context of the larger text. Removing this state yields this diagram. (b) Otherwise, requiring that each noun appears at least twice would remove the states \texttt{basket} and \texttt{groceries}, leaving this diagram.}
    \label{fig:noun_pruning}
\end{figure}

This design allows the mechanism that determines the candidate nouns for removal to be customised. We provide one such method by filtering nouns that do not appear with a minimum required frequency in the text. Using tagging information supplied by the coreference resolver, we can determine the frequency with which each noun occurs before the diagrams are constructed. Then, nouns are included in the removal list if they occur less often than the supplied parameter. Figure \ref{fig:noun_pruning} shows examples of noun filtering.

%%%%%%%%%%%
%%%%%%%%%%%

% NK: This section is not necessary IMO and is a relatively minor detail.
% %%%%%%%%%%%
% \subsection{Reindexing}
% %%%%%%%%%%%

% After rewrites, word indices may differ from original tokens. We create a map to virtually update their indexes 

% %%%%%%%%%%%
% \nk{TODO: Nikhil
% \begin{itemize}
%     \item Describe how reindexer is a dictionary that provides virtual indices from the original tokens to their updated position after pruning/rewrites/spacy processing 
% \end{itemize}}
%%%%%%%%%%%

%%%%%%%%%%%
\section{Converting frames into unitaries}\label{subsec:sandwich}
%%%%%%%%%%%

The process we reviewed in Section \ref{sec:text2diagram} delivers a diagram in which states, boxes and wires can be directly mapped to elements of a quantum circuit. However, it does not natively specify how frames should be represented in a quantum setting. In order to provide a corresponding quantum circuit, we need to define a representation of frames in terms of parameterised unitary operators that respects the nested components. 

% For this process we rely on the ``sandwich'' construction described in \citep{Laakkonen_2024}. In order to avoid introducing postselections or discards into the resultant quantum circuits, this representation decomposes frames as a sequence of boxes, interleaved with the contents of the frame, sequentially composed. In our work we extend this decomposition further.

% \nk{TODO: Nikhil
% Talk about Kraus decomposition equivalent to quantum supermaps. Will use to motivate our implementation of the sandwiches}

% As mentioned earlier, the semantic functor that takes {\DCC} diagrams to quantum circuits does not have a direct analogy for frames or quantum supermaps. In order to provide a corresponding quantum circuit we need to define a representation of frames in terms of parameterised unitary operators that respects the nested components. 

The convention we choose to adopt in this paper is to layer these ``sub-circuits'' between pairs of unitaries that are functorially determined from the initial frame. This method is based on the ``sandwich'' construction defined in \citep{Laakkonen_2024}. Explicitly, each frame is assigned a pair of parameterised unitary operators that act on the frame's input states. Then, each nested sub-circuit of the frame is tensored with identity wires for each unused state and sandwiched between the pair of unitaries. This sandwiching operation is repeated for each component in the frame and then composed to give the final circuit.

\begin{figure}[]
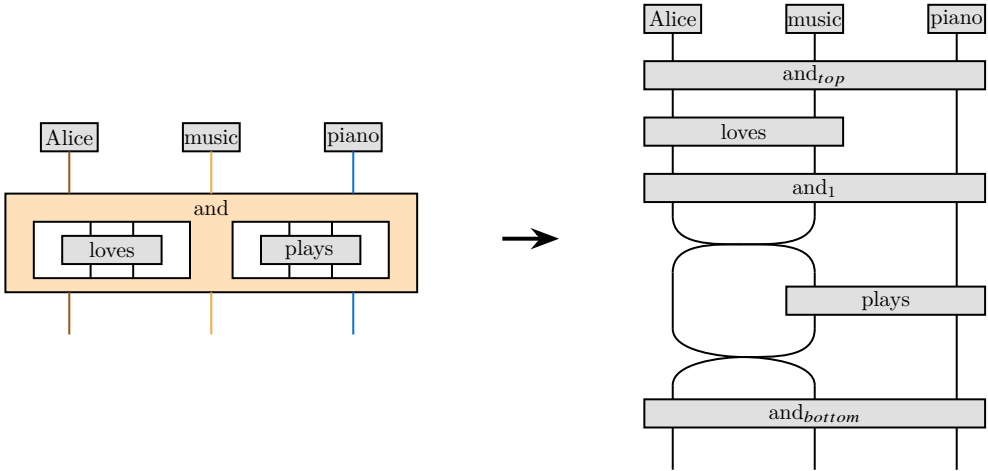

\hspace{-2cm}
\centering
\includestandalone[width=0.9\textwidth]{rewrites/sandwich}
\caption{Example of the sandwich functor being applied to a frame with two nested sub-circuits.}
    \label{fig:sandwich}
\end{figure}

As illustrated in Figure \ref{fig:sandwich}, swaps are applied to internal wires to supply the correct input states to sub-circuits. In the procedure described above, each frame is assigned two parameterised operators whose weights are learned through the training task. Optionally, we also supply an alternative \texttt{foliated\textunderscore frames} convention where every layer in the sandwich is assigned its own independent operator. In specific circumstances, this setup may be preferable as it allows the model to decide how frames modify their components. 

Unlike with frames, each layer in the sandwich requires a bijective correspondence between input and output states. Consequently, when creating the sandwiched sub-circuit, we need to adapt the procedure in Section \ref{subsec:processing-frames} to satisfy this stricter requirement. The first step is to determine the frame node's pregroup type, which can be recovered (up to broken loops) by tensoring \texttt{node.type} with the type attributes of the node's children. After retrieving the node's complete pregroup type, the input and output states can be obtained by tracing the pregroup tree according to the type signature. Conceptually, this corresponds to following paths in the pregroup string diagram back to their source nouns. 

%%%%%%%%%%%
\section{Creating the quantum circuit}\label{subsec:PQCs}
%%%%%%%%%%%

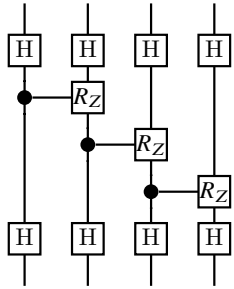
\begin{figure}[]
\centering
\tikzset{every picture/.style={line width=0.75pt}, anchor=base} %set default line width to 0.75pt

\tikzstyle{box}=[-, fill={rgb,255: red,255; green,255; blue,255}, line width=1.0pt]
\tikzstyle{black_wire}=[-, draw={rgb,255: red,0; green,0; blue,0}, line width=1.0pt]
\tikzstyle{control_dot}=[circle, fill=black, minimum size=3pt, inner sep=0.5ex]

% When embedding into a *.tex file, uncomment and include the following lines:
\pgfdeclarelayer{nodelayer}
\pgfdeclarelayer{edgelayer}
\pgfdeclarelayer{labellayer}
\pgfsetlayers{nodelayer, edgelayer, labellayer}
\begin{tikzpicture}
	\begin{pgfonlayer}{nodelayer}
		\node [] (0) at (0, -1) {};
		\node [] (1) at (-0.25, 1.75) {};
		\node [] (2) at (0.25, 1.75) {};
		\node [] (3) at (0.25, 2.25) {};
		\node [] (4) at (-0.25, 2.25) {};
		\node [] (6) at (0.75, 1.75) {};
		\node [] (7) at (1.25, 1.75) {};
		\node [] (8) at (1.25, 2.25) {};
		\node [] (9) at (0.75, 2.25) {};
		\node [] (11) at (1.75, 1.75) {};
		\node [] (12) at (2.25, 1.75) {};
		\node [] (13) at (2.25, 2.25) {};
		\node [] (14) at (1.75, 2.25) {};
		\node [] (16) at (2.75, 1.75) {};
		\node [] (17) at (3.25, 1.75) {};
		\node [] (18) at (3.25, 2.25) {};
		\node [] (19) at (2.75, 2.25) {};
		\node [style={control_dot}] (21) at (0, 1.25) {};
		\node [] (22) at (0.75, 1) {};
		\node [] (23) at (1.25, 1) {};
		\node [] (24) at (1.25, 1.5) {};
		\node [] (25) at (0.75, 1.5) {};
		\node [] (27) at (0, 1) {};
		\node [] (28) at (0, 1.5) {};
		\node [] (29) at (0.75, 1.25) {};
		\node [style={control_dot}] (30) at (1, 0.5) {};
		\node [] (31) at (1.75, 0.25) {};
		\node [] (32) at (2.25, 0.25) {};
		\node [] (33) at (2.25, 0.75) {};
		\node [] (34) at (1.75, 0.75) {};
		\node [] (36) at (1, 0.25) {};
		\node [] (37) at (1, 0.75) {};
		\node [] (38) at (1.75, 0.5) {};
		\node [style={control_dot}] (39) at (2, -0.25) {};
		\node [] (40) at (2.75, -0.5) {};
		\node [] (41) at (3.25, -0.5) {};
		\node [] (42) at (3.25, 0) {};
		\node [] (43) at (2.75, 0) {};
		\node [] (45) at (2, -0.5) {};
		\node [] (46) at (2, 0) {};
		\node [] (47) at (2.75, -0.25) {};
		\node [] (48) at (-0.25, -1.25) {};
		\node [] (49) at (0.25, -1.25) {};
		\node [] (50) at (0.25, -0.75) {};
		\node [] (51) at (-0.25, -0.75) {};
		\node [] (53) at (0.75, -1.25) {};
		\node [] (54) at (1.25, -1.25) {};
		\node [] (55) at (1.25, -0.75) {};
		\node [] (56) at (0.75, -0.75) {};
		\node [] (58) at (1.75, -1.25) {};
		\node [] (59) at (2.25, -1.25) {};
		\node [] (60) at (2.25, -0.75) {};
		\node [] (61) at (1.75, -0.75) {};
		\node [] (63) at (2.75, -1.25) {};
		\node [] (64) at (3.25, -1.25) {};
		\node [] (65) at (3.25, -0.75) {};
		\node [] (66) at (2.75, -0.75) {};
		\node [] (68) at (0, 2.75) {};
		\node [] (69) at (0, 2.25) {};
		\node [] (70) at (1, 2.75) {};
		\node [] (71) at (1, 2.25) {};
		\node [] (72) at (2, 2.75) {};
		\node [] (73) at (2, 2.25) {};
		\node [] (74) at (3, 2.75) {};
		\node [] (75) at (3, 2.25) {};
		\node [] (76) at (0, 1.75) {};
		\node [] (77) at (1, 1.75) {};
		\node [] (78) at (1, 1.5) {};
		\node [] (79) at (1, 1) {};
		\node [] (80) at (2, 1.75) {};
		\node [] (81) at (2, 0.75) {};
		\node [] (82) at (2, 0.25) {};
		\node [] (83) at (3, 1.75) {};
		\node [] (84) at (3, 0) {};
		\node [] (85) at (0, -0.75) {};
		\node [] (86) at (1, -0.75) {};
		\node [] (87) at (2, -0.75) {};
		\node [] (88) at (3, -0.5) {};
		\node [] (89) at (3, -0.75) {};
		\node [] (90) at (0, -1.25) {};
		\node [] (91) at (0, -1.75) {};
		\node [] (92) at (1, -1.25) {};
		\node [] (93) at (1, -1.75) {};
		\node [] (94) at (2, -1.25) {};
		\node [] (95) at (2, -1.75) {};
		\node [] (96) at (3, -1.25) {};
		\node [] (97) at (3, -1.75) {};
	\end{pgfonlayer}
	\begin{pgfonlayer}{edgelayer}
		\draw [style=box] (1.center)
			 to (2.center)
			 to (3.center)
			 to (4.center)
			 to cycle;
		\draw [style=box] (6.center)
			 to (7.center)
			 to (8.center)
			 to (9.center)
			 to cycle;
		\draw [style=box] (11.center)
			 to (12.center)
			 to (13.center)
			 to (14.center)
			 to cycle;
		\draw [style=box] (16.center)
			 to (17.center)
			 to (18.center)
			 to (19.center)
			 to cycle;
		\draw [style=box] (22.center)
			 to (23.center)
			 to (24.center)
			 to (25.center)
			 to cycle;
		\draw [style={black_wire}, in=90, out=-90] (27.center) to (28.center);
		\draw [style={black_wire}, in=180, out=0] (21) to (29.center);
		\draw [style=box] (31.center)
			 to (32.center)
			 to (33.center)
			 to (34.center)
			 to cycle;
		\draw [style={black_wire}, in=90, out=-90] (36.center) to (37.center);
		\draw [style={black_wire}, in=180, out=0] (30) to (38.center);
		\draw [style=box] (40.center)
			 to (41.center)
			 to (42.center)
			 to (43.center)
			 to cycle;
		\draw [style={black_wire}, in=90, out=-90] (45.center) to (46.center);
		\draw [style={black_wire}, in=180, out=0] (39) to (47.center);
		\draw [style=box] (48.center)
			 to (49.center)
			 to (50.center)
			 to (51.center)
			 to cycle;
		\draw [style=box] (53.center)
			 to (54.center)
			 to (55.center)
			 to (56.center)
			 to cycle;
		\draw [style=box] (58.center)
			 to (59.center)
			 to (60.center)
			 to (61.center)
			 to cycle;
		\draw [style=box] (63.center)
			 to (64.center)
			 to (65.center)
			 to (66.center)
			 to cycle;
		\draw [style={black_wire}, in=90, out=-90] (68.center) to (69.center);
		\draw [style={black_wire}, in=90, out=-90] (70.center) to (71.center);
		\draw [style={black_wire}, in=90, out=-90] (72.center) to (73.center);
		\draw [style={black_wire}, in=90, out=-90] (74.center) to (75.center);
		\draw [style={black_wire}, in=90, out=-90] (76.center) to (28.center);
		\draw [style={black_wire}, in=90, out=-90] (77.center) to (78.center);
		\draw [style={black_wire}, in=90, out=-90] (79.center) to (37.center);
		\draw [style={black_wire}, in=90, out=-90] (80.center) to (81.center);
		\draw [style={black_wire}, in=90, out=-90] (82.center) to (46.center);
		\draw [style={black_wire}, in=90, out=-90] (83.center) to (84.center);
		\draw [style={black_wire}, in=90, out=-90] (27.center) to (85.center);
		\draw [style={black_wire}, in=90, out=-90] (36.center) to (86.center);
		\draw [style={black_wire}, in=90, out=-90] (45.center) to (87.center);
		\draw [style={black_wire}, in=90, out=-90] (88.center) to (89.center);
		\draw [style={black_wire}, in=90, out=-90] (90.center) to (91.center);
		\draw [style={black_wire}, in=90, out=-90] (92.center) to (93.center);
		\draw [style={black_wire}, in=90, out=-90] (94.center) to (95.center);
		\draw [style={black_wire}, in=90, out=-90] (96.center) to (97.center);
	\end{pgfonlayer}
	\begin{pgfonlayer}{labellayer}
		\node [] (5) at (0, 1.9) {H};
		\node [] (10) at (1, 1.9) {H};
		\node [] (15) at (2, 1.9) {H};
		\node [] (20) at (3, 1.9) {H};
		\node [] (26) at (1, 1.15) {$R_Z$};
		\node [] (35) at (2, 0.4) {$R_Z$};
		\node [] (44) at (3, -0.35) {$R_Z$};
		\node [] (52) at (0, -1.1) {H};
		\node [] (57) at (1, -1.1) {H};
		\node [] (62) at (2, -1.1) {H};
		\node [] (67) at (3, -1.1) {H};
	\end{pgfonlayer}
\end{tikzpicture}
\caption{IQP on four qubits from \citep{Havlek2018SupervisedLW}. }
    \label{fig:iqpAnsatz}
\end{figure}

The ``sandwiched'' string diagram generated in the previous section can now be directly converted into a quantum circuit by the application of an ansatz. The \texttt{lambeq} library accommodates several ansatz designs that are configurable based on the number of qubits per state and layers per box. For example, the \texttt{IQPAnsatz} class applies the instantaneous quantum polynomial ansatz \citep{Havlek2018SupervisedLW} (pictured in Figure \ref{fig:iqpAnsatz}) to the diagram. This configuration alternates between layers of Hadamard gates and diagonal unitaries constructed from \texttt{num\textunderscore layers - 1} adjacent $CR_Z$ gates. 

% The final step is to convert sandwiched diagrams into parameterised quantum circuits. 

% This is achieved using the \textit{semantic functor} from \cite{Laakkonen_2024} and our variant of the sandwich functor defined in Section \ref{subsec:sandwich}. The semantic functor permits the flexibility to select a choice of ansatz to replace boxes and states in the diagram. 

% The Lambeq library accommodates several ansatz designs that are configurable based on the number of qubits per state and layers per box. For example, the option \texttt{Sim4Ansatz} implements Circuit 4 (pictured in Figure \ref{fig:sim4circuit}) from \cite{Sim_2019}. This architecture consists of successive layers of $R_X$ and $R_Z$ gates followed by $CR_X$ gates on adjacent wires.

%%%%%%%%%%%
%%%%%%%%%%%

\begin{figure}[]
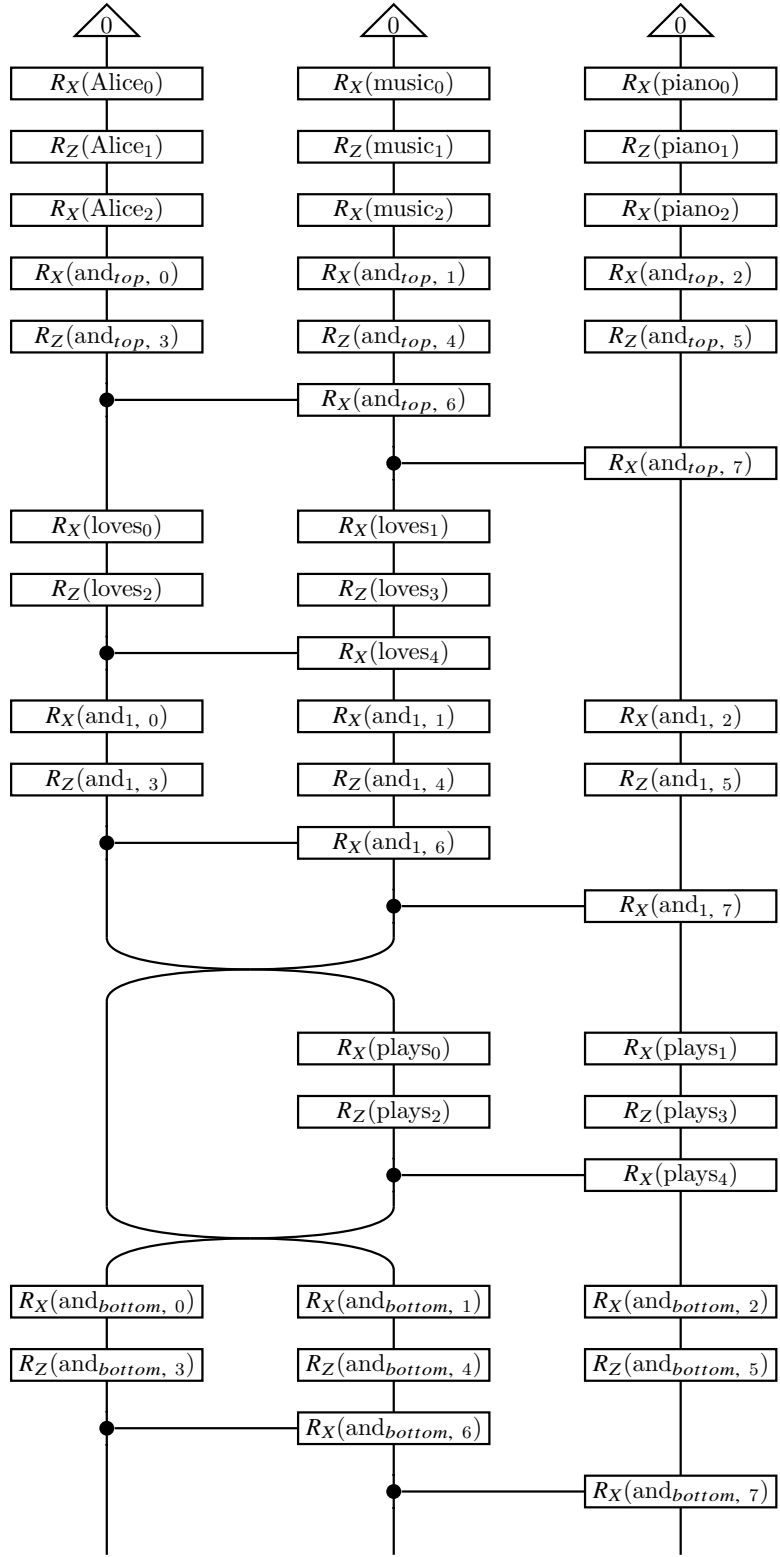

\centering
\includestandalone[width=0.7\textwidth]{rewrites/sim4circuit}
\caption{Sim4 ansatz applied to the text ``Alice loves music and plays piano''. Here \texttt{num\textunderscore layers} is set to one and a single qubit is assigned to each state. }
    \label{fig:sim4circuit}
\end{figure}

Alternatively, the \texttt{Sim4Ansatz} class implements Circuit 4 from \citep{Sim_2019}. This architecture consists of successive layers of $R_X$ and $R_Z$ gates followed by $CR_X$ gates on adjacent wires. Figure \ref{fig:sim4circuit} illustrates this arrangement using the sandwiched diagram from Figure \ref{fig:sandwich} as input.

%%%%%%%%%%%
%%%%%%%%%%%

%%%%%%%%%%%
\section{Performance tests}\label{sec:results}
\graphicspath{ {./tests}}

In this section, we test the performance of our method in two aspects: (a) the width of the coverage it can achieve on standard text based on the number of short documents that can be parsed successfully, and (b) the size of text it can parse. As a baseline, we compare against the \citep{jono} pipeline, on which the first complete set of experiments with DisCoCirc models \citep{Duneau2024ScalableAI} was based. For all the experiments, we used the default Bobcat parser of the \texttt{lambeq} package for getting the original sentence-level pregroup diagrams.

\subsection{Coverage tests}\label{subsec:coverage}

For the coverage tests, we generated two datasets based on the Simplebooks-92 dataset \citep{simplebooks}. The first dataset (\textit{3sents-5to10words}) consists of texts containing three sentences each, with each sentence comprising 5 to 10 words. The second dataset (\textit{5sents-10to30words}) contains texts with five sentences each, where each sentence consists of 10 to 30 words. For both datasets, we created 10 batches of 500 texts each. The sentences are unique across the texts, and the texts are constructed as contiguous sequences of sentences from the original Simplebooks-92 dataset, preserving the original sentence order.

\begin{figure}[h!]
\vspace{1cm}
\centering
\begin{subfigure}{0.32\textwidth}
    \includegraphics[width=\textwidth]{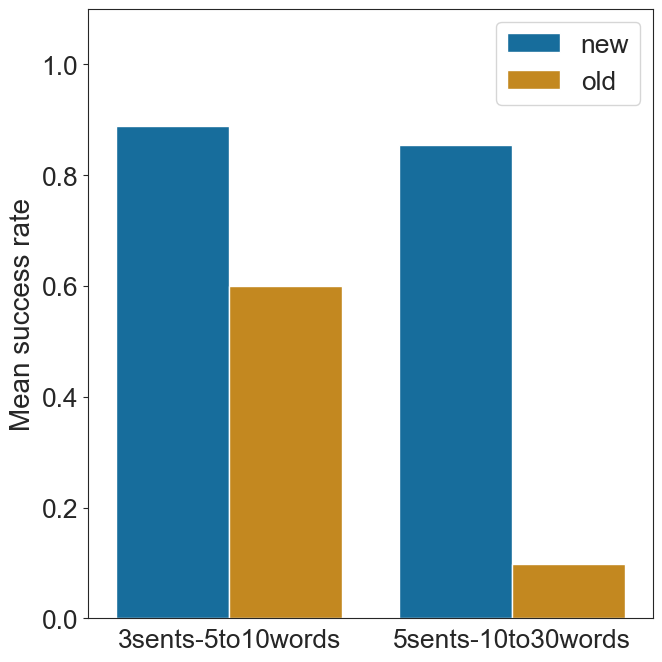}
    \caption{}
\end{subfigure}%
\hspace*{0.01\textwidth}
\begin{subfigure}{0.32\textwidth}
    \includegraphics[width=\textwidth]{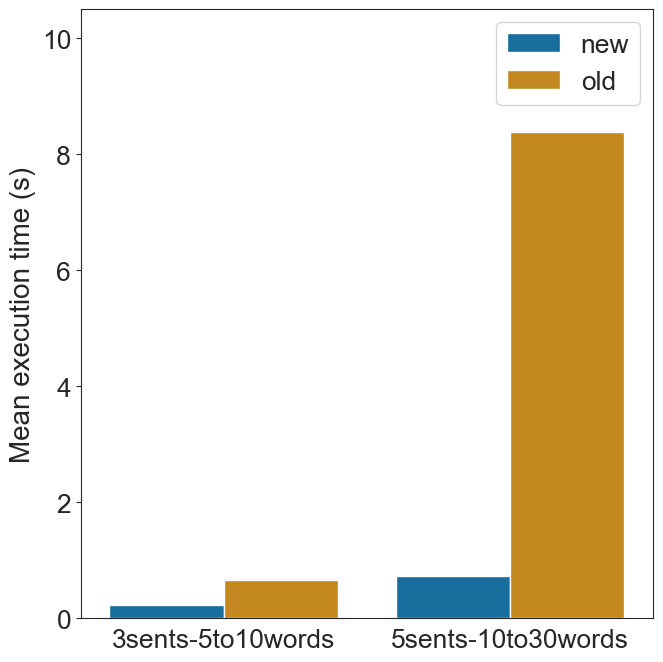}
    \caption{}
\end{subfigure}%
\hspace*{0.01\textwidth}
\begin{subfigure}{0.32\textwidth}
    \includegraphics[width=\textwidth]{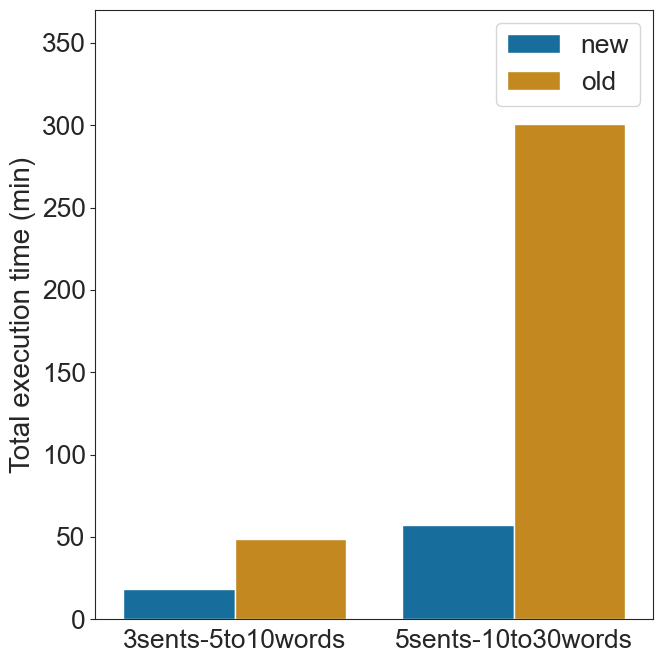}
    \caption{}
\end{subfigure}%
    \caption{Coverage test results. (a) The new pipeline significantly outperforms the old pipeline in terms of mean success rate across both datasets. While the coverage for the old pipeline drops sharply by 84\% with the longer texts in the \textit{5sents-10to30words} dataset, the new pipeline shows only a modest decrease of around 4\%. (b, c) The new pipeline is also substantially faster than the old pipeline for both datasets, demonstrating up to an 11x improvement in processing speed per text and a 5x improvement in overall processing speed per dataset, as observed with the \textit{5sents-10to30words} dataset.}
    \label{fig:coverage}
\end{figure}

We applied both our and the \citep{jono} pipeline to generate diagrams for the texts in the two datasets. For evaluation, we calculated the success rate of the conversion for each batch and reported the mean and standard deviation values. Additionally, we measured the mean execution time per text for successful conversions within each batch, along with the corresponding mean and standard deviation values. Finally, we recorded the total execution time across all batches. The results are summarized in Table \ref{table:coverage} and visualized in Figure \ref{fig:coverage}. All coverage tests were conducted on a 64-core hyper-threaded Intel\textsuperscript{\textregistered} Xeon\textsuperscript{\textregistered} Gold 6326 CPU running at 2.90GHz, with 251GB of RAM.

%% TODO: Fix table width
\begin{table}
    \centering
    \begin{tabularx}{0.8\textwidth}{ccccc} 
      \toprule
       {Pipeline} & {Dataset} & {Mean} & {Mean} & {Total} \\ 
       & & success rate & execution time & execution \\
       & & & per text (s) & time (min) \\
       \midrule
         Liu et al. & \textit{3sents-5to10words} & 0.599$\pm$0.028 & 0.661$\pm$0.582 & 49 \\
         & \textit{5sents-10to30words} & 0.098$\pm$0.023 & 8.373$\pm$6.892 & 301 \\ \midrule
         This work & \textit{3sents-5to10words} & \textbf{0.888}$\pm$0.026 & 0.222$\pm$0.192 & 18.2 \\
         & \textit{5sents-10to30words} & \textbf{0.855}$\pm$0.041 & 0.726$\pm$0.506 & 57.2 \\ \bottomrule
    \end{tabularx}
    \caption{Coverage test results. Both mean success rate and mean execution time per text were computed for each batch and then mean and standard deviation values are reported, while the total execution time was obtained for the entire dataset (either \textit{3sents-5to10words} or \textit{5sents-10to30words}).}
    \label{table:coverage}
\end{table}

\subsection{Crash tests}\label{subsec:crash}

To evaluate the robustness of the new pipeline with significantly larger texts, we created a different type of task, also based on Simplebooks-92. Three datasets were generated, each consisting of one batch of texts. The first dataset, \textit{40sents-min25words}, contains texts with 40 sentences, each having a minimum of 25 words. The second dataset, \textit{100sents-min10words}, includes texts with 100 sentences, each containing a minimum of 10 words. Similarly, the third dataset, \textit{150sents-min10words}, comprises texts with 150 sentences, each with at least 10 words. No maximum word count was imposed on sentences, as doing so would have limited the ability to generate text samples from the original dataset. In all datasets, texts are unique.

For the crash test metrics, we reported the success ratio for each dataset, the word count of the longest text successfully converted, and the corresponding execution time. Additionally, we recorded the mean execution time per text for successful conversions. Results were reported only for the new pipeline, as the old pipeline failed for all datasets. The crash tests were conducted on the same machine used for the coverage tests, and are summarised in Table \ref{table:crash}.

%% TODO: Where to mention that almost all of the errors for both the coverage and crash tests were due to BobcatParser failing and that it's not a bug with the pregroup tree conversion algorithm?

\begin{table}
    \centering
    \begin{tabularx}{0.8\linewidth}{ccccc} 
        \toprule
        {Dataset} & {Success /} & {Mean} & {Maximum} & {Maximum} \\ 
        & Total batch & execution & words & {execution}\\
        & size & time per text & & time (s) \\
        \midrule
        \textit{40sents-min25words} & 2 / 3 & 72.2$\pm$10.1 & 2073 & 82.3 \\
        \textit{100sents-min10words} & 22 / 78 & 1091$\pm$697 & 5109 & 2088 \\
        \textit{150sents-min10words} & 1 / 21 & -- & 6410 & 6088 \\ 
        \bottomrule
    \end{tabularx}
    \caption{Crash test results. The performance of the new pipeline declines when processing texts significantly longer (40-–120 times) than those used in the coverage tests, with the lowest success ratio observed at 1 in 21 sentences for the \textit{150sents-min10words} dataset. Interestingly, the longest text successfully processed by the pipeline was also from this dataset, containing 6,410 words (roughly corresponding to a 13-paged document). Execution time per text shows increasing variability as text length grows. Notably, nearly all errors in both the coverage and crash tests stemmed from Bobcat parser failures in generating the sentence-level pregroup diagrams, which form the starting point of the new pipeline. Reducing the failure rate at this stage would undoubtedly enhance overall performance.}
    \label{table:crash}
\end{table}
%%%%%%%%%%%

\section{Future work and conclusions}
\label{sec:conclusion}

In this paper, we introduced an efficient algorithm for encoding large texts into quantum circuits based on the DisCoCirc model, \db{making it the first practically useful algorithm to do so.} Our implementation demonstrates significant advancements over previous approaches, achieving improved coverage and supporting longer documents -- up to 13 standard 500-word pages in our tests. By integrating this method into the open-source QNLP toolkit, \texttt{lambeq}, we make this promising technology widely accessible to the community. Transforming complete texts into quantum circuits introduces an innovative approach that fundamentally differs from traditional hybrid QML techniques, where only a limited portion of the neural network’s already processed output is passed to the quantum hardware. 

This method represents a significant step toward interpretable models, as it preserves the full compositional structure of the input data within the quantum framework. By doing so, it enables a more transparent mapping between input, processing, and output, allowing for clearer explanations of model behavior. Additionally, this approach facilitates the use of diagrammatic reasoning and structural analysis, enhancing the model’s explainability through well-defined mathematical transformations. This work is ongoing, and future research directions will be shaped by further experimentation on real-world NLP tasks. We hope for, and actively encourage, the participation of the QNLP community in these efforts.

From an implementation standpoint, a key area for future research involves developing more efficient rewriting methods. As demonstrated, text beyond a certain length can result in diagrams that are challenging to interpret and process. Potential solutions include advanced noun-filtering mechanisms driven by statistical models, which prioritize the most prominent semantic entities from the text while omitting less relevant ones. Additionally, linguistically inspired rewriting approaches, such as those proposed in \citep{coecke_2021b, vincent}, \db{in part based on earlier work for DisCoCat \citep{FrobMeanI, Kartsaklis13reasoningabout, CLM}}, offer a promising path to enhance the practicality of DisCoCirc models on real quantum hardware.

% \ck{Acknowledgements:The authors would like to thank Ian Fan and Ragunath Chandrasekharan for their contributions to visualisation tools for {\DCC}.}

\subsection*{Acknowledgements}

The authors would like to thank Ian Fan for his contributions on an early version of this project, as well as Ragunath Chandrasekharan for his work on the visualisation tools for DisCoCirc diagrams. We are also grateful to Anna Pearson and her team for their valuable feedback during the development of this work, to Hamza Waseem and Jonathon Liu for their comments on a draft version of this paper, and to Vincent Wang-Mascianica for useful discussions. The concept of pregroup trees was initially developed as part of a separate, ongoing project by the last two authors, which is yet to be published. 

\indent\hspace{.25in} 
%\nocite{*}
\bibliography{refs} 
\bibliographystyle{plainnat}

\begin{appendices}
% !TEX root = ../main.tex

%%%%%%%%%%%
\section{Conjunction problem}
\label{sec:appendix-conjuction}
%%%%%%%%%%%
In conjunctive sentences, a single subject phrase may apply to multiple objects, delimited by a conjunction like ``and'', as in the example below.
%%%%%%%%%%%
\begin{figure}[H]
\vspace{-3cm}
\centering
\scalebox{0.8}{\tikzset{every picture/.style={line width=0.75pt}, anchor=base} %set default line width to 0.75pt

\tikzstyle{black_wire}=[-, draw={rgb,255: red,0; green,0; blue,0}, line width=1.0pt]
\tikzstyle{box}=[-, fill={rgb,255: red,255; green,255; blue,255}, line width=1.0pt]

% When embedding into a *.tex file, uncomment and include the following lines:
\pgfdeclarelayer{nodelayer}
\pgfdeclarelayer{edgelayer}
\pgfdeclarelayer{labellayer}
\pgfsetlayers{nodelayer, edgelayer, labellayer}
\begin{tikzpicture}
	\begin{pgfonlayer}{nodelayer}
		\node [] (0) at (0, 6.5) {};
		\node [] (1) at (4, 1.5) {};
		\node [] (2) at (4, 1.25) {};
		\node [] (3) at (6, 1.5) {};
		\node [] (4) at (6, 1.25) {};
		\node [] (5) at (11.5, 1.5) {};
		\node [] (6) at (11.5, 1.25) {};
		\node [] (7) at (13.5, 1.5) {};
		\node [] (8) at (13.5, 1.25) {};
		\node [] (9) at (9.21429, 1.5) {};
		\node [] (10) at (9.21429, 1.25) {};
		\node [] (11) at (10.5, 1.5) {};
		\node [] (12) at (10.5, 1.25) {};
		\node [] (13) at (8.92857, 1.5) {};
		\node [] (14) at (8.92857, 1) {};
		\node [] (15) at (11, 1.5) {};
		\node [] (16) at (11, 1) {};
		\node [] (17) at (3.5, 1.5) {};
		\node [] (18) at (3.5, 1) {};
		\node [] (19) at (7.78571, 1.5) {};
		\node [] (20) at (7.78571, 1) {};
		\node [] (21) at (3, 1.5) {};
		\node [] (22) at (3, 1) {};
		\node [] (23) at (8.07143, 1.5) {};
		\node [] (24) at (8.07143, 1) {};
		\node [] (25) at (1, 1.5) {};
		\node [] (26) at (1, 1) {};
		\node [] (27) at (8.35714, 1.5) {};
		\node [] (28) at (8.35714, 1) {};
		\node [] (29) at (8.64286, 1.5) {};
		\node [] (30) at (8.64286, -2) {};
		\node [] (31) at (5, 0.925) {};
		\node [] (32) at (12.5, 1) {};
		\node [] (33) at (9.85714, 1) {};
		\node [] (34) at (9.96429, 0.425) {};
		\node [] (35) at (5.64286, 0.4) {};
		\node [] (36) at (5.53571, -0.05) {};
		\node [] (37) at (4.67857, -0.6) {};
		\node [] (38) at (0, 1.5) {};
		\node [] (39) at (2, 1.5) {};
		\node [] (40) at (2, 2) {};
		\node [] (41) at (1, 2.1) {};
		\node [] (42) at (0, 2) {};
		\node [] (44) at (2.5, 1.5) {};
		\node [] (45) at (4.5, 1.5) {};
		\node [] (46) at (4.5, 2) {};
		\node [] (47) at (3.5, 2.1) {};
		\node [] (48) at (2.5, 2) {};
		\node [] (50) at (5, 1.5) {};
		\node [] (51) at (7, 1.5) {};
		\node [] (52) at (7, 2) {};
		\node [] (53) at (6, 2.1) {};
		\node [] (54) at (5, 2) {};
		\node [] (56) at (7.5, 1.5) {};
		\node [] (57) at (9.5, 1.5) {};
		\node [] (58) at (9.5, 2) {};
		\node [] (59) at (8.5, 2.1) {};
		\node [] (60) at (7.5, 2) {};
		\node [] (62) at (10, 1.5) {};
		\node [] (63) at (12, 1.5) {};
		\node [] (64) at (12, 2) {};
		\node [] (65) at (11, 2.1) {};
		\node [] (66) at (10, 2) {};
		\node [] (68) at (12.5, 1.5) {};
		\node [] (69) at (14.5, 1.5) {};
		\node [] (70) at (14.5, 2) {};
		\node [] (71) at (13.5, 2.1) {};
		\node [] (72) at (12.5, 2) {};
	\end{pgfonlayer}
	\begin{pgfonlayer}{edgelayer}
		\draw [style={black_wire}, in=90, out=-90] (1.center) to (2.center);
		\draw [style={black_wire}, in=90, out=-90] (3.center) to (4.center);
		\draw [style={black_wire}, in=90, out=-90] (5.center) to (6.center);
		\draw [style={black_wire}, in=90, out=-90] (7.center) to (8.center);
		\draw [style={black_wire}, in=90, out=-90] (9.center) to (10.center);
		\draw [style={black_wire}, in=90, out=-90] (11.center) to (12.center);
		\draw [style={black_wire}, in=90, out=-90] (13.center) to (14.center);
		\draw [style={black_wire}, in=90, out=-90] (15.center) to (16.center);
		\draw [style={black_wire}, in=90, out=-90] (17.center) to (18.center);
		\draw [style={black_wire}, in=90, out=-90] (19.center) to (20.center);
		\draw [style={black_wire}, in=90, out=-90] (21.center) to (22.center);
		\draw [style={black_wire}, in=90, out=-90] (23.center) to (24.center);
		\draw [style={black_wire}, in=90, out=-90] (25.center) to (26.center);
		\draw [style={black_wire}, in=90, out=-90] (27.center) to (28.center);
		\draw [style={black_wire}, in=90, out=-90] (29.center) to (30.center);
		\draw [style={black_wire}, in=180, out=-90, looseness=0.75] (2.center) to (31.center);
		\draw [style={black_wire}, in=0, out=-90, looseness=0.75] (4.center) to (31.center);
		\draw [style={black_wire}, in=180, out=-90, looseness=0.75] (6.center) to (32.center);
		\draw [style={black_wire}, in=0, out=-90, looseness=0.75] (8.center) to (32.center);
		\draw [style={black_wire}, in=180, out=-90] (10.center) to (33.center);
		\draw [style={black_wire}, in=0, out=-90] (12.center) to (33.center);
		\draw [style={black_wire}, in=180, out=-90, looseness=1.25] (14.center) to (34.center);
		\draw [style={black_wire}, in=0, out=-90, looseness=1.25] (16.center) to (34.center);
		\draw [style={black_wire}, in=180, out=-90, looseness=0.75] (18.center) to (35.center);
		\draw [style={black_wire}, in=0, out=-90, looseness=0.75] (20.center) to (35.center);
		\draw [style={black_wire}, in=180, out=-90] (22.center) to (36.center);
		\draw [style={black_wire}, in=0, out=-90] (24.center) to (36.center);
		\draw [style={black_wire}, in=180, out=-90] (26.center) to (37.center);
		\draw [style={black_wire}, in=0, out=-90] (28.center) to (37.center);
		\draw [style=box] (38.center)
			 to (39.center)
			 to (40.center)
			 to (41.center)
			 to (42.center)
			 to cycle;
		\draw [style=box] (44.center)
			 to (45.center)
			 to (46.center)
			 to (47.center)
			 to (48.center)
			 to cycle;
		\draw [style=box] (50.center)
			 to (51.center)
			 to (52.center)
			 to (53.center)
			 to (54.center)
			 to cycle;
		\draw [style=box] (56.center)
			 to (57.center)
			 to (58.center)
			 to (59.center)
			 to (60.center)
			 to cycle;
		\draw [style=box] (62.center)
			 to (63.center)
			 to (64.center)
			 to (65.center)
			 to (66.center)
			 to cycle;
		\draw [style=box] (68.center)
			 to (69.center)
			 to (70.center)
			 to (71.center)
			 to (72.center)
			 to cycle;
	\end{pgfonlayer}
    \begin{pgfonlayer}{labellayer}
        \node [] (43) at (1, 1.65) {Alice};
		\node [] (49) at (3.5, 1.65) {loves};
		\node [] (55) at (6, 1.65) {music};
		\node [] (61) at (8.5, 1.65) {and};
		\node [] (67) at (11, 1.65) {plays};
		\node [] (73) at (13.5, 1.65) {piano};
    \end{pgfonlayer}
\end{tikzpicture}}
\end{figure}
%%%%%%%%%%%

For this example pregroup diagram, Algorithm \ref{alg:tree2frame} produces a frame diagram as follows.

\begin{figure}[H]
\centering
\scalebox{0.8}{\tikzset{every picture/.style={line width=0.75pt}} %set default line width to 0.75pt

\tikzstyle{gray_box}=[-, fill={rgb,255: red,224; green,224; blue,224}, line width=1.0pt]
\tikzstyle{fee1ba_box}=[-, fill={rgb,255: red,254; green,225; blue,186}, line width=1.0pt]
\tikzstyle{box}=[-, fill={rgb,255: red,255; green,255; blue,255}, line width=1.0pt]
\tikzstyle{9c540e_wire}=[-, draw={rgb,255: red,156; green,84; blue,14}, line width=1.0pt]
\tikzstyle{f4a940_wire}=[-, draw={rgb,255: red,244; green,169; blue,64}, line width=1.0pt]
\tikzstyle{066ee2_wire}=[-, draw={rgb,255: red,6; green,110; blue,226}, line width=1.0pt]
\tikzstyle{black_wire}=[-, draw={rgb,255: red,0; green,0; blue,0}, line width=1.0pt]

% When embedding into a *.tex file, uncomment and include the following lines:
\pgfdeclarelayer{nodelayer}
\pgfdeclarelayer{edgelayer}
\pgfdeclarelayer{labellayer}
\pgfsetlayers{nodelayer, edgelayer, labellayer}
\begin{tikzpicture}
	\begin{pgfonlayer}{nodelayer}
		\node [] (1) at (0.875, 3.25) {};
		\node [] (2) at (1.875, 3.25) {};
		\node [] (3) at (1.875, 3.75) {};
		\node [] (4) at (0.875, 3.75) {};
		\node [] (6) at (3.375, 3.25) {};
		\node [] (7) at (4.375, 3.25) {};
		\node [] (8) at (4.375, 3.75) {};
		\node [] (9) at (3.375, 3.75) {};
		\node [] (11) at (5.875, 3.25) {};
		\node [] (12) at (6.875, 3.25) {};
		\node [] (13) at (6.875, 3.75) {};
		\node [] (14) at (5.875, 3.75) {};
		\node [] (16) at (0.25, 0.75) {};
		\node [] (17) at (7.5, 0.75) {};
		\node [] (18) at (7.5, 2.5) {};
		\node [] (19) at (0.25, 2.5) {};
		\node [] (21) at (0.75, 1) {};
		\node [] (22) at (3.5, 1) {};
		\node [] (23) at (3.5, 2) {};
		\node [] (24) at (0.75, 2) {};
		\node [] (25) at (1.25, 1.25) {};
		\node [] (26) at (3, 1.25) {};
		\node [] (27) at (3, 1.75) {};
		\node [] (28) at (1.25, 1.75) {};
		\node [] (30) at (4.25, 1) {};
		\node [] (31) at (7, 1) {};
		\node [] (32) at (7, 2) {};
		\node [] (33) at (4.25, 2) {};
		\node [] (34) at (4.75, 1.25) {};
		\node [] (35) at (6.5, 1.25) {};
		\node [] (36) at (6.5, 1.75) {};
		\node [] (37) at (4.75, 1.75) {};
		\node [] (39) at (1.375, 3.25) {};
		\node [] (40) at (1.375, 2.5) {};
		\node [] (41) at (3.875, 3.25) {};
		\node [] (42) at (3.875, 2.5) {};
		\node [] (43) at (6.375, 3.25) {};
		\node [] (44) at (6.375, 2.5) {};
		\node [] (45) at (1.75, 2) {};
		\node [] (46) at (1.75, 1.75) {};
		\node [] (47) at (2.5, 2) {};
		\node [] (48) at (2.5, 1.75) {};
		\node [] (49) at (1.75, 1.25) {};
		\node [] (50) at (1.75, 1) {};
		\node [] (51) at (2.5, 1.25) {};
		\node [] (52) at (2.5, 1) {};
		\node [] (53) at (5.25, 2) {};
		\node [] (54) at (5.25, 1.75) {};
		\node [] (55) at (6, 2) {};
		\node [] (56) at (6, 1.75) {};
		\node [] (57) at (5.25, 1.25) {};
		\node [] (58) at (5.25, 1) {};
		\node [] (59) at (6, 1.25) {};
		\node [] (60) at (6, 1) {};
		\node [] (61) at (1.375, 0.75) {};
		\node [] (62) at (1.375, 0) {};
		\node [] (63) at (3.875, 0.75) {};
		\node [] (64) at (3.875, 0) {};
		\node [] (65) at (6.375, 0.75) {};
		\node [] (66) at (6.375, 0) {};
	\end{pgfonlayer}
	\begin{pgfonlayer}{edgelayer}
		\draw [style={gray_box}] (1.center)
			 to (2.center)
			 to (3.center)
			 to (4.center)
			 to cycle;
		\draw [style={gray_box}] (6.center)
			 to (7.center)
			 to (8.center)
			 to (9.center)
			 to cycle;
		\draw [style={gray_box}] (11.center)
			 to (12.center)
			 to (13.center)
			 to (14.center)
			 to cycle;
		\draw [style={fee1ba_box}] (16.center)
			 to (17.center)
			 to (18.center)
			 to (19.center)
			 to cycle;
		\draw [style=box] (21.center)
			 to (22.center)
			 to (23.center)
			 to (24.center)
			 to cycle;
		\draw [style={gray_box}] (25.center)
			 to (26.center)
			 to (27.center)
			 to (28.center)
			 to cycle;
		\draw [style=box] (30.center)
			 to (31.center)
			 to (32.center)
			 to (33.center)
			 to cycle;
		\draw [style={gray_box}] (34.center)
			 to (35.center)
			 to (36.center)
			 to (37.center)
			 to cycle;
		\draw [style={9c540e_wire}, in=90, out=-90] (39.center) to (40.center);
		\draw [style={f4a940_wire}, in=90, out=-90] (41.center) to (42.center);
		\draw [style={066ee2_wire}, in=90, out=-90] (43.center) to (44.center);
		\draw [style={black_wire}, in=90, out=-90] (45.center) to (46.center);
		\draw [style={black_wire}, in=90, out=-90] (47.center) to (48.center);
		\draw [style={black_wire}, in=90, out=-90] (49.center) to (50.center);
		\draw [style={black_wire}, in=90, out=-90] (51.center) to (52.center);
		\draw [style={black_wire}, in=90, out=-90] (53.center) to (54.center);
		\draw [style={black_wire}, in=90, out=-90] (55.center) to (56.center);
		\draw [style={black_wire}, in=90, out=-90] (57.center) to (58.center);
		\draw [style={black_wire}, in=90, out=-90] (59.center) to (60.center);
		\draw [style={9c540e_wire}, in=90, out=-90] (61.center) to (62.center);
		\draw [style={f4a940_wire}, in=90, out=-90] (63.center) to (64.center);
		\draw [style={066ee2_wire}, in=90, out=-90] (65.center) to (66.center);
	\end{pgfonlayer}
    \begin{pgfonlayer}{labellayer}
        \node [] (5) at (1.375, 3.5) {Alice};
		\node [] (10) at (3.875, 3.5) {music};
		\node [] (15) at (6.375, 3.5) {piano};
		\node [] (20) at (3.875, 2.25) {and};
		\node [] (29) at (2.125, 1.5) {loves};
		\node [] (38) at (5.6, 1.5) {plays};
    \end{pgfonlayer}
\end{tikzpicture}}
\end{figure}

While this diagram is a valid text circuit, it is possible for us to decompose the conjunction to produce a more explicit form, through the use of existing semantic rewrites for pregroup diagrams implemented in \texttt{lambeq}. In particular, we may employ the \texttt{coordination} rewrite rule \citep{kartsaklis2016coordination}, which replaces ``and'' pregroup boxes with explicit internal structure composed of spiders (Frobenius operators) and wiring. Applying the coordination rewrite on the above diagram yields the following diagram.
%%%%%%%%%%%
\begin{figure}[H]
\centering
\scalebox{0.8}{\tikzset{every picture/.style={line width=0.75pt}, anchor=base} %set default line width to 0.75pt

\tikzstyle{black_wire}=[-, draw={rgb,255: red,0; green,0; blue,0}, line width=1.0pt]
\tikzstyle{box}=[-, fill={rgb,255: red,255; green,255; blue,255}, line width=1.0pt]
\tikzstyle{dotted_box}=[-, dash pattern={on 0.84pt off 2.51pt}, fill={rgb,255: red,255; green,255; blue,255}, line width=1.0pt]
\tikzstyle{spider}=[circle, fill=black, minimum size=3pt, inner sep=0.5ex]

% When embedding into a *.tex file, uncomment and include the following lines:
\pgfdeclarelayer{nodelayer}
\pgfdeclarelayer{edgelayer}
\pgfdeclarelayer{labellayer}
\pgfsetlayers{nodelayer, edgelayer, labellayer}
\begin{tikzpicture}
	\begin{pgfonlayer}{nodelayer}
		\node [] (1) at (4, 1.5) {};
		\node [] (2) at (4, 1.25) {};
		\node [] (3) at (6, 1.5) {};
		\node [] (4) at (6, 1.25) {};
		\node [] (5) at (12.5, 1.5) {};
		\node [] (6) at (12.5, 1.25) {};
		\node [] (7) at (14.5, 1.5) {};
		\node [] (8) at (14.5, 1.25) {};
		\node [] (9) at (10.2143, 2.1) {};
		\node [] (10) at (10.2143, 1.25) {};
		\node [] (11) at (11.5, 1.5) {};
		\node [] (12) at (11.5, 1.25) {};
		\node [] (13) at (9.80357, 1.825) {};
		\node [] (14) at (9.80357, 1) {};
		\node [] (15) at (12, 1.5) {};
		\node [] (16) at (12, 1) {};
		\node [] (17) at (3.5, 1.5) {};
		\node [] (18) at (3.5, 1) {};
		\node [] (19) at (7.78571, 2.1) {};
		\node [] (20) at (7.78571, 1) {};
		\node [] (21) at (3, 1.5) {};
		\node [] (22) at (3, 1) {};
		\node [] (23) at (8.19643, 1.825) {};
		\node [] (24) at (8.19643, 1) {};
		\node [] (25) at (1, 1.5) {};
		\node [] (26) at (1, 1) {};
		\node [] (27) at (8.60714, 1.7) {};
		\node [] (28) at (8.60714, 1) {};
		\node [] (29) at (9.39286, 1.7) {};
		\node [] (30) at (9.39286, -2) {};
		\node [] (31) at (5, 0.925) {};
		\node [] (32) at (13.5, 1) {};
		\node [] (33) at (10.8571, 1) {};
		\node [] (34) at (10.9643, 0.425) {};
		\node [] (35) at (5.64286, 0.4) {};
		\node [] (36) at (5.53571, -0.05) {};
		\node [] (37) at (4.67857, -0.6) {};
		\node [] (38) at (0, 1.5) {};
		\node [] (39) at (2, 1.5) {};
		\node [] (40) at (2, 2) {};
		\node [] (41) at (1, 2.1) {};
		\node [] (42) at (0, 2) {};
		\node [] (44) at (2.5, 1.5) {};
		\node [] (45) at (4.5, 1.5) {};
		\node [] (46) at (4.5, 2) {};
		\node [] (47) at (3.5, 2.1) {};
		\node [] (48) at (2.5, 2) {};
		\node [] (50) at (5, 1.5) {};
		\node [] (51) at (7, 1.5) {};
		\node [] (52) at (7, 2) {};
		\node [] (53) at (6, 2.1) {};
		\node [] (54) at (5, 2) {};
		\node [] (56) at (7.5, 1.5) {};
		\node [] (57) at (10.5, 1.5) {};
		\node [] (58) at (10.5, 3) {};
		\node [] (59) at (9, 3.1) {};
		\node [] (60) at (7.5, 3) {};
		\node [] (62) at (11, 1.5) {};
		\node [] (63) at (13, 1.5) {};
		\node [] (64) at (13, 2) {};
		\node [] (65) at (12, 2.1) {};
		\node [] (66) at (11, 2) {};
		\node [] (68) at (13.5, 1.5) {};
		\node [] (69) at (15.5, 1.5) {};
		\node [] (70) at (15.5, 2) {};
		\node [] (71) at (14.5, 2.1) {};
		\node [] (72) at (13.5, 2) {};
		\node [] (75) at (9.61071, 2.05) {};
		\node [] (76) at (8.38929, 2.05) {};
		\node [] (77) at (9, 2.1) {};
		\node [] (78) at (8.39286, 2.3) {};
		\node [] (79) at (9.60714, 2.3) {};
	\end{pgfonlayer}
	\begin{pgfonlayer}{edgelayer}
		\draw [style={dotted_box}] (56.center)
			 to (57.center)
			 to (58.center)
			 to (59.center)
			 to (60.center)
			 to cycle;
		\draw [style={black_wire}, in=90, out=-90] (1.center) to (2.center);
		\draw [style={black_wire}, in=90, out=-90] (3.center) to (4.center);
		\draw [style={black_wire}, in=90, out=-90] (5.center) to (6.center);
		\draw [style={black_wire}, in=90, out=-90] (7.center) to (8.center);
		\draw [style={black_wire}, in=90, out=-90] (9.center) to (10.center);
		\draw [style={black_wire}, in=90, out=-90] (11.center) to (12.center);
		\draw [style={black_wire}, in=90, out=-90] (13.center) to (14.center);
		\draw [style={black_wire}, in=90, out=-90] (15.center) to (16.center);
		\draw [style={black_wire}, in=90, out=-90] (17.center) to (18.center);
		\draw [style={black_wire}, in=90, out=-90] (19.center) to (20.center);
		\draw [style={black_wire}, in=90, out=-90] (21.center) to (22.center);
		\draw [style={black_wire}, in=90, out=-90] (23.center) to (24.center);
		\draw [style={black_wire}, in=90, out=-90] (25.center) to (26.center);
		\draw [style={black_wire}] (27.center) to (28.center);
		\draw [style={black_wire}, in=90, out=-90] (29.center) to (30.center);
		\draw [style={black_wire}, in=180, out=-90, looseness=0.75] (2.center) to (31.center);
		\draw [style={black_wire}, in=0, out=-90, looseness=0.75] (4.center) to (31.center);
		\draw [style={black_wire}, in=180, out=-90, looseness=0.75] (6.center) to (32.center);
		\draw [style={black_wire}, in=0, out=-90, looseness=0.75] (8.center) to (32.center);
		\draw [style={black_wire}, in=180, out=-90] (10.center) to (33.center);
		\draw [style={black_wire}, in=0, out=-90] (12.center) to (33.center);
		\draw [style={black_wire}, in=180, out=-90, looseness=1.25] (14.center) to (34.center);
		\draw [style={black_wire}, in=0, out=-90, looseness=1.25] (16.center) to (34.center);
		\draw [style={black_wire}, in=180, out=-90, looseness=0.75] (18.center) to (35.center);
		\draw [style={black_wire}, in=0, out=-90, looseness=0.75] (20.center) to (35.center);
		\draw [style={black_wire}, in=180, out=-90] (22.center) to (36.center);
		\draw [style={black_wire}, in=0, out=-90] (24.center) to (36.center);
		\draw [style={black_wire}, in=180, out=-90] (26.center) to (37.center);
		\draw [style={black_wire}, in=0, out=-90] (28.center) to (37.center);
		\draw [style=box] (38.center)
			 to (39.center)
			 to (40.center)
			 to (41.center)
			 to (42.center)
			 to cycle;
		\draw [style=box] (44.center)
			 to (45.center)
			 to (46.center)
			 to (47.center)
			 to (48.center)
			 to cycle;
		\draw [style=box] (50.center)
			 to (51.center)
			 to (52.center)
			 to (53.center)
			 to (54.center)
			 to cycle;
		\draw [style=box] (62.center)
			 to (63.center)
			 to (64.center)
			 to (65.center)
			 to (66.center)
			 to cycle;
		\draw [style=box] (68.center)
			 to (69.center)
			 to (70.center)
			 to (71.center)
			 to (72.center)
			 to cycle;
		\draw [style={black_wire}, in=180, out=90, looseness=1.25] (29.center) to (75.center);
		\draw [style={black_wire}, in=0, out=90] (13.center) to (75.center);
		\draw [style={black_wire}, in=180, out=90] (23.center) to (76.center);
		\draw [style={black_wire}, in=0, out=90, looseness=1.25] (27.center) to (76.center);
		\draw [style={black_wire}, in=180, out=90, looseness=0.75] (19.center) to (78.center);
		\draw [style={black_wire}, in=0, out=90, looseness=0.75] (77.center) to (78.center);
		\draw [style={black_wire}, in=180, out=90, looseness=0.75] (77.center) to (79.center);
		\draw [style={black_wire}, in=0, out=90, looseness=0.75] (9.center) to (79.center);
		\draw [style={black_wire}, in=0, out=-90] (77.center) to (27.center);
		\draw [style={black_wire}, in=180, out=-90] (77.center) to (29.center);
	\end{pgfonlayer}
    \begin{pgfonlayer}{labellayer}
        \node [style={spider}] (81) at (8.60714, 1.7) {};
		\node [style={spider}] (82) at (9.39286, 1.7) {};
		\node [] (43) at (1, 1.65) {Alice};
		\node [] (49) at (3.5, 1.65) {loves};
		\node [] (55) at (6, 1.65) {music};
		\node [] (67) at (12, 1.65) {plays};
		\node [] (73) at (14.5, 1.65) {piano};
		\node [] (80) at (9, 2.65) {and};
    \end{pgfonlayer}
\end{tikzpicture}}
\end{figure}
%%%%%%%%%%%
Then, interpreting the Frobenius co-multiplication as a copy-map, we may push the ``Alice'' box through the co-multiplication, obtaining two copies, as in the diagram below.
%%%%%%%%%%%
\begin{figure}[H]
\centering
\scalebox{0.8}{\tikzset{every picture/.style={line width=0.75pt}, anchor=base} %set default line width to 0.75pt

\tikzstyle{black_wire}=[-, draw={rgb,255: red,0; green,0; blue,0}, line width=1.0pt]
\tikzstyle{box}=[-, fill={rgb,255: red,255; green,255; blue,255}, line width=1.0pt]
\tikzstyle{dotted_box}=[-, dash pattern={on 0.84pt off 2.51pt}, fill={rgb,255: red,255; green,255; blue,255}, line width=1.0pt]
\tikzstyle{spider}=[circle, fill=black, minimum size=3pt, inner sep=0.5ex]

% When embedding into a *.tex file, uncomment and include the following lines:
\pgfdeclarelayer{nodelayer}
\pgfdeclarelayer{edgelayer}
\pgfdeclarelayer{labellayer}
\pgfsetlayers{nodelayer, edgelayer, labellayer}
\begin{tikzpicture}
	\begin{pgfonlayer}{nodelayer}
		\node [] (1) at (4.5, 1.5) {};
		\node [] (2) at (4.5, 1.25) {};
		\node [] (3) at (6.5, 1.5) {};
		\node [] (4) at (6.5, 1.25) {};
		\node [] (16) at (13, 1) {};
		\node [] (17) at (4, 1.5) {};
		\node [] (18) at (4, 1) {};
		\node [] (25) at (1.5, 1.5) {};
		\node [] (26) at (1.5, 1.25) {};
		\node [] (27) at (3.50714, 1.5) {};
		\node [] (28) at (3.50714, 1.25) {};
		\node [] (29) at (8.5, 0.25) {};
		\node [] (30) at (8.5, -1.5) {};
		\node [] (31) at (5.5, 0.925) {};
		\node [] (34) at (8.5, 0.25) {};
		\node [] (35) at (8.5, 0.25) {};
		\node [] (37) at (2.50357, 0.925) {};
		\node [] (38) at (0.5, 1.5) {};
		\node [] (39) at (2.5, 1.5) {};
		\node [] (40) at (2.5, 2) {};
		\node [] (41) at (1.5, 2.1) {};
		\node [] (42) at (0.5, 2) {};
		\node [] (44) at (3, 1.5) {};
		\node [] (45) at (5, 1.5) {};
		\node [] (46) at (5, 2) {};
		\node [] (47) at (4, 2.1) {};
		\node [] (48) at (3, 2) {};
		\node [] (50) at (5.5, 1.5) {};
		\node [] (51) at (7.5, 1.5) {};
		\node [] (52) at (7.5, 2) {};
		\node [] (53) at (6.5, 2.1) {};
		\node [] (54) at (5.5, 2) {};
		\node [] (81) at (13.5, 1.5) {};
		\node [] (82) at (13.5, 1.25) {};
		\node [] (83) at (15.5, 1.5) {};
		\node [] (84) at (15.5, 1.25) {};
		\node [] (87) at (13, 1.5) {};
		\node [] (88) at (13, 1) {};
		\node [] (89) at (10.5, 1.5) {};
		\node [] (90) at (10.5, 1.25) {};
		\node [] (91) at (12.5071, 1.5) {};
		\node [] (92) at (12.5071, 1.25) {};
		\node [] (93) at (14.5, 0.925) {};
		\node [] (94) at (11.5036, 0.925) {};
		\node [] (95) at (9.5, 1.5) {};
		\node [] (96) at (11.5, 1.5) {};
		\node [] (97) at (11.5, 2) {};
		\node [] (98) at (10.5, 2.1) {};
		\node [] (99) at (9.5, 2) {};
		\node [] (100) at (12, 1.5) {};
		\node [] (101) at (14, 1.5) {};
		\node [] (102) at (14, 2) {};
		\node [] (103) at (13, 2.1) {};
		\node [] (104) at (12, 2) {};
		\node [] (105) at (14.5, 1.5) {};
		\node [] (106) at (16.5, 1.5) {};
		\node [] (107) at (16.5, 2) {};
		\node [] (108) at (15.5, 2.1) {};
		\node [] (109) at (14.5, 2) {};
	\end{pgfonlayer}
	\begin{pgfonlayer}{edgelayer}
		\draw [style={black_wire}, in=90, out=-90] (1.center) to (2.center);
		\draw [style={black_wire}, in=90, out=-90] (3.center) to (4.center);
		\draw [style={black_wire}, in=90, out=-90] (17.center) to (18.center);
		\draw [style={black_wire}, in=90, out=-90] (25.center) to (26.center);
		\draw [style={black_wire}] (27.center) to (28.center);
		\draw [style={black_wire}, in=90, out=-90] (29.center) to (30.center);
		\draw [style={black_wire}, in=180, out=-90, looseness=0.75] (2.center) to (31.center);
		\draw [style={black_wire}, in=0, out=-90, looseness=0.75] (4.center) to (31.center);
		\draw [style={black_wire}, in=0, out=-90, looseness=0.50] (16.center) to (34.center);
		\draw [style={black_wire}, in=180, out=-90, looseness=0.50] (18.center) to (35.center);
		\draw [style={black_wire}, in=180, out=-90, looseness=0.75] (26.center) to (37.center);
		\draw [style={black_wire}, in=0, out=-90, looseness=0.75] (28.center) to (37.center);
		\draw [style=box] (38.center)
			 to (39.center)
			 to (40.center)
			 to (41.center)
			 to (42.center)
			 to cycle;
		\draw [style=box] (44.center)
			 to (45.center)
			 to (46.center)
			 to (47.center)
			 to (48.center)
			 to cycle;
		\draw [style=box] (50.center)
			 to (51.center)
			 to (52.center)
			 to (53.center)
			 to (54.center)
			 to cycle;
		\draw [style={black_wire}, in=90, out=-90] (81.center) to (82.center);
		\draw [style={black_wire}, in=90, out=-90] (83.center) to (84.center);
		\draw [style={black_wire}, in=90, out=-90] (87.center) to (88.center);
		\draw [style={black_wire}, in=90, out=-90] (89.center) to (90.center);
		\draw [style={black_wire}] (91.center) to (92.center);
		\draw [style={black_wire}, in=180, out=-90, looseness=0.75] (82.center) to (93.center);
		\draw [style={black_wire}, in=0, out=-90, looseness=0.75] (84.center) to (93.center);
		\draw [style={black_wire}, in=180, out=-90, looseness=0.75] (90.center) to (94.center);
		\draw [style={black_wire}, in=0, out=-90, looseness=0.75] (92.center) to (94.center);
		\draw [style=box] (95.center)
			 to (96.center)
			 to (97.center)
			 to (98.center)
			 to (99.center)
			 to cycle;
		\draw [style=box] (100.center)
			 to (101.center)
			 to (102.center)
			 to (103.center)
			 to (104.center)
			 to cycle;
		\draw [style=box] (105.center)
			 to (106.center)
			 to (107.center)
			 to (108.center)
			 to (109.center)
			 to cycle;
	\end{pgfonlayer}
    \begin{pgfonlayer}{labellayer}
        \node [style={spider}] (35) at (8.5, 0.25) {};
        \node [] (43) at (1.5, 1.65) {Alice};
		\node [] (49) at (4, 1.65) {loves};
		\node [] (55) at (6.5, 1.65) {music};
		\node [] (110) at (10.5, 1.65) {Alice};
		\node [] (111) at (13, 1.65) {plays};
		\node [] (112) at (15.5, 1.65) {piano};
    \end{pgfonlayer}
\end{tikzpicture}}
\end{figure}
%%%%%%%%%%%
This diagram represents the Frobenius multiplication of the two coordinated phrases, which can be seen as the equal combination of information from two sources \citep{kartsaklis2016coordination}. For conversion to a text circuit, it is possible to individually prepare a circuit for each coordinated phrase, and compose the resulting circuits as described in Section \ref{subsec:composing_circuits}, adding coreference information for the multiple subject copies introduced by the coordination rewrite.

\begin{figure}[H]
    \centering
    \scalebox{0.80}{% When embedding into a *.tex file, uncomment and include the following lines:
\pgfdeclarelayer{nodelayer}
\pgfdeclarelayer{edgelayer}
\pgfdeclarelayer{labellayer}
\pgfsetlayers{nodelayer, edgelayer, labellayer}
\begin{tikzpicture}[]
\begin{pgfonlayer}{nodelayer}
\node (0) at (0, 0) {};
\node [] (1) at (-0.5, 2.25) {};
\node [] (2) at (0.5, 2.25) {};
\node [] (3) at (0.5, 2.75) {};
\node [] (4) at (-0.5, 2.75) {};
\node [] (6) at (2.0, 2.25) {};
\node [] (7) at (3.0, 2.25) {};
\node [] (8) at (3.0, 2.75) {};
\node [] (9) at (2.0, 2.75) {};
\node [] (11) at (4.5, 2.25) {};
\node [] (12) at (5.5, 2.25) {};
\node [] (13) at (5.5, 2.75) {};
\node [] (14) at (4.5, 2.75) {};
\node [] (16) at (-0.5, 1.25) {};
\node [] (17) at (3.0, 1.25) {};
\node [] (18) at (3.0, 1.75) {};
\node [] (19) at (-0.5, 1.75) {};
\node [] (21) at (2.0, -0.75) {};
\node [] (22) at (5.5, -0.75) {};
\node [] (23) at (5.5, -0.25) {};
\node [] (24) at (2.0, -0.25) {};
\node [] (26) at (0.0, 2.25) {};
\node [] (27) at (0.0, 1.75) {};
\node [] (29) at (2.5, 2.25) {};
\node [] (30) at (2.5, 1.75) {};
\node [] (32) at (0.0, 1.25) {};
\node [] (33) at (0.0, 0.75) {};
\node [] (35) at (2.5, 1.25) {};
\node [] (36) at (2.5, 0.75) {};
\node [] (38) at (2.5, 0.25) {};
\node [] (39) at (2.5, -0.25) {};
\node [] (41) at (5.0, 2.25) {};
\node [] (42) at (5.0, -0.25) {};
\node [] (44) at (0.0, 0.25) {};
\node [] (45) at (0.0, -1.25) {};
\node [] (47) at (2.5, -0.75) {};
\node [] (48) at (2.5, -1.25) {};
\node [] (50) at (0.0, -1.75) {};
\node [] (51) at (0.0, -2.5) {};
\node [] (53) at (2.5, -1.75) {};
\node [] (54) at (2.5, -2.5) {};
\node [] (56) at (5.0, -0.75) {};
\node [] (57) at (5.0, -2.5) {};
\node [] (59) at (1.25, 0.5) {};
\node [] (60) at (1.25, -1.5) {};
\end{pgfonlayer}
\begin{pgfonlayer}{edgelayer}
\draw [-, fill={rgb,255: red,224; green,224; blue,224}, line width=1.0pt] (1.center) to (2.center) to (3.center) to (4.center) to (1.center);
\draw [-, fill={rgb,255: red,224; green,224; blue,224}, line width=1.0pt] (6.center) to (7.center) to (8.center) to (9.center) to (6.center);
\draw [-, fill={rgb,255: red,224; green,224; blue,224}, line width=1.0pt] (11.center) to (12.center) to (13.center) to (14.center) to (11.center);
\draw [-, fill={rgb,255: red,224; green,224; blue,224}, line width=1.0pt] (16.center) to (17.center) to (18.center) to (19.center) to (16.center);
\draw [-, fill={rgb,255: red,224; green,224; blue,224}, line width=1.0pt] (21.center) to (22.center) to (23.center) to (24.center) to (21.center);
\draw [in=90, out=-90, -, draw={rgb,255: red,156; green,84; blue,14}, line width=1.0pt] (26.center) to (27.center);
\draw [in=90, out=-90, -, draw={rgb,255: red,244; green,169; blue,64}, line width=1.0pt] (29.center) to (30.center);
\draw [in=90, out=-90, -, draw={rgb,255: red,156; green,84; blue,14}, line width=1.0pt] (32.center) to (33.center);
\draw [in=90, out=-90, -, draw={rgb,255: red,244; green,169; blue,64}, line width=1.0pt] (35.center) to (36.center);
\draw [in=90, out=-90, -, draw={rgb,255: red,156; green,84; blue,14}, line width=1.0pt] (38.center) to (39.center);
\draw [in=90, out=-90, -, draw={rgb,255: red,6; green,110; blue,226}, line width=1.0pt] (41.center) to (42.center);
\draw [in=90, out=-90, -, draw={rgb,255: red,244; green,169; blue,64}, line width=1.0pt] (44.center) to (45.center);
\draw [in=90, out=-90, -, draw={rgb,255: red,156; green,84; blue,14}, line width=1.0pt] (47.center) to (48.center);
\draw [in=90, out=-90, -, draw={rgb,255: red,156; green,84; blue,14}, line width=1.0pt] (50.center) to (51.center);
\draw [in=90, out=-90, -, draw={rgb,255: red,6; green,110; blue,226}, line width=1.0pt] (56.center) to (57.center);
\draw [in=90, out=-90, -, draw={rgb,255: red,244; green,169; blue,64}, line width=1.0pt] (53.center) to (54.center);
\draw [in=90, out=180, -, draw={rgb,255: red,244; green,169; blue,64}, line width=1.0pt, looseness=0.4118] (59.center) to (44.center);
\draw [in=90, out=0, -, draw={rgb,255: red,156; green,84; blue,14}, line width=1.0pt, looseness=0.4118] (59.center) to (38.center);
\draw [in=180, out=-90, -, draw={rgb,255: red,156; green,84; blue,14}, line width=1.0pt, looseness=0.4118] (33.center) to (59.center);
\draw [in=0, out=-90, -, draw={rgb,255: red,244; green,169; blue,64}, line width=1.0pt, looseness=0.4118] (36.center) to (59.center);
\draw [in=90, out=180, -, draw={rgb,255: red,156; green,84; blue,14}, line width=1.0pt, looseness=0.4118] (60.center) to (50.center);
\draw [in=90, out=0, -, draw={rgb,255: red,244; green,169; blue,64}, line width=1.0pt, looseness=0.4118] (60.center) to (53.center);
\draw [in=180, out=-90, -, draw={rgb,255: red,244; green,169; blue,64}, line width=1.0pt, looseness=0.4118] (45.center) to (60.center);
\draw [in=0, out=-90, -, draw={rgb,255: red,156; green,84; blue,14}, line width=1.0pt, looseness=0.4118] (48.center) to (60.center);
\end{pgfonlayer}
\begin{pgfonlayer}{labellayer}
\node (5) at (0, 2.5) {Alice};
\node  (10) at (2.5, 2.5) {music};
\node  (15) at (5.0, 2.5) {piano};
\node  (20) at (1.25, 1.5) {loves};
\node  (25) at (3.75, -0.5) {plays};
\end{pgfonlayer}
\end{tikzpicture}}
\end{figure}

% !TEX root = ../main.tex

%%%%%%%%%%%%%%%%%%%
\section{Training DisCoCirc models with lambeq\\and Pennylane}
\label{sec:appendix-training}
\graphicspath{ {./AppendixExperiments}}
%%%%%%%%%%%%%%%%%%%

In this section, we apply a standard \texttt{lambeq} training pipeline on DisCoCirc diagrams to solve a simple meaning classification task on paragraphs. In \texttt{lambeq}, DisCocirc diagrams can be trained as any other circuit using the standard trainers and models provided by the \texttt{training} module. For this task, we take advantage of \texttt{lambeq}'s integration with the popular QML package Pennylane \citep{pennylane} and use the \texttt{PennylaneModel} class for the training, backed by \texttt{PyTorchTrainer}. 

The task we attempt is a variation of the meaning classification task introduced in \citep{Lorenz_2023}, the goal of which was to classify simple sentences as related to IT or food. For the purposes of testing discourse diagrams, however, we modify the dataset to consist of 300 small paragraphs of 3 sentences each. For example:

\begin{center}
\begin{tabular}{lll}
Man prepares tasty lunch. & ~~~~~~~~ & Woman writes efficient program.\\
He used to be a good chef. & ~~~~~~~~ & She is a great programmer.\\
He likes trying new recipes. & ~~~~~~~~ & She thrives in solving hard problems.\\
\end{tabular}
\end{center}

The training pipeline consist of the following steps:

\begin{enumerate}
 \item{Convert the text into DisCoCirc diagrams by using the \texttt{DisCoCircReader} class from the \texttt{lambeq.experimental} module.}
 \item{Convert the diagrams into quantum circuits by the application of an ansatz.}
 \item{Create the \texttt{lambeq} model and trainer to be used by providing the necessary information such as optimizer, training rate, and loss function.}
\end{enumerate}

\begin{figure}[h]
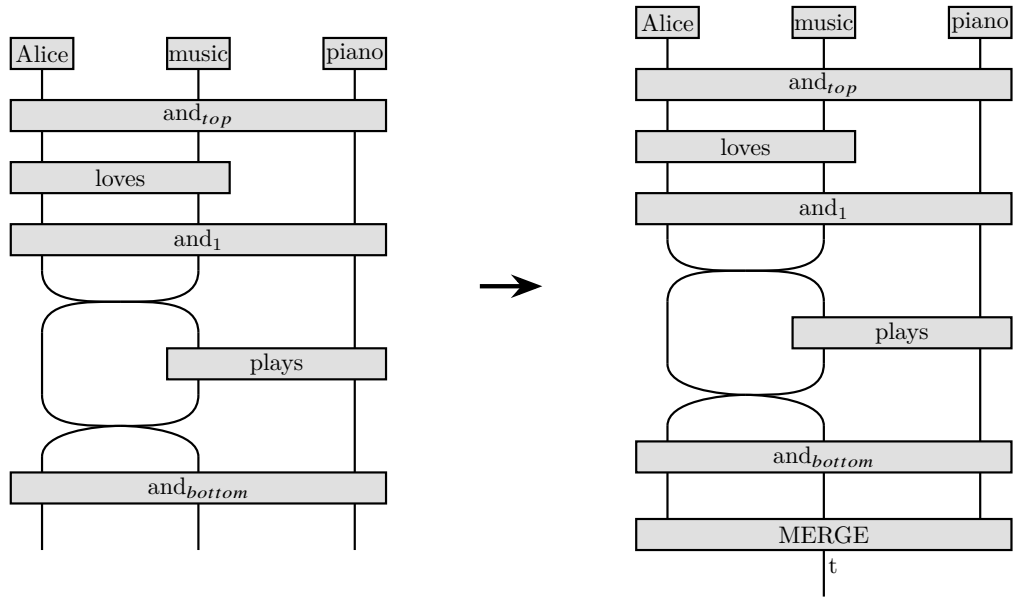

\vspace{1cm}
\centering
\includestandalone[width=0.9\textwidth]{AppendixExperiments/mergeBox}
\caption{Combining the noun wires of a text into a single state.}
\label{fig:merge-box}
\end{figure}

As the output of a DisCoCirc diagram is a set of wires corresponding to the nouns in the text, for our classification task we need a means to combine these noun wires and output a single wire/state for the entire text that we will use for the prediction. This is done by adding at the bottom of the diagram a unitary box with domain the current diagram output and codomain a single wire of type \texttt{t} for ``text'' (Figure \ref{fig:merge-box}). Note that there are multiple versions of this box corresponding to different number of codomain wires in the original diagram.

\begin{figure}[h]
\vspace{1cm}
\centering
\includegraphics[width=0.9\textwidth]{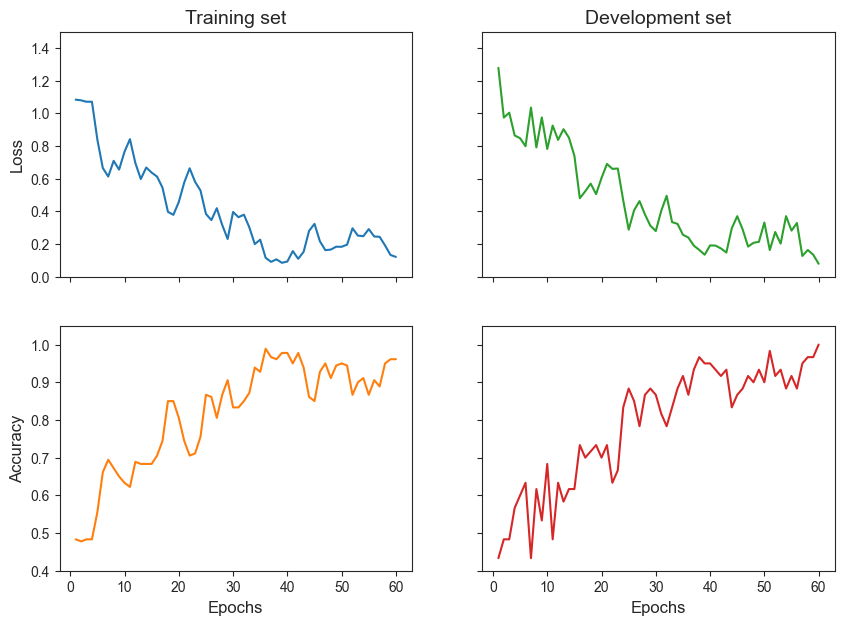}
\caption{Convergence and performance of the DisCoCirc model on the paragraph dataset.}
\label{fig:exp-plots}
\end{figure}

For this experiment, we convert the DisCoCirc diagrams into quantum circuits by applying the \texttt{Sim4Ansatz} (Figure \ref{fig:sim4circuit}) from \citep{Sim_2019}. As mentioned above, we use \texttt{lambeq}'s PennyLane/PyTorch backend for training, with Adam optimiser and a standard binary cross entropy loss function. We train for 60 epochs, with batch size of 10 and learning rate set to 0.01. The model converges smoothly with accuracy on the test set 0.93, and the results are presented in the plots of Figure \ref{fig:exp-plots}.\footnote{For more details, you can check the \texttt{lambeq} tutorial on DisCoCirc training at \url{https://docs.quantinuum.com/lambeq/tutorials/discocirc-mc-task.html}.}

\end{appendices}

\end{document}